\documentclass[a4paper,12pt]{article}
\usepackage{jheppub_modified_green}
\usepackage{caption}

\usepackage{placeins}
\usepackage{longtable}
\usepackage{yfonts}

\usepackage{multirow} 
\usepackage{simpler-wick} 
\usepackage{epsfig}
\usepackage[makeroom]{cancel}
\usepackage{amsfonts}
\usepackage{amssymb,esint}
\usepackage{enumitem}
\usepackage{comment}
\usepackage[normalem]{ulem}
\usepackage{amsthm}
\usepackage{subfig}
\usepackage{bm}
\usepackage{hyperref}
\usepackage{mathrsfs}
\usepackage{bbm}
\usepackage{cancel}
\usepackage[dvipsnames]{xcolor}
\usepackage[table,dvipsnames]{xcolor}
\usepackage{relsize}
\usepackage{float, slashed, graphicx, amssymb, amsmath}
\usepackage{pgfplots}
\usepackage{tikz-feynman}
\usepackage{tikz}
\usepackage{silence}
\def\dsqcup{\sqcup\mathchoice{\mkern-3mu}{\mkern-3mu}{\mkern-3.2mu}{\mkern-3.8mu}\sqcup}
\pgfplotsset{compat=1.18}
\tikzfeynmanset{compat=1.1.0}
\tikzfeynmanset{warn luatex=false}
\tikzfeynmanset{graviton/.style={circle, draw=green!60, fill=green!5, very thick, minimum size=7mm}}
\tikzfeynmanset{codot/.style={/tikz/shape=circle,/tikz/fill=white,/tikz/minimum size=0.1cm,/tikz/inner sep=1.8pt}}
\tikzfeynmanset{myblob/.style={/tikz/shape=rectangle,/tikz/fill=red,/tikz/minimum size=0.2cm,/tikz/inner sep=1.8pt} }
\tikzfeynmanset{ghc/.style={/tikz/shape=circle,/tikz/fill=white,/tikz/minimum size=0.05cm,} }
\tikzfeynmanset{HV/.style=
{/tikz/shape=circle,
/tikz/fill={rgb:black,1;white,2},
 /tikz/minimum size=0.01cm,/tikz/inner sep=3.0pt
 } }
\tikzfeynmanset{GR/.style={/tikz/shape=ellipse,/tikz/fill={rgb:black,1;white,2},/tikz/minimum size=0.3cm,} }
\tikzfeynmanset{GR2/.style={/tikz/shape=ellipse,/tikz/fill={rgb:black,1;white,2},/tikz/minimum width = 1.2cm, 
    /tikz/minimum height = 3.4cm} }
\tikzset{box/.pic={\filldraw[fill=black]  (0,0) circle (2.5pt); \filldraw [fill=black] (0.5,0) circle (2.5pt); \draw [line width=5pt] (0,0) -- (0.5,0);}}
 \tikzfeynmanset{myblob2/.style=
{/tikz/shape=rectangle,
/tikz/fill=black,
 /tikz/minimum width=0.1cm,/tikz/inner sep=1.8pt
 } }
 \tikzfeynmanset{sb/.style=
{/tikz/shape=circle,
/tikz/fill=black,
 /tikz/minimum size=0.3cm,/tikz/inner sep=1.pt
 } }
 \tikzfeynmanset{myblob/.style=
{/tikz/shape=ellipse,
/tikz/fill=red,
 /tikz/minimum width=0.5cm,
 } }

\usetikzlibrary{arrows.meta} 
\usetikzlibrary{calc}
\usetikzlibrary{decorations.pathmorphing}
\usetikzlibrary{decorations.pathreplacing}
\usetikzlibrary{decorations.markings}

\newcommand{\ComptonBlob}[1]{%
\vcenter{\hbox{%
\begin{tikzpicture}[thick,scale=0.75]
    \draw (-1.5,0) -- (-0.6,0);
    \draw (0.6,0) -- (1.5,0);

    \node[above] at (-1.05,0) {$p$};
    \node[above] at (1.05,0) {$p'$};

    \draw[fill=white] (0,0) circle (0.6);
    \node at (0,0) {$#1$};

    \draw[decorate, decoration={snake, amplitude=0.5mm, segment length=2mm}]
        (-0.35,-0.45) -- (-0.9,-1.25) node[below] {$q_1$};

    \draw[decorate, decoration={snake, amplitude=0.5mm, segment length=2mm}]
        (0.35,-0.45) -- (0.9,-1.25) node[below] {$q_n$};

    \node at (0,-1.15) {$\cdots$};
\end{tikzpicture}%
}}%
}
\tikzset{
   vector2/.style={decorate, decoration={snake, amplitude=1pt, segment length=6pt}, draw,double},
   vector/.style={decorate, decoration={snake, amplitude=1pt, segment length=6pt}, draw},
	provector/.style={decorate, decoration={snake,amplitude=2.5pt}, draw},
	antivector/.style={decorate, decoration={snake,amplitude=-2.5pt}, draw},
    fermion/.style={draw=black, postaction={decorate},
        decoration={markings,mark=at position .55 with {\arrow[draw=black]{>}}}},
    fermionbar/.style={draw=black, postaction={decorate},
        decoration={markings,mark=at position .55 with {\arrow[draw=black]{<}}}},
    fermionnoarrow/.style={draw=black},
    gluon/.style={decorate, draw=black,
        decoration={coil,amplitude=4pt, segment length=5pt}},
    scalar/.style={dashed,draw=black, postaction={decorate},
        decoration={markings,mark=at position .55 with {\arrow[draw=black]{>}}}},
    scalarbar/.style={dashed,draw=black, postaction={decorate},
        decoration={markings,mark=at position .55 with {\arrow[draw=black]{<}}}},
    scalarnoarrow/.style={dashed,draw=black},
    electron/.style={draw=black, postaction={decorate},
        decoration={markings,mark=at position .55 with {\arrow[draw=black]{>}}}},
	bigvector/.style={decorate, decoration={snake,amplitude=4pt}, draw},
massive/.style={},
	R/.style={blue},
	A/.style={green},
	cut/.style={postaction={decorate, decoration={markings, mark=at position 0.5 with {
					\draw[opacity=1,red] (0,#1*-2pt) -- (0,#1*2pt);
	}}}},
	cut/.default=1,
	Rnew/.style={postaction={decorate, decoration={markings, mark=at position #1 with {
					\arrow[xshift=3pt]{Latex}
	}}}},
    Rnew/.default=0.5,
	Anew/.style={postaction={decorate, decoration={markings, mark=at position #1 with {
					\arrowreversed[xshift=-3pt]{Latex}
	}}}},
    Anew/.default=0.5,
    thetaLineOld/.style={postaction={decorate, decoration={markings,
			mark=at position 0 with {\node (s) at (0,1pt*#1){};},
			mark=at position 0.999  with {\node (e) at (0,1pt*#1){};},
			mark=at position 1 with {\draw[red] (s.center) -- (e.center);}
		},
	}},
	thetaLineOld/.default=1,
	bothLines/.style={decoration={markings,
				mark=at position 0 with {\node (s1) at (0,0.5pt*#1){};\node (s2) at (0,-0.5pt*#1){};},
				mark=at position 0.999  with {\node (e1) at (0,0.5pt*#1){};\node (e2) at (0,-0.5pt*#1){};},
				mark=at position 1 with {\draw[red] (s1.center) -- (e1.center); \draw[black] (s2.center)-- (e2.center);} 
			},
	},
	bothLines/.default=1,
    causArrow/.style={postaction={decorate}, decoration={markings, mark=at position #1 with{\arrow[xshift=3pt,scale=0.7]{Latex}}}},
    causArrow/.default=0.5,
    causArrowR/.style={postaction={decorate}, decoration={markings, mark=at position #1 with{\arrow[xshift=-3pt,scale=-0.7]{Latex}}}},
    causArrowR/.default=0.5,
    noncausArrow/.style={},
    thetaLine/.style={thin, red, causArrow=#1},
    thetaLine/.default=0.5
}
\tikzset{cross/.style={cross out, draw, 
         minimum size=2*(#1-\pgflinewidth), 
         inner sep=0pt, outer sep=0pt}}

\tikzstyle{block} = [draw, rectangle, 
    minimum height=3em, minimum width=6em]

\usepackage[T1]{fontenc} 

\def\sc#1{\overline{#1}}
\makeatletter
\newcommand \UPlus {\mathop {\operator@font \uplus }\limits }
\makeatother
\makeatletter
\newcommand \Bigcup {\mathop {\operator@font \bigcup }\limits }
\makeatother
\def\LabelNote#1{}
\def\Label#1{\label{#1}%
\smash{\hbox to\phipt{\raise1ex\hbox{\tiny[#1]}\hss}}}

\newcommand{\shuffle}{\,{\dsqcup}\,}

\definecolor{bananayellow}{rgb}{1.0, 0.88, 0.21}
\definecolor{amber}{rgb}{1.0, 0.75, 0.0}

\newcommand{\cL}{\mathcal{L}}
\newcommand{\cO}{\mathcal{O}}

\newcommand{\eps}{\epsilon}
\newcommand{\vareps}{\varepsilon}

\newcommand{\vb}{\bar v}
\newcommand{\pb}{\bar p}

\newcommand{\pbar}{\bar{p}}

\newcommand{\seee}{\beta}

\DeclareMathOperator{\LE}{LE}

\def\nn{\nonumber}

\def\spa#1.#2{\left\langle#1\,#2\right\rangle}
\def\spb#1.#2{\left[#1\,#2\right]}

\def\be{\begin{equation}}
\def\ee{\end{equation}}
\def\bea{\begin{eqnarray}}
\def\eea{\end{eqnarray}}  
\allowdisplaybreaks

\usepackage{amssymb,amsmath}
\usepackage{mathtools} 
\usepackage{cancel} 
\usepackage{graphicx} 

\usepackage{hyperref}
\hypersetup{
	colorlinks=true,
	linktoc=page,
	citecolor=americanrose,
	linkcolor=cadmiumgreen,
	urlcolor=blue} 
\definecolor{americanrose}{rgb}{1.0, 0.01, 0.24}
\definecolor{cadmiumgreen}{rgb}{0.0, 0.42, 0.24}

\usepackage[colorinlistoftodos]{todonotes}

\newcommand{\Cdot}{{\cdot}} 
\makeatletter
\newcommand*{\bigcdot}{}
\DeclareRobustCommand*{\bigcdot}{%
  \mathbin{\mathpalette\bigcdot@{}}%
}
\newcommand*{\bigcdot@scalefactor}{.6}
\newcommand*{\bigcdot@widthfactor}{1.25}
\newcommand*{\bigcdot@}[2]{%
  \sbox0{$#1\vcenter{}$}
  \sbox2{$#1\cdot\m@th$}%
  \hbox to \bigcdot@widthfactor\wd2{%
    \hfil
    \raise\ht0\hbox{%
      \scalebox{\bigcdot@scalefactor}{%
        \lower\ht0\hbox{$#1\bullet\m@th$}%
      }%
    }%
    \hfil
  }%
}
\makeatother
\def\nn{\nonumber}

\newcommand{\sgn}{\text{sgn}}

\newcommand{\cmark}{\textcolor{green!60!black}{\checkmark}}
\newcommand{\xmark}{\textcolor{red!75!black}{\times}}
\newcommand{\shadecell}{\cellcolor{gray!20}}

\title{
The classical limit of the Magnus expansion}
\author{Andreas Brandhuber$\mbox{}^{a}$,}
\author{Graham R.~Brown$\mbox{}^{b}$,}
\author{S\'{e}bastien MacDonald$\mbox{}^{a}$,}
\author{\\Paolo Pichini$\mbox{}^{a}$,}
\author{Gabriele Travaglini$\mbox{}^{a}$}
\author{and Pablo Vives Matasan$\mbox{}^{a}$}
\affiliation{$\mbox{}^{a}$Centre for Theoretical Physics and Astronomy, Department of Physics and Astronomy, \\
Queen Mary University of London, Mile End Road, London E1 4NS, United Kingdom}
\affiliation{$\mbox{}^{b}$Higgs Centre for Theoretical Physics, School of Physics and Astronomy, \\
The University of Edinburgh, Edinburgh EH9 3JZ, Scotland, United Kingdom}
\emailAdd{a.brandhuber@qmul.ac.uk}
\emailAdd{graham.brown@ed.ac.uk}
\emailAdd{s.t.macdonald@qmul.ac.uk}
\emailAdd{p.pichini@qmul.ac.uk}
\emailAdd{g.travaglini@qmul.ac.uk}
\emailAdd{p.vivesmatasan@qmul.ac.uk}
\begin{document}
\vspace*{-1.5cm}
\begin{flushright}
	QMUL-PH-26-31
\end{flushright}

\abstract{We uncover a connection between the classical limit of the Magnus expansion in quantum field theory and the Malvenuto-Reutenauer Hopf algebra of permutations. We study a cubic scalar theory describing a massive particle coupled to massless scalar quanta, which captures the propagator structure relevant to classical  scattering in gravity and scalar QED. Using a Schwinger parametrisation of the massive propagators, 
we compute the classical limit of Magnus amplitudes at tree level in this theory. First, we prove that all hyperclassical contributions cancel at the level of the  Schwinger proper-time integrand, before integration.
Then, we show that the classical Magnus amplitude precisely matches expectations from worldline quantum field theory. 
Both results  rely on cancellations of disconnected diagrams 
which we relate to the Hopf algebra, specifically to the action of the 
adjoint first Eulerian projector, which annihilates the shuffle products associated with disconnected contributions. 
Moreover, the worldline diagrams include generic directed-tree topologies, whereas the field-theory diagrams only include directed chains. The correspondence we establish between the two therefore leads to a new formula for generic Murua coefficients in terms of simpler directed chain coefficients. 
Finally, we 
observe a suggestive  connection between Magnus amplitudes and  the web-mixing matrices governing the exponentiation of multiple Wilson lines.
}
\vspace{-2.6cm}

\maketitle

\flushbottom
 \tableofcontents
\newpage 


\section{Introduction and motivation}
\label{sec:intro}

The $S$-matrix is a central object encoding the observable content of a quantum field theory. Traditionally, perturbative calculations of the $S$-matrix rely on the Dyson series~\cite{Dyson:1949bp}, which expresses the time-evolution operator as a sum of time-ordered products of the interaction Hamiltonian $H_I$. A less explored alternative formulation is provided by the Magnus expansion~\cite{Magnus:1954zz},
in which the same operator is written as 
\begin{equation}
\label{eq:intro_Nop}
  \hat S=\exp(i\hat{N}/\hbar)\, ,
\end{equation}
where the  exponent, or $N$-operator, is constructed from nested commutators of $H_I$.%
\footnote{The Magnus expansion has a long history in quantum mechanics and chemical physics \cite{Pechukas1966,Blanes2009,Blanes_2010}, and has also found applications in the study of Feynman integrals \cite{Argeri:2014qva}.}
 The two
expansions already differ at second order: while the Dyson series contains
the time-ordered product $T\big\{H_I(t_1)H_I(t_2)\big\}$, the Magnus expansion for the $N$-operator contains
$\theta(t_1-t_2)\,\big[H_I(t_1),H_I(t_2)\big]$, trading time-ordering for commutators
that combine with $\theta$-functions into retarded and advanced propagators.

This organisation has two immediate advantages. 
First,  since $\hat N$ is Hermitian at any finite order,
unitarity is manifest term by term. 
 Second, each commutator of interaction Hamiltonians is of order $\hbar$. 
 This suggests  on general grounds that the $N$-operator is free of terms with negative powers of $\hbar$, known as hyperclassical contributions. Such terms encode iterations of lower-order classical interactions rather than genuinely new classical information, and appear in the conventional amplitude-based approaches. The classical contributions 
 to observables, such as impulses, scattering angles and waveforms, are then extracted through a careful $\hbar\to0$ expansion and 
emerge only after intricate cancellations of hyperclassical terms among many individual diagrams \cite{Kosower:2018adc,Kalin:2019rwq}. 
The nested commutator structure of the $N$-operator instead provides an organisation in which these iterative contributions are 
absent in the
final result. 
$N$-matrix elements, which we call \emph{Magnus amplitudes}, therefore provide a natural starting point for isolating classical information directly \cite{Brandhuber:2025igz,Kim:2024svw}.
 In particular, the relation between the radial action and the four-point scattering amplitude, which  was made precise in \cite{Bern:2021dqo},  can be extracted directly from the four-point $N$-matrix element \cite{Damgaard:2021ipf,Damgaard:2023ttc}.  
 This establishes a close relation between the Magnus description, the gravitational eikonal \cite{DiVecchia:2023frv}, and worldline quantum field theory (WQFT) \cite{Mogull:2020sak,Jakobsen:2021smu,Jakobsen:2022psy,Haddad:2025cmw}.%
 \footnote{The Magnus and eikonal constructions are nevertheless conceptually distinct: the eikonal phase is obtained by taking the logarithm of an $S$-matrix element, whereas an $N$-matrix element is obtained by taking a matrix element of $\log\hat S$.}
 The Magnus expansion has also been recently applied to the computation of gravitational Compton scattering at high loop orders in WQFT \cite{Bautista:2026qse,Brunello:2026rdk,Brunello:2026lzf}.

At the diagrammatic level, however, this cancellation is non-trivial. Individual field theory contributions contain hyperclassical terms, and it is far from manifest how these terms cancel.  
Understanding this cancellation is the first aim of our work.
We study this question in the simplest non-trivial setting: the $(n+2)$-point $N$-matrix element for the emission of $n$ massless scalar quanta from a massive scalar line, which we refer to as a Magnus Compton amplitude.
This toy model strips away spin, colour and polarisation degrees of freedom, while its interaction vertices carry no momentum dependence: all non-trivial kinematic information in the Magnus Compton amplitude is therefore encoded in the propagators  along the massive line.

Magnus amplitudes were constructed 
in relativistic quantum field theory in \cite{Brandhuber:2025igz}, where tree-level matrix elements of the $N$-operator were organised in terms of diagrams containing retarded and advanced propagators, weighted by combinatorial coefficients determined by a formula due to Murua \cite{Murua_2006} and extended by \cite{Kim:2024svw} (a particularly efficient algorithm can be found in \cite{Ochirov:2026ixx}). This construction was motivated by the earlier WQFT analysis of \cite{Kim:2024svw,Kim:2025olv}, where the Magnus expansion of the radial action was represented in terms of directed trees%
\footnote{A directed tree  is a tree in which every edge has been assigned an orientation, with the corresponding arrow specifying whether the propagator is retarded or advanced.} 
carrying the same combinatorial weights  (see also the recent work \cite{Gonzo:2026yha}).  We refer to these weights as Murua coefficients. At loop level, the structure of Magnus amplitudes was derived in \cite{Brandhuber:2025igz} and in the works \cite{Kim:2025ebl, Guo:2026xaw} a recursive formula for all corresponding Murua coefficients was derived. Schematically, the resulting relation between conventional tree-level amplitudes and Magnus amplitudes, namely the corresponding $T$- and $N$-matrix elements, takes the form, in the case of Compton scattering,
\begin{equation} iT_{n+2} = \sum_{\hat g} \ComptonBlob{\hat g}, \quad \qquad iN_{n+2} = \sum_{\hat g} \sum_{g\in A(\hat g)} \omega(g) \ComptonBlob{g}\, . 
\label{eq:intro-T-N-tree-expansions} 
\end{equation} 
Here $\hat g$ denotes an undirected tree built from Feynman propagators, while $ A(\hat g)$ is the set of its admissible causal orientations. Each $g\in A(\hat g)$ is a directed tree built from retarded and advanced propagators, with its arrows recording the flow of causality. Thus, the Magnus contribution associated with a fixed undirected topology $\hat g$ is obtained by summing over its admissible causal orientations, each weighted by the corresponding Murua coefficient $\omega(g)$.

 Our model  provides a particularly convenient setting in which to isolate the interplay between the classical expansion, retarded and advanced prescriptions, and Murua coefficients.
 Its simplicity  also makes the $\hbar$ power counting in the classical limit  transparent. For a fixed ordering of $n$ emissions, the massive line contains $n-1$ internal propagators. In the classical regime, emitted and transferred momenta  scale as $\hbar$, while the  massive momentum is of order one. Relative to the classical contribution, each individual ordering begins at order $\hbar^{-n+1}$ and therefore contains the hyperclassical powers%
 \footnote{For this work we will ignore the additional powers of $\hbar$ that appear with the coupling $\lambda$. Since these are simply overall factors, they do not change the structure of the expansion.} 
$\hbar^{-n+1},\ldots,\hbar^{-1}$. 
These terms are present in individual diagrams, even though the general commutator argument predicts their absence from the complete $N$-matrix element.

We find that the appropriate framework for making the cancellation of hyperclassical terms manifest is a Schwinger parametrisation of the massive propagators in the Magnus Compton amplitudes, following an approach inspired by \cite{Capatti:2024bid}. 
Assigning a Schwinger proper time $\tau_i$ to each emission vertex allows contributions with different emission orderings and causal prescriptions to be combined under a common integration measure.
The resulting soft phase can then be expanded to produce monomials in the 
proper-time differences
$\tau_{ij}\coloneqq\tau_i-\tau_j$, which can be represented as multigraphs,%
\footnote{A multigraph is a graph in which more than one edge may join the same pair of vertices. In the present setting, repeated factors of the same proper-time difference correspond to repeated edges.} with each factor corresponding to an edge joining two emission vertices. In this representation, the cancellations can be studied before the proper-time integrations are performed.

These amplitudes also suggest a second central question. In the field theory description, the massless quanta are emitted successively along the massive line, so the underlying Magnus diagrams have chain topology. Consequently, only the Murua coefficients of directed chains appear explicitly. By contrast, arbitrary directed tree topologies occur in WQFT. How can the propagator structures and Murua coefficients of general worldline trees already be encoded in the field-theory chain sector?
 The multigraph expansion makes this relation visible: although the original field-theory diagrams are chains, the monomials generated by their soft phases define more general graph topologies. Furthermore, the order in the soft expansion, and hence the power of $\hbar$, is determined by the number of edges in the corresponding multigraph.

This graphical representation reveals that all hyperclassical terms are associated with disconnected graphs. Disconnected graphs can also appear at classical and higher orders when some edges are repeated. After the proper-time integrations, such repeated-edge terms would produce cubic or higher powers of linearised propagators, structures that are absent from the corresponding WQFT description. We will show that all disconnected contributions vanish through the same mechanism, thereby accounting both for the cancellation of hyperclassical terms and for the absence of structures with no counterpart in WQFT.
Specifically, for a disconnected graph, the ordering of the vertices within each connected component may be fixed independently, while the relative ordering of vertices belonging to different components remains unconstrained. One must therefore sum over all total orderings that preserve the chosen ordering within each component. For example, if one component has the internal ordering $(1\,2)$ and a second component consists only of the vertex $(3)$, the compatible total orderings are $(1\,2\,3)$, $(1\,3\,2)$ and $(3\,1\,2)$. Their sum is the shuffle product $(1\,2)\shuffle (3)$. The cancellation of both hyperclassical terms and disconnected contributions at the classical (and quantum) level  can therefore be reduced to a single algebraic identity for shuffle sums of permutations.

The key to these cancellations is a connection with a particular Hopf algebra, known as the Malvenuto-Reutenauer algebra of permutations \cite{Malvenuto:1993,MalvenutoReutenauer:1995}.%
\footnote{The relevance of this algebra in the context of the Magnus expansion was also noticed in \cite{Arnal_2018}.}
 The first clue to this  
 is a coincidence of coefficients. When the sum over causal assignments in a Magnus Compton amplitude is decomposed into fully ordered proper-time regions, these regions are labelled by permutations. The term labelled by the inverse permutation $\rho^{-1}$ of the external massless particles carries the signed weight 
 \begin{equation}
 \frac{(-1)^{d(\rho)}} {n\binom{n-1}{d(\rho)}} = (-1)^{d(\rho)} \omega\!\left(\ell_n^{(d(\rho))}\right), \label{eq:intro-Eulerian-coefficients} \end{equation} 
 where $d(\rho)$ is the descent number of the permutation $\rho$, namely the number of positions $i\in\{1,\ldots,n-1\}$ for which $\rho(i)>\rho(i+1)$, and $\ell_n^{(d(\rho))}$ denotes a directed chain with $d(\rho)$ reversed arrows (compared to a reference configuration with all retarded propagators).  Here, the magnitude of the weight is precisely the corresponding chain Murua coefficient, while the factor $(-1)^{d(\rho)}$ arises from a product of the causal signs.%
 \footnote{We also note that in \cite{Arnal_2018,Arnal:2020xpt} a rewriting of the general perturbative expansion for the $N$-operator was found that features the same coefficients, see also (2.15) in \cite{Brandhuber:2025igz}.
 }
 The same weighted sum over permutations defines a known projector in the Malvenuto-Reutenauer algebra, called the adjoint first Eulerian idempotent: 
\begin{equation} 
e_1^\ast = \sum_{\rho\in S_n} \frac{(-1)^{d(\rho)}} {n\binom{n-1}{d(\rho)}}\, \rho^{-1}. 
\label{eq:intro-adjoint-Eulerian-projector} 
\end{equation} 
This coefficient-level correspondence identifies the Murua-weighted causal sum with the action of the adjoint Eulerian projector on proper-time orderings. A key property of this projector is that it annihilates shuffle products. The shuffle sums associated with disconnected graphs therefore vanish identically, providing a unified explanation for the cancellation of all hyperclassical contributions and, more generally, of all disconnected contributions at classical and higher orders. Importantly, these cancellations take place before the proper-time integrations are performed.

At the classical order, the soft expansion contains $n-1$ edges on $n$ emission vertices. Since every disconnected contribution vanishes, the surviving graphs must be connected. A connected graph on $n$ vertices with $n-1$ edges is necessarily a spanning tree,  namely a connected, cycle-free graph containing all the emission vertices.
We show that the sign-weighted word polynomials associated with these connected trees have a non-vanishing Lie component and hence survive the Eulerian projection. After proper-time integration, each edge produces a squared retarded or advanced linearised propagator. Thus the field theory calculation reproduces the connected propagator structure expected from WQFT, while the cubic and higher powers of linearised propagators generated at intermediate stages cancel identically. 

The same algebraic structure also explains how the Murua coefficients for  general directed trees emerge from directed chains. This yields one of our main results: an explicit, non-recursive formula for the Murua coefficient of an arbitrary directed tree. To state the formula, we first describe the proper-time orderings compatible with the tree. A directed tree $g$ defines a partial ordering of the Schwinger proper times $\tau_i$.
We denote by $\operatorname{LE}(g)$ the set of its linear extensions, namely the total orderings compatible with all the causal arrows of $g$. For example, if $g$ imposes $\tau_1<\tau_2$ and $\tau_1<\tau_3$, then $\operatorname{LE}(g)=\{123,132\}$. Each $f\in\operatorname{LE}(g)$ is a permutation of the $n$ vertex labels and therefore represents a fully ordered proper-time region, or equivalently a directed chain. After labelling the vertices so that one chosen linear extension is $12\cdots n$, we show that the Murua coefficient of $g$ is reconstructed by summing the signed chain coefficients associated with all its linear extensions: 
\begin{equation} 
\omega(g) = \sum_{f\in\operatorname{LE}(g)} (-1)^{d(f)} \omega\!\left(\ell_n^{(d(f))}\right) = \sum_{f\in\operatorname{LE}(g)} \frac{(-1)^{d(f)}} {n\binom{n-1}{d(f)}}\, . \label{eq:intro-general-Murua-linear-extensions} 
\end{equation}
The complete tree-level Murua data required by the worldline description are therefore already contained in the Eulerian projector built from the chain coefficients of the field theory Magnus expansion.

The construction also extends beyond trees. At higher orders in the soft expansion, the same tree-level field theory amplitudes generate connected multigraphs containing undirected loops. For their acyclic causal orientations, we derive an explicit, non-recursive formula for the Murua coefficients as a sign-weighted sum over all permutations of the vertices. The weights involve the same universal descent coefficients that govern the tree formula, together with signs determined by the graph. This extends the reconstruction of Murua coefficients to loop topologies and makes the reduction of pairs of repeated edges manifest.

We also find a suggestive parallel with the exponentiation of multiple Wilson lines. In that setting, idempotent web-mixing matrices project onto the fully connected colour combinations entering the exponent of a Wilson line correlator \cite{Gardi:2013ita,Dukes:2013gea,Dukes:2013wa}. Their diagonal entries are expressed as descent-weighted sums over linear extensions of precisely the same form as our formula \eqref{eq:intro-general-Murua-linear-extensions} for the Murua coefficients of directed trees. 
Connectedness is realised differently in the two settings: web-mixing matrices select connected colour factors, whereas the adjoint Eulerian projector removes disconnected proper-time structures in Magnus amplitudes. The appearance of the same projector coefficients and linear-extension formula suggests a common Hopf-algebraic origin for these two exponentiation mechanisms.

Summarising, our  results 
provide   a unified algebraic explanation of the classical limit of Magnus Compton amplitudes: the adjoint Eulerian projector removes the shuffle products associated with disconnected contributions, leaving at classical order precisely the spanning trees expected from WQFT. The same projector also reconstructs the Murua coefficients of arbitrary directed trees from those of directed chains, thereby establishing a direct correspondence between the field-theory and worldline descriptions.

The rest of the paper is organised as follows.
In Section~\ref{sec:SchwingerMagnusAmplitudes} we introduce the scalar toy model and develop the Schwinger representation of the four- and five-point Magnus Compton amplitudes.
In Section~\ref{sec:GeneralClassicalLimit} we derive the general $(n+2)$-point expression, analyse its classical expansion and show how disconnected monomials organise into shuffle sums. In Section~\ref{Section Hopf Algebras} we introduce the relevant word and Hopf-algebra structures, identify the dressed theta function with the adjoint first Eulerian projector, and prove the general shuffle cancellation.
In Section~\ref{Section new Murua for general} we derive explicit formulae for the Murua coefficients of arbitrary directed trees and their extension to acyclic orientations of multigraphs containing loops. We also obtain identities among these coefficients and discuss their recursive determination.
We conclude with a discussion of the relation to WQFT, Wilson line exponentiation and possible extensions.

Appendix~\ref{app:notation} summarises our notation and conventions, while Appendix~\ref{app:Murua-chain} derives the closed formula for the Murua coefficients of directed chains. This is followed by Appendix~\ref{spanningtreesnotshufflessection}, where we demonstrate that spanning trees are non-zero after projection. An extension of the eikonal identity is presented in Appendix~\ref{section: New Eikonal Identity}. Appendix~\ref{sec:Dyson-Magnus-Eulerian}
  explains the relation between the Dyson and Magnus expansions through the Eulerian projector. Finally, Appendix~\ref{app:EulerianId} proves an Eulerian number identity, while Appendix~\ref{app: recursionProof} establishes the recursive determination of tree-level Murua coefficients.

{\bf Note added:}
After this work was completed, \cite{Guo:2026hyi} appeared which has some overlap with our results.




\section{Schwinger representation of Magnus amplitudes}\label{sec:SchwingerMagnusAmplitudes}
We work in the scalar toy model with Lagrangian 
\begin{equation}
  \cL
  =
  \frac12(\partial\phi)^2-\frac12m^2\phi^2
  +\frac12(\partial h)^2
  +\frac{\lambda}{2}\phi^2h \,,
\end{equation}
with no $\phi^2h^2$ contact vertex.  The only field theory  Compton diagrams are
therefore the ordered cubic insertions on the heavy line.  For the emission of $n$ massless $h$-particles with outgoing momenta $q_i$ from an incoming massive particle with momentum $p$, we use the symmetric HEFT parametrisation of the heavy momenta \cite{Landshoff:1969yyn,Parra-Martinez:2020dzs}
\begin{equation}
  q^\mu=\sum_{i=1}^n q_i^\mu\,,\qquad
  p^\mu=\pb^\mu+\frac{q^\mu}{2}\,,\qquad
  p'^\mu=\pb^\mu-\frac{q^\mu}{2}\,,\qquad
  \pb\cdot q=0\, .
\end{equation}
The heavy momentum $\pb^\mu=\bar m\,\vb^\mu$ is hard, $\pbar=\cO(1)$, whereas the momenta of the emitted massless quanta are soft, $q_i=\cO(\hbar)$ and $q_i^2=0$.  The relevant kinematic invariants are
\begin{align}
\label{eq:invariants-1}
  D_i&\coloneqq 2\,\pbar\Cdot q_i=\cO(\hbar)\,,
  \\ 
  \label{eq:invariants-2}
  c_{ij}& \coloneqq q_i\Cdot q_j=\cO(\hbar^2)\,,
\end{align}
with
\begin{equation}
\label{eq:invariant_relations}
  \sum_i D_i=2\,\pbar\Cdot q=0\,.
\end{equation}
Furthermore, if $K$ is a sum of soft momenta, then 
\begin{equation}
  \sgn(p^0-K^0)=+1\,,
\end{equation}
so the retarded and advanced prescriptions for the heavy propagators reduce to fixed signs,
\begin{equation}
  \frac{i}{A+i\sigma\vareps}\,,\qquad \sigma=\pm 1 \,,
\end{equation}
where $A$ denotes the relevant kinematic denominator, 
cf.~\eqref{eq:firstpropagator}. 

We represent the propagators using Schwinger parameters:
\begin{equation}
\label{eq:Schw}
    \frac{i}{A+i\sigma\vareps} = \sigma\int\!d\tau \, \theta(-\sigma\tau) \, e^{-iA\tau}\, , 
\end{equation}
where $\sigma=\pm1$ for retarded and advanced propagators respectively. 
Following \cite{Capatti:2024bid}, we  
assign a proper time $\tau_i$ to each vertex, so that each propagator is associated with the proper-time difference $\tau_{ij}=\tau_i - \tau_j$. We then perform a change of variables that allows us to combine contributions from different massless emission orderings under a common integration measure. After combining these contributions, we expand in the soft, or equivalently heavy-mass, limit and only then carry out the proper-time integrals to obtain explicit classical expressions. 
The theta functions encode the $i\vareps$ prescriptions. 
Although elementary, \eqref{eq:Schw} is the key relation underlying the remainder of this work. It converts retarded and advanced propagator prescriptions into a product of signs and proper-time ordering constraints. These are precisely the structures that later reappear in the permutation sums governing the classical limit.

\subsection{Four-point Magnus amplitude}
The four-point Magnus Compton amplitude is built from four diagrams,
\begin{align}
iN_4=
    \begin{tikzpicture}[baseline={([yshift=-0.5ex]current bounding box.center)},thick,scale=0.8]
    \draw  (0,0) -- ++(0.75,0);
    \draw (1.75,0) -- ++(0.75,0);
    \draw[vector] (1.25,0) -- ++(-105:1.3)node[below]{$q_1$};
    \draw[vector] (1.25,0) -- ++(-75:1.3)node[below]{$q_2$};
    \draw[causArrow] (0.1,0.2) -- ++(0.6,0) node[midway, above]{$p$};
    \draw[causArrow] (1.85,0.2) -- ++(0.6,0) node[midway, above]{$p'$};
    \draw[fill=white] (1.25,0) circle(0.5);
    \end{tikzpicture}
    \,&=
    \omega(\begin{tikzpicture}[baseline={([yshift=-.5ex]current bounding box.center)}, thick]
    \draw[causArrow] (0,0) -- (0.5,0);
    \draw[fill=white] (0,0) circle(3pt);
    \draw[fill=white] (0.5,0) circle(3pt);
    \end{tikzpicture})
    \left(
    \begin{tikzpicture}[baseline={([yshift=-0.5ex]current bounding box.center)},thick,scale=0.8]
    \draw             (0,0) -- ++(0.6,0);
    \draw[causArrow]  (0.6,0) -- ++(0.8,0);
    \draw             (1.4,0) -- ++(0.6,0);
    \draw[vector]     (0.6,0) -- ++(-90:1)node[below]{$q_1$};
    \draw[vector]     (1.4,0) -- ++(-90:1)node[below]{$q_2$};
    \node[above] at (0.6,0) {$\tau_1$};
    \node[above] at (1.4,0) {$\tau_2$};
    \end{tikzpicture}
    +
    \begin{tikzpicture}[baseline={([yshift=-0.5ex]current bounding box.center)},thick,scale=0.8]
    \draw             (0,0) -- ++(0.6,0);
    \draw[causArrow]  (1.4,0) -- ++(-0.8,0);
    \draw             (1.4,0) -- ++(0.6,0);
    \draw[vector]     (0.6,0) -- ++(-90:1)node[below]{$q_1$};
    \draw[vector]     (1.4,0) -- ++(-90:1)node[below]{$q_2$};
    \node[above] at (0.6,0) {$\tau_1$};
    \node[above] at (1.4,0) {$\tau_2$};
    \end{tikzpicture}
    +
    \begin{tikzpicture}[baseline={([yshift=-0.5ex]current bounding box.center)},thick,scale=0.8]
    \draw             (0,0) -- ++(0.6,0);
    \draw[causArrow]  (0.6,0) -- ++(0.8,0);
    \draw             (1.4,0) -- ++(0.6,0);
    \draw[vector]     (1.4,0) -- (0.6,-1)node[below]{$q_1$};
    \draw[fill=white, draw=none] (1.05,-0.5) circle(0.1);
    \draw[vector]     (0.6,0) -- (1.4,-1)node[below]{$q_2$};
    \node[above] at (1.4,0) {$\tau_1$};
    \node[above] at (0.6,0) {$\tau_2$};
    \end{tikzpicture}
    +
    \begin{tikzpicture}[baseline={([yshift=-0.5ex]current bounding box.center)},thick,scale=0.8]
    \draw             (0,0) -- ++(0.6,0);
    \draw[causArrow]  (1.4,0) -- ++(-0.8,0);
    \draw             (1.4,0) -- ++(0.6,0);
    \draw[vector]     (1.4,0) -- (0.6,-1)node[below]{$q_1$};
    \draw[fill=white, draw=none] (1.05,-0.5) circle(0.1);
    \draw[vector]     (0.6,0) -- (1.4,-1)node[below]{$q_2$};
    \node[above] at (1.4,0) {$\tau_1$};
    \node[above] at (0.6,0) {$\tau_2$};
    \end{tikzpicture}\right),
\end{align}
where the factor of 
$\omega(\begin{tikzpicture}[baseline={([yshift=-.5ex]current bounding box.center)}, thick]
    \draw[causArrow] (0,0) -- (0.5,0);
    \draw[fill=white] (0,0) circle(3pt);
    \draw[fill=white] (0.5,0) circle(3pt);
    \end{tikzpicture})=1/2$ is the Murua coefficient, and we associate  a canonical time to the vertex of each massless emission. 
We have also used that $\omega(\begin{tikzpicture}[baseline={([yshift=-.5ex]current bounding box.center)}, thick]
    \draw[causArrow] (0.5,0) -- (0,0);
    \draw[fill=white] (0,0) circle(3pt);
    \draw[fill=white] (0.5,0) circle(3pt);
    \end{tikzpicture})=\omega(\begin{tikzpicture}[baseline={([yshift=-.5ex]current bounding box.center)}, thick]
    \draw[causArrow] (0,0) -- (0.5,0);
    \draw[fill=white] (0,0) circle(3pt);
    \draw[fill=white] (0.5,0) circle(3pt);
    \end{tikzpicture})$.  
The propagator for the first diagram gives
\begin{align}
    &\frac{i}{(p-q_1)^2-m^2+i\sgn(p^0-q_1^0)\vareps} \nn \\
    &=\frac{i}{(p-q_1)^2-m^2+i\vareps} =\frac{i}{-2p\cdot q_1 + i\vareps} = \frac{i}{-2\pbar \cdot q_1 -q_1\cdot q_2 +i\vareps} = \frac{i}{-D_1-c_{12}+i\vareps}\,,
    \label{eq:firstpropagator}
\end{align}
where $D_i$ and $c_{ij}$ are defined in \eqref{eq:invariants-1} and \eqref{eq:invariants-2}.
Using the Schwinger parametrisation  \eqref{eq:Schw} and with $\tau_{ij} \coloneq \tau_i - \tau_j$,
\begin{equation}
    \frac{i}{-D_1-c_{12}+i\vareps} = \int\!d\tau_{12} \,\theta(-\tau_{12})e^{iD_1\tau_{12}}e^{ic_{12}\tau_{12}}\, . 
\end{equation}
 This 
 can be rewritten as
\begin{equation}
    \int\!d\tau_{12} \ \theta(-\tau_{12})\, e^{i\sum_{i=1}^{2} D_i\tau_{i}}e^{ic_{12}\tau_{12}}\,,
\end{equation}
using the fact that $D_1+D_2=0$ by momentum conservation \eqref{eq:invariant_relations}, so the first exponential is symmetric under permutations of the massless legs.
Therefore, the  sum of the uncrossed diagrams gives
\begin{align}
\label{eq:uncro}
    \omega(\begin{tikzpicture}[baseline={([yshift=-.5ex]current bounding box.center)}, thick]
    \draw[causArrow] (0,0) -- (0.5,0);
    \draw[fill=white] (0,0) circle(3pt);
    \draw[fill=white] (0.5,0) circle(3pt);
    \end{tikzpicture})\left(
    \begin{tikzpicture}[baseline={([yshift=-0.5ex]current bounding box.center)},thick,scale=0.8]
    \draw             (0,0) -- ++(0.3,0);
    \draw[causArrow]  (0.3,0) -- ++(0.5,0);
    \draw             (0.8,0) -- ++(0.3,0);
    \draw[vector]     (0.3,0) -- ++(-90:0.7);
    \draw[vector]     (0.8,0) -- ++(-90:0.7);
    \end{tikzpicture}
    +
    \begin{tikzpicture}[baseline={([yshift=-0.5ex]current bounding box.center)},thick,scale=0.8]
    \draw             (0,0) -- ++(0.3,0);
    \draw[causArrow]  (0.8,0) -- ++(-0.5,0);
    \draw             (0.8,0) -- ++(0.3,0);
    \draw[vector]     (0.3,0) -- ++(-90:0.7);
    \draw[vector]     (0.8,0) -- ++(-90:0.7);
    \end{tikzpicture}
    \right)
    =
    \frac{(i\lambda)^2}{2}\int\!d\tau_{12} \, e^{i\sum_{i=1}^{2} D_i\tau_{i}}e^{ic_{12}\tau_{12}} \left[ \,\theta(-\tau_{12}) - \theta(\tau_{12}) \right]\, ,
\end{align}
while summing the two crossed diagrams gives  
\begin{align}
\label{eq:cro}
    \omega(\begin{tikzpicture}[baseline={([yshift=-.5ex]current bounding box.center)}, thick]
    \draw[causArrow] (0,0) -- (0.5,0);
    \draw[fill=white] (0,0) circle(3pt);
    \draw[fill=white] (0.5,0) circle(3pt);
    \end{tikzpicture})
    \left(
    \begin{tikzpicture}[baseline={([yshift=-0.5ex]current bounding box.center)},thick,scale=0.8]
    \draw             (0,0) -- ++(0.3,0);
    \draw[causArrow]  (0.3,0) -- ++(0.6,0);
    \draw             (0.9,0) -- ++(0.3,0);
    \draw[vector]     (0.9,0) -- (0.3,-0.7);
    \draw[fill=white, draw=none] (0.6,-0.35) circle(0.1);
    \draw[vector]     (0.3,0) -- (0.9,-0.7);
    \end{tikzpicture}
    +
    \begin{tikzpicture}[baseline={([yshift=-0.5ex]current bounding box.center)},thick,scale=0.8]
    \draw             (0,0) -- ++(0.3,0);
    \draw[causArrow]  (0.9,0) -- ++(-0.6,0);
    \draw             (0.9,0) -- ++(0.3,0);
    \draw[vector]     (0.9,0) -- (0.3,-0.7);
    \draw[fill=white, draw=none] (0.6,-0.35) circle(0.1);
    \draw[vector]     (0.3,0) -- (0.9,-0.7);
    \end{tikzpicture}
    \right)
    =
    \frac{(i\lambda)^2}{2}\int\!d\tau_{21} \, e^{i\sum_{i=1}^{2} D_i\tau_{i}}e^{ic_{12}\tau_{21}} \left[ \,\theta(-\tau_{21}) - \theta(\tau_{21}) \right]\, .
\end{align}
Rewriting \eqref{eq:cro} in terms of $\tau_{12}= - \tau_{21}$ 
gives
\begin{align}
\label{eq:com}
    \omega(\begin{tikzpicture}[baseline={([yshift=-.5ex]current bounding box.center)}, thick]
    \draw[causArrow] (0,0) -- (0.5,0);
    \draw[fill=white] (0,0) circle(3pt);
    \draw[fill=white] (0.5,0) circle(3pt);
    \end{tikzpicture})
    \left(
    \begin{tikzpicture}[baseline={([yshift=-0.5ex]current bounding box.center)},thick,scale=0.8]
    \draw             (0,0) -- ++(0.3,0);
    \draw[causArrow]  (0.3,0) -- ++(0.6,0);
    \draw             (0.9,0) -- ++(0.3,0);
    \draw[vector]     (0.9,0) -- (0.3,-0.7);
    \draw[fill=white, draw=none] (0.6,-0.35) circle(0.1);
    \draw[vector]     (0.3,0) -- (0.9,-0.7);
    \end{tikzpicture}
    +
    \begin{tikzpicture}[baseline={([yshift=-0.5ex]current bounding box.center)},thick,scale=0.8]
    \draw             (0,0) -- ++(0.3,0);
    \draw[causArrow]  (0.9,0) -- ++(-0.6,0);
    \draw             (0.9,0) -- ++(0.3,0);
    \draw[vector]     (0.9,0) -- (0.3,-0.7);
    \draw[fill=white, draw=none] (0.6,-0.35) circle(0.1);
    \draw[vector]     (0.3,0) -- (0.9,-0.7);
    \end{tikzpicture}
    \right)
    =
    -\frac{(i\lambda)^2}{2}\int\!d\tau_{12} \, e^{i\sum_{i=1}^{2} D_i\tau_{i}}e^{-ic_{12}\tau_{12}} \big[ \,\theta(-\tau_{12}) - \theta(\tau_{12}) \big].
\end{align} 
Combining all contributions, the four-point Magnus Compton amplitude is
\begin{align}
\label{eq:C4c}
i N_4 & =     \frac{(i\lambda)^2}{2}\int\!d\tau_{12} \, e^{i\sum_{i=1}^{2} D_i\tau_{i}} \big[  \,\theta(-\tau_{12}) - \theta(\tau_{12}) \big]\Big(  e^{ic_{12}\tau_{12}} - e^{-ic_{12}\tau_{12}}\Big) .
\end{align}
Now recall that $c_{12} = q_1\Cdot q_2\sim \cO(\hbar^2)$. Expanding the exponentials, the leading hyperclassical term cancels,%
\footnote{The proper times $\tau_i$ are conjugate to the $D_i=2 \bar{p}\Cdot q_i\sim\cO(\hbar)$ variables and hence $\tau_i\sim \cO(\hbar^{-1})$. Therefore the first term in the classical expansion in \eqref{eq:C4c}, which happens to vanish, would have been $\cO(\hbar^{-1})$.}
while the classical term gives
\begin{align}
\label{eq:cl4}
  i (i\lambda) ^2 c_{12} \int\!d\tau_{12} \, \tau_{12}\,  e^{i\sum_{i=1}^{2} D_i\tau_{i}} \big[  \,\theta(-\tau_{12}) - \theta(\tau_{12}) \big]\,  .
\end{align}
We can integrate \eqref{eq:cl4} to arrive at the classical expression using
\begin{align}
  \int\! d \tau\,
  \theta(\sigma\tau)\, e^{i\tau x}
  &= i\sigma \frac{1}{x+i\sigma\vareps}\,,\\
  \int\!d\tau\,
  \tau\,\theta(\sigma\tau)\, e^{i\tau x}
  &= -\sigma \frac{1}{(x+i\sigma\vareps)^2}
  \label{eq:back-integration}\, . 
\end{align}
The result for the classical Magnus amplitude (after undoing the change of variables $\sum_{i=1}^2 D_i \tau_i = \tau_{12}  D_1$)  is then 
\begin{align}
    iN_4 = i(i\lambda)^2\,c_{12}\left( \frac{1}{(D_1+i\vareps)^2} + \frac{1}{(D_1-i\vareps)^2} \right)\, , 
\end{align}
where we recall that 
$D_i$ and $c_{ij}$ are defined in \eqref{eq:invariants-1} and \eqref{eq:invariants-2}.
 Note that $i N_4 \sim \cO(\hbar^{2-2}) = \cO(1)$ as expected.  
Finally, we observe that with the conventional Feynman prescription the same calculation gives 
\begin{align}
i T_4 = i N_4 - \lambda^2\pi  \delta ( \bar{p}\Cdot q_1)\, , 
\end{align}
showing that the $N$-matrix element has no discontinuity.

\subsection{Five-point Magnus amplitude}
For the five-point Magnus Compton amplitude, we must compute 
\begin{align}\label{eq: 5pointComptonDiagrams}
iN_5=
    \begin{tikzpicture}[baseline={([yshift=-0.5ex]current bounding box.center)},thick,scale=0.8]
    \draw  (0,0) -- ++(0.75,0);
    \draw (1.75,0) -- ++(0.75,0);
    \draw[vector] (1.25,0) -- ++(-120:1.3)node[below]{$q_1$};
    \draw[vector] (1.25,0) -- ++(-90:1.3)node[below]{$q_2$};
    \draw[vector] (1.25,0) -- ++(-60:1.3)node[below]{$q_3$};
    \draw[causArrow] (0.1,0.2) -- ++(0.6,0) node[midway, above]{$p$};
    \draw[causArrow] (1.85,0.2) -- ++(0.6,0) node[midway, above]{$p'$};
    \draw[fill=white] (1.25,0) circle(0.5);
    \end{tikzpicture}
    \,&=\sum_{S_3}\left(
    \omega(\begin{tikzpicture}[baseline={([yshift=-.5ex]current bounding box.center)}, thick]
    \draw[causArrow] (0,0) -- (0.5,0);
    \draw[causArrow] (0.5,0) -- (1,0);
    \draw[fill=white] (0,0) circle(3pt);
    \draw[fill=white] (0.5,0) circle(3pt);
    \draw[fill=white] (1,0) circle(3pt);
    \end{tikzpicture})
    \begin{tikzpicture}[baseline={([yshift=-0.5ex]current bounding box.center)},thick,scale=0.8]
    \draw             (0,0) -- ++(0.6,0);
    \draw[causArrow]  (0.6,0) -- ++(0.8,0);
    \draw[causArrow]  (1.4,0) -- ++(0.8,0);
    \draw             (2.2,0) -- ++(0.6,0);
    \draw[vector]     (0.6,0) -- ++(-90:1)node[below]{$q_1$};
    \draw[vector]     (1.4,0) -- ++(-90:1)node[below]{$q_2$};
    \draw[vector]     (2.2,0) -- ++(-90:1)node[below]{$q_3$};
    \node[above] at (0.6,0) {$\tau_1$};
    \node[above] at (1.4,0) {$\tau_2$};
    \node[above] at (2.2,0) {$\tau_3$};
    \end{tikzpicture}
    + (\text{perms of arrows})\right),
\end{align}
where the sum is over all permutations of the massless legs $1,2,3$. 
The relevant Murua coefficients are those of an $n$-vertex chain ($n=3$) with $k(\vec \sigma)$ reversed arrows (compared to a fixed ordering with all retarded propagators). 
There is a simple closed formula for these coefficients,%
\footnote{This formula also appeared in the recent work \cite{Bautista:2026fcp}. }
proved in Appendix~\ref{app:Murua-chain}: 
\begin{equation}
\boxed{
\label{eq:muruaclosed}
  \omega(\vec \sigma)
  =
  \frac{1}{n\binom{n-1}{k(\vec\sigma)}}
  }\, , 
\end{equation} 
where 
$\vec{\sigma}= (\sigma_1, \ldots , \sigma_{n-1})$ labels a causal assignment of the propagators, with $\sigma_i=\pm 1$ corresponding to a retarded/advanced propagator.   
At  five points,  this gives 
\begin{align} 
\begin{aligned} \omega\!\left( \begin{tikzpicture}[baseline={([yshift=-.5ex]current bounding box.center)}, thick] \draw[causArrow] (0,0) -- (0.5,0); \draw[causArrow] (0.5,0) -- (1.0,0); \draw[fill=white] (0,0) circle(3pt); \draw[fill=white] (0.5,0) circle(3pt); \draw[fill=white] (1.0,0) circle(3pt); 
\end{tikzpicture} \right) &= 
\omega(+,+) =   \frac{1}{3}\,, 
\\[1em] 
\omega\!\left( \begin{tikzpicture}[baseline={([yshift=-.5ex]current bounding box.center)}, thick] \draw[causArrow] (0.5,0) -- (0,0); \draw[causArrow] (0.5,0) -- (1.0,0); \draw[fill=white] (0,0) circle(3pt); \draw[fill=white] (0.5,0) circle(3pt); \draw[fill=white] (1.0,0) circle(3pt); \end{tikzpicture} \right) &= 
\omega (-,+) = \frac{1}{6}\,, 
\\[1em] 
\omega\!\left( \begin{tikzpicture}[baseline={([yshift=-.5ex]current bounding box.center)}, thick] \draw[causArrow] (0,0) -- (0.5,0); \draw[causArrow] (1.0,0) -- (0.5,0); \draw[fill=white] (0,0) circle(3pt); \draw[fill=white] (0.5,0) circle(3pt); \draw[fill=white] (1.0,0) circle(3pt); \end{tikzpicture} \right) &= 
\omega(+,-) = \frac{1}{6}\,, 
\\[1em] 
\omega\!\left( \begin{tikzpicture}[baseline={([yshift=-.5ex]current bounding box.center)}, thick] \draw[causArrow] (0.5,0) -- (0,0); \draw[causArrow] (1.0,0) -- (0.5,0); \draw[fill=white] (0,0) circle(3pt); \draw[fill=white] (0.5,0) circle(3pt); \draw[fill=white] (1.0,0) circle(3pt); \end{tikzpicture} \right) &= 
\omega (-,-) = \frac{1}{3}\,. 
\end{aligned} 
\label{eq:Murua-5-coeff} 
\end{align}
For instance, the diagram with ordering $(1\,2\,3)$ of massless particle emissions and both propagators retarded ($\sigma_1 = \sigma_2 = +1$) gives
\begin{equation}
    \begin{tikzpicture}[baseline={([yshift=-0.5ex]current bounding box.center)},thick,scale=0.8]
    \draw             (0,0) -- ++(0.6,0);
    \draw[causArrow]  (0.6,0) -- ++(0.8,0);
    \draw[causArrow]  (1.4,0) -- ++(0.8,0);
    \draw             (2.2,0) -- ++(0.6,0);
    \draw[vector]     (0.6,0) -- ++(-90:1)node[below]{$q_1$};
    \draw[vector]     (1.4,0) -- ++(-90:1)node[below]{$q_2$};
    \draw[vector]     (2.2,0) -- ++(-90:1)node[below]{$q_3$};
     \node[above] at (0.6,0) {$\tau_1$};
    \node[above] at (1.4,0) {$\tau_2$};
    \node[above] at (2.2,0) {$\tau_3$};
    \end{tikzpicture} =(i\lambda)^3\int\!d\tau_{12}\,d\tau_{23}\, \theta(-\tau_{12})\,\theta(-\tau_{23})e^{i\sum_{i=1}^3D_i\tau_i}e^{i(c_{12}\tau_{12}+c_{13}\tau_{13}+c_{23}\tau_{23})}.
\end{equation}
As another example, the diagram with ordering $(2\,1\,3)$ and both propagators retarded ($\sigma_1 = \sigma_2 = +1$) gives
\begin{equation}
    \label{eq:invertedexample}
    \begin{tikzpicture}[baseline={([yshift=-0.5ex]current bounding box.center)},thick,scale=0.8]
    \draw             (0,0) -- ++(0.6,0);
    \draw[causArrow]  (0.6,0) -- ++(0.8,0);
    \draw[causArrow]  (1.4,0) -- ++(0.8,0);
    \draw             (2.2,0) -- ++(0.6,0);
    \draw[vector]     (0.6,0) -- ++(-90:1)node[below]{$q_2$};
    \draw[vector]     (1.4,0) -- ++(-90:1)node[below]{$q_1$};
    \draw[vector]     (2.2,0) -- ++(-90:1)node[below]{$q_3$};
     \node[above] at (0.6,0) {$\tau_2$};
    \node[above] at (1.4,0) {$\tau_1$};
    \node[above] at (2.2,0) {$\tau_3$};
    \end{tikzpicture} =(i\lambda)^3\int\!d\tau_{21}\,d\tau_{13}\, \theta(-\tau_{21})\,\theta(-\tau_{13})e^{i\sum_{i=1}^3D_i\tau_i}e^{i(-c_{12}\tau_{12}+c_{13}\tau_{13}+c_{23}\tau_{23})}\, . 
\end{equation}
Note the change of sign in the $c_{12}\tau_{12}$ term in the exponent. This follows from the symmetry of $c_{ij}$ and the antisymmetry of $\tau_{ij}$ under swapping indices. Summing over all permutations $\rho = (\rho_1\,\rho_2\,\rho_3)$ and all trees, the five-point Magnus Compton amplitude is
\begin{equation}\label{eq:l3}
  iN_5= (i\lambda)^3\sum_{\rho\in S_3}\int\! d\tau_{\rho_1\rho_2}d\tau_{\rho_2\rho_3} e^{i\sum_{i=1}^3 D_i\tau_i}\mathcal{X}_\rho\ ,
\end{equation}
where we have defined
\begin{equation}
    \mathcal{X}_\rho \coloneq e^{i\sum_{i<j}c_{\rho_i\rho_j}\tau_{\rho_i\rho_j}}\sum_{\sigma_1,\sigma_2=\pm}
  \omega (\sigma_1,\sigma_2)\sigma_1\sigma_2\,
  \theta(-\sigma_1\tau_{\rho_1\rho_2})
  \theta(-\sigma_2\tau_{\rho_2\rho_3})  =  e^{i\seee_\rho}\Theta_\rho^{\rm M}\ ,
\end{equation}
also introducing the notation 
\begin{equation}
\label{def:phipi}
    \seee_\rho \coloneq  \sum_{i<j}c_{\rho_i\rho_j}\tau_{\rho_i\rho_j}\,,
    \end{equation}
    and 
    \begin{align}
        \Theta_\rho^{\rm M}= \sum_{\sigma_1,\sigma_2=\pm}
 \sigma_1\sigma_2\, \omega (\sigma_1,\sigma_2)\, 
  \theta(-\sigma_1\tau_{\rho_1\rho_2})
  \theta(-\sigma_2\tau_{\rho_2\rho_3})\,.
\end{align}
The  expression for $\Theta^{\rm M}_\rho$  for arbitrary $n$  is given  later in \eqref{eq:ThetaMagnusGeneral}, and 
in the five-point case is 
\begin{align}
  \Theta_\rho^{\rm M}
  = 
\frac13\theta_{\rho_1\rho_2}\theta_{\rho_2\rho_3} + \frac13\theta_{\rho_2\rho_1}\theta_{\rho_3\rho_2}
 -\frac16\theta_{\rho_1\rho_2}\theta_{\rho_3\rho_2}
  -\frac16\theta_{\rho_2\rho_1}\theta_{\rho_2\rho_3}
  \, , 
  \label{eq:path-theta-combo}
\end{align} 
with 
$\theta_{ab}\coloneq\theta(\tau_{ab})$. The measures $d\tau_{\rho_1\rho_2}d\tau_{\rho_2\rho_3}$ can always be transformed into $d\tau_{12}d\tau_{23}$ with unit Jacobian. This provides a common integration measure under which we can sum  $\Theta_\rho^{\rm M}$ over permutations. 

\subsubsection{Hyperclassical contributions}
The leading hyperclassical contribution, of order $\mathcal{O}(\hbar^{-2})$, is obtained by expanding the soft exponentials $e^{ic_{ij}\tau_{ij}}$ 
only up to their leading terms, $e^{ic_{ij}\tau_{ij}}\simeq1+\cO(\hbar)$, leaving
\begin{equation}
  e^{i\sum_{i=1}^3 D_i\tau_i}
\sum_{\rho\in S_3}
\Theta_\rho^{\rm M}\, . 
\end{equation}
Using the identities 
\begin{align}
\begin{split}
\theta_{ij}\theta_{jk}+\theta_{ik}\theta_{kj}+\theta_{ki}\theta_{ij}
  &=\theta_{ij}\,,\\
  \theta_{ij}\theta_{jk}+\theta_{ik}\theta_{kj}
  &=\theta_{ik}\theta_{ij}\,,
  \label{eq:zeng-theta-identities}
  \end{split}
\end{align}
we find that
\begin{equation}
    \label{eq:leadinghyper5point}
    \sum_{\rho\in S_3}\Theta_\rho^{\rm M}=0\, ,
\end{equation}
and hence the leading hyperclassical $\mathcal{O}(\hbar^{-2})$ contribution vanishes. This is our first explicit cancellation of hyperclassical contributions that depends on the Murua coefficients. For the hyperclassical $\mathcal{O}(\hbar^{-1})$ contribution, we retain the next term in the expansion of the soft exponentials $e^{ic_{ij}\tau_{ij}}$. For example,
\begin{align}
    \omega(\begin{tikzpicture}[baseline={([yshift=-.5ex]current bounding box.center)}, thick]
    \draw[causArrow] (0,0) -- (0.5,0);
    \draw[causArrow] (0.5,0) -- (1,0);
    \draw[fill=white] (0,0) circle(3pt);
    \draw[fill=white] (0.5,0) circle(3pt);
    \draw[fill=white] (1,0) circle(3pt);
    \end{tikzpicture})
    \begin{tikzpicture}[baseline={([yshift=-0.5ex]current bounding box.center)},thick,scale=0.8]
    \draw             (0,0) -- ++(0.6,0);
    \draw[causArrow]  (0.6,0) -- ++(0.8,0);
    \draw[causArrow]  (1.4,0) -- ++(0.8,0);
    \draw             (2.2,0) -- ++(0.6,0);
    \draw[vector]     (0.6,0) -- ++(-90:1)node[below]{$q_1$};
    \draw[vector]     (1.4,0) -- ++(-90:1)node[below]{$q_2$};
    \draw[vector]     (2.2,0) -- ++(-90:1)node[below]{$q_3$};
     \node[above] at (0.6,0) {$\tau_1$};
    \node[above] at (1.4,0) {$\tau_2$};
    \node[above] at (2.2,0) {$\tau_3$};
    \end{tikzpicture} = (i\lambda)^3\omega(\begin{tikzpicture}[baseline={([yshift=-.5ex]current bounding box.center)}, thick]
    \draw[causArrow] (0,0) -- (0.5,0);
    \draw[causArrow] (0.5,0) -- (1,0);
    \draw[fill=white] (0,0) circle(3pt);
    \draw[fill=white] (0.5,0) circle(3pt);
    \draw[fill=white] (1,0) circle(3pt);
    \end{tikzpicture})\int d\tau_{12}d\tau_{23}\ \theta_{21}\theta_{32}e^{i\sum_{i=1}^3D_i\tau_i}e^{i\sum_{i < j}c_{ij}\tau_{ij}}\,,
    \label{eq:nonflippedfivepoint}
\end{align}
and, after our usual change of variables, we can show that the flipped diagram
\begin{align}
    \omega(\begin{tikzpicture}[baseline={([yshift=-.5ex]current bounding box.center)}, thick]
    \draw[causArrow] (0.5,0) -- (0,0);
    \draw[causArrow] (1,0) -- (0.5,0);
    \draw[fill=white] (0,0) circle(3pt);
    \draw[fill=white] (0.5,0) circle(3pt);
    \draw[fill=white] (1,0) circle(3pt);
    \end{tikzpicture})
    \begin{tikzpicture}[baseline={([yshift=-0.5ex]current bounding box.center)},thick,scale=0.8]
    \draw             (0,0) -- ++(0.6,0);
    \draw[causArrow] (1.4,0) -- (0.6,0);
    \draw[causArrow] (2.2,0) -- (1.4,0);
    \draw             (2.2,0) -- ++(0.6,0);
    \draw[vector]     (0.6,0) -- ++(-90:1)node[below]{$q_3$};
    \draw[vector]     (1.4,0) -- ++(-90:1)node[below]{$q_2$};
    \draw[vector]     (2.2,0) -- ++(-90:1)node[below]{$q_1$};
     \node[above] at (0.6,0) {$\tau_3$};
    \node[above] at (1.4,0) {$\tau_2$};
    \node[above] at (2.2,0) {$\tau_1$};
    \end{tikzpicture} = (i\lambda)^3\omega(\begin{tikzpicture}[baseline={([yshift=-.5ex]current bounding box.center)}, thick]
    \draw[causArrow] (0.5,0) -- (0,0);
    \draw[causArrow] (1,0) -- (0.5,0);
    \draw[fill=white] (0,0) circle(3pt);
    \draw[fill=white] (0.5,0) circle(3pt);
    \draw[fill=white] (1,0) circle(3pt);
    \end{tikzpicture})\int d\tau_{12}d\tau_{23}\ \theta_{21}\theta_{32}e^{i\sum_{i=1}^3D_i\tau_i}e^{-i\sum_{i < j}c_{ij}\tau_{ij}}\,.
    \label{eq:flippedfivepoint}
\end{align}
The only difference between \eqref{eq:nonflippedfivepoint} and \eqref{eq:flippedfivepoint}  is the change of sign in the exponent of the final exponential. Summing the contributions from these two flipped diagrams gives
\begin{align}
\label{eq:5co}
    (i\lambda)^3\omega(\begin{tikzpicture}[baseline={([yshift=-.5ex]current bounding box.center)}, thick]
    \draw[causArrow] (0,0) -- (0.5,0);
    \draw[causArrow] (0.5,0) -- (1,0);
    \draw[fill=white] (0,0) circle(3pt);
    \draw[fill=white] (0.5,0) circle(3pt);
    \draw[fill=white] (1,0) circle(3pt);
    \end{tikzpicture})
\int d\tau_{12}d\tau_{23}\ \theta_{21}\theta_{32}e^{i\sum_{i=1}^3D_i\tau_i}\Big(e^{i\sum_{i < j}c_{ij}\tau_{ij}}+e^{-i\sum_{i < j}c_{ij}\tau_{ij}}\Big)\,,
\end{align}
where we used that $\omega(\begin{tikzpicture}[baseline={([yshift=-.5ex]current bounding box.center)}, thick]
    \draw[causArrow] (0,0) -- (0.5,0);
    \draw[causArrow] (0.5,0) -- (1,0);
    \draw[fill=white] (0,0) circle(3pt);
    \draw[fill=white] (0.5,0) circle(3pt);
    \draw[fill=white] (1,0) circle(3pt);
    \end{tikzpicture}) = \omega(\begin{tikzpicture}[baseline={([yshift=-.5ex]current bounding box.center)}, thick]
    \draw[causArrow] (0.5,0) -- (0,0);
    \draw[causArrow] (1,0) -- (0.5,0);
    \draw[fill=white] (0,0) circle(3pt);
    \draw[fill=white] (0.5,0) circle(3pt);
    \draw[fill=white] (1,0) circle(3pt);
    \end{tikzpicture})\, . $
Therefore, expanding \eqref{eq:5co} we find, for the  subleading $\mathcal{O}(\hbar^{-1})$ hyperclassical  contribution from summing these flipped diagrams, 
\begin{align}
    i(i\lambda)^3\omega(\begin{tikzpicture}[baseline={([yshift=-.5ex]current bounding box.center)}, thick]
    \draw[causArrow] (0,0) -- (0.5,0);
    \draw[causArrow] (0.5,0) -- (1,0);
    \draw[fill=white] (0,0) circle(3pt);
    \draw[fill=white] (0.5,0) circle(3pt);
    \draw[fill=white] (1,0) circle(3pt);
    \end{tikzpicture})
\int d\tau_{12}d\tau_{23}\ \theta_{21}\theta_{32}e^{i\sum_{i=1}^3D_i\tau_i}\underbrace{\Big(\sum_{i < j}c_{ij}\tau_{ij}-\sum_{i < j}c_{ij}\tau_{ij}\Big)}_{0}\, . 
\end{align}
Similar pairwise cancellations occur for all other causality assignments and particle orderings. Each diagram labelled by $(\rho, \vec{\sigma})$ is paired with one obtained by simultaneously reversing the ordering of the particles along the massive line and all causality arrows. The contributions of each such pair cancel, and hence the subleading hyperclassical term vanishes.

\subsubsection{Classical contributions}

We now turn to the classical contributions to the five-point Magnus Compton amplitude. As we have seen, the hyperclassical contributions have cancelled before the proper-time integrals were carried out, so we will work with the permutation-dependent part of the Schwinger integrand of the Magnus Compton amplitude,
\begin{align}
    \mathcal{X}_3 \coloneq  \sum_{\rho \in S_3}\mathcal{X}_\rho=\sum_{\rho \in S_3}\Theta_\rho^{\rm M} e^{i\seee_\rho}\ .
\end{align}
The ordering-dependent phase can be written as
\begin{equation}
    \seee_\rho
    =
    \sum_{i<j}
    c_{\rho_i\rho_j}\tau_{\rho_i\rho_j}
    =\sum_{i<j}
    s_\rho(i,j)c_{ij}\tau_{ij}\,,
    \label{eq:fivepoint-phase-signs}
\end{equation}
with
\begin{equation}
    s_\rho(i,j)
    \coloneq
    \begin{cases}
        +1, & \text{if $i$ appears before $j$ in $\rho$}\,,\\
        -1, & \text{if $j$ appears before $i$ in $\rho$}\,.
    \end{cases}
    \label{eq:fivepoint-ordering-sign}
\end{equation}
Thus $s_\rho(i,j)$ simply records whether the relative ordering of the
vertices $i$ and $j$ agrees with the canonical orientation $i\to j$.

At five points the classical contribution is obtained from the
second-order term in the expansion of $e^{i\seee_\rho}$,
\begin{equation}
    \mathcal X_3^{(2)}
    =
    -\frac{1}{2}
    \sum_{\rho\in S_3}
    \seee_\rho^2\,
    \Theta_\rho^{\rm M}\,.
    \label{eq:fivepoint-classical-second-order}
\end{equation}
Using $s_\rho(i,j)^2=1$, its expansion can be written explicitly as
\begin{align}
    \mathcal X_3^{(2)}
    ={}&
    -\frac{1}{2}
    \left(
        c_{12}^2\tau_{12}^2
        +c_{13}^2\tau_{13}^2
        +c_{23}^2\tau_{23}^2
    \right)
    \sum_{\rho\in S_3}\Theta_\rho^{\rm M}
    \nonumber\\
    &-
    c_{12}c_{13}\tau_{12}\tau_{13}
    \sum_{\rho\in S_3}
    s_\rho(1,2)s_\rho(1,3)\Theta_\rho^{\rm M}
    \nonumber\\
    &-
    c_{12}c_{23}\tau_{12}\tau_{23}
    \sum_{\rho\in S_3}
    s_\rho(1,2)s_\rho(2,3)\Theta_\rho^{\rm M}
    \nonumber\\
    &-
    c_{13}c_{23}\tau_{13}\tau_{23}
    \sum_{\rho\in S_3}
    s_\rho(1,3)s_\rho(2,3)\Theta_\rho^{\rm M}\,.
    \label{eq:fivepoint-classical-expanded}
\end{align}
The terms in the first line have no ordering dependence; each is proportional to
\begin{equation}
    \sum_{\rho\in S_3}\Theta_\rho^{\rm M}=0\,,
\end{equation}
the same mechanism responsible for the leading hyperclassical cancellation \eqref{eq:leadinghyper5point}. Consequently all terms of the form
$c_{ij}^2\tau_{ij}^2$ cancel. After back-integration these terms would have
produced linear propagators raised to cubic powers, which are absent in the
corresponding worldline description.

We can represent monomials in $c_{ij}\tau_{ij}$ as graphs, with vertices representing massless emissions and $\tau_{ij}$ representing edges connecting vertices $i$ and $j$. These graphs, when combined with propagator prescriptions, will be exactly the diagrams appearing in WQFT, as shown explicitly in Section~\ref{Section new Murua for general}. The surviving terms involve products associated with two distinct
edges. With three vertices, every such pair of edges connects all vertices and therefore defines a spanning tree.%
\footnote{A spanning tree is  a connected tree graph whose vertex set includes all of $\{\tau_1,\dots,\tau_n\}$.} For example,
\begin{align}
    c_{12}c_{23}\tau_{12}\tau_{23} \qquad \begin{tikzpicture}[baseline={([yshift=-1ex]current bounding box.center)}, thick]
    \node[draw,circle,fill=black,inner sep=1.8pt,label=above:$\tau_1$] (v1) at (0,0) {};
    \node[draw,circle,fill=black,inner sep=1.8pt,label=above:$\tau_2$] (v2) at (1.2,0) {};
    \node[draw,circle,fill=black,inner sep=1.8pt,label=above:$\tau_3$] (vn) at (2.4,0) {};
    \draw (v1) -- (v2);
    \draw (v2) -- (vn);
\end{tikzpicture}\label{eq:firstmonomial5point}
\end{align}
and similarly for the other two choices of central vertex. In contrast, a term which would give rise to a cubic propagator is, for example,
\begin{align}
    \quad c_{12}^2\tau_{12}^2 \qquad \qquad\begin{tikzpicture}[baseline={([yshift=-1ex]current bounding box.center)}, thick]
    \node[draw,circle,fill=black,inner sep=1.8pt,label=above:$\tau_1$] (v1) at (0,0) {};
    \node[draw,circle,fill=black,inner sep=1.8pt,label=above:$\tau_2$] (v2) at (1.2,0) {};
    \node[draw,circle,fill=black,inner sep=1.8pt,label=above:$\tau_3$] (vn) at (2.4,0) {};
    \draw (v1) to[out=60,in=120] (v2);
    \draw (v1) to[out=-60,in=-120] (v2);
\end{tikzpicture}
\end{align}
with the vertex $3$ left disconnected.%
\footnote{In a worldline calculation, this process would have two $hz$ vertices and one $hz^2$ vertex; the two $z$ propagators must connect the $hz^2$ vertex to the two other vertices to connect the graph.} This is the
first indication of the general structure discussed in
Section~\ref{sec:generalcancellationviashuffles}: disconnected structures cancel, whereas spanning trees survive.

To make the ordering structure of a surviving tree explicit, consider the tree $g$ with edges $\{a,b\}$ and $\{b,c\}$ and choose the reference orientation $(a\,b\,c)$.
Its ordering sign is
\begin{equation}
    \epsilon_{g}(\rho)
    \coloneq
    s_\rho(a,b)s_\rho(b,c)\,.
\end{equation}
The contribution to $iN_5$ associated with this tree is therefore
\begin{equation}
    -(i\lambda)^3c_{ab}c_{bc}
    \int d\tau_{ab}\,d\tau_{bc}\,
    \tau_{ab}\tau_{bc}\,
    e^{i\sum_{i=1}^3D_i\tau_i}
    \sum_{\rho\in S_3}
    \epsilon_{g}(\rho)\,
    \Theta_\rho^{\rm M}\,.
    \label{eq:fivepoint-tree-sign-sum}
\end{equation}
The sign-weighted sum simplifies to
\begin{equation}
    \sum_{\rho\in S_3}
    \epsilon_{g}(\rho)\,
    \Theta_\rho^{\rm M}
    =
    4\,\Theta_{abc}^{\rm M}\,.
    \label{eq:fivepoint-tree-four}
\end{equation}
Using \eqref{eq:fivepoint-tree-four}, the contribution of the tree with
central vertex $b$ becomes
\begin{align}
    -4(i\lambda)^3c_{ab}c_{bc}
    \int d\tau_{ab}\,d\tau_{bc}\,
    \tau_{ab}\tau_{bc}\,
    e^{i\sum_{i=1}^3D_i\tau_i}\,
    \Theta_{abc}^{\rm M}\,.
    \label{eq:fivepoint-tree-theta}
\end{align}
Momentum conservation, $D_a+D_b+D_c=0$, allows us to write
\begin{equation}
    \sum_{i=1}^3D_i\tau_i
    =
    D_a\tau_{ab}
    +
    D_{ab}\tau_{bc}\,,
\end{equation}
where $D_{ab}=D_a+D_b$. Using the definition of the dressed theta function $\Theta_\rho^{\rm M}$,
we therefore obtain
\begin{align}
    -4(i\lambda)^3c_{ab}c_{bc}
    \int d\tau_{ab}d\tau_{bc}\,
    \tau_{ab}\tau_{bc}\,
    e^{iD_a\tau_{ab}}
    e^{iD_{ab}\tau_{bc}}
    \sum_{\sigma_a,\sigma_b=\pm}
    \omega (\sigma_a,\sigma_b)
    \sigma_a\sigma_b\,
    \theta(-\sigma_a\tau_{ab})
    \theta(-\sigma_b\tau_{bc})\,.
    \label{eq:fivepoint-tree-causal}
\end{align}
The proper-time integrations now factorise, one for each edge of the
spanning tree. For fixed
$\sigma_a,\sigma_b=\pm1$,
\begin{align}
    &\sigma_a\sigma_b
    \int d\tau_{ab}\,d\tau_{bc}\,
    \tau_{ab}\tau_{bc}\,
    \theta(-\sigma_a\tau_{ab})
    \theta(-\sigma_b\tau_{bc})
    e^{iD_a\tau_{ab}+iD_{ab}\tau_{bc}}
    \nonumber\\
    &\hspace{2cm}
    =
    \frac{1}{(-D_a+i\sigma_a\vareps)^2}
    \frac{1}{(-D_{ab}+i\sigma_b\vareps)^2}\,.
    \label{eq:FivePointCentralBackIntegrationB}
\end{align}
Thus the contribution associated with the tree whose central vertex is
$b$ is 
\begin{equation}
    -4(i\lambda)^3c_{ab}c_{bc}
    \sum_{\sigma_a,\sigma_b=\pm}
    \omega (\sigma_a,\sigma_b)
    \frac{1}{(-D_a+i\sigma_a\vareps)^2}
    \frac{1}{(-D_{ab}+i\sigma_b\vareps)^2}\,.
    \label{eq:FivePointFixedCentralVertexB}
\end{equation}
The relevant chain Murua coefficients are 
\begin{equation}
    \omega (+,+)=\omega (-,-)=\frac13\,,
    \qquad
    \omega (+,-)=\omega (-,+)=\frac16\,.
\end{equation}
For example, choosing
\begin{equation}
    (a\,b\,c)=(1\,2\,3)\,,
\end{equation}
the central vertex is $b=2$, and the corresponding spanning tree is
associated with the monomial in
\eqref{eq:firstmonomial5point}. Since
\begin{equation}
    \sum_{i=1}^3D_i\tau_i
    =
    D_1\tau_{12}
    +
    D_{12}\tau_{23}\,,
\end{equation}
its contribution is
\begin{align}
    \left. iN_5\right|_{b=2}
    ={}&
    -4(i\lambda)^3c_{12}c_{23}
    \Bigg[
        \frac13
        \frac{1}{(D_1+i\vareps)^2}
        \frac{1}{(D_{12}+i\vareps)^2}
        +
        \frac13
        \frac{1}{(D_1-i\vareps)^2}
        \frac{1}{(D_{12}-i\vareps)^2}
        \nonumber\\
        &\hspace{1.8cm}
        +
        \frac16
        \frac{1}{(D_1+i\vareps)^2}
        \frac{1}{(D_{12}-i\vareps)^2}
        +
        \frac16
        \frac{1}{(D_1-i\vareps)^2}
        \frac{1}{(D_{12}+i\vareps)^2}
    \Bigg]\,.
    \label{eq:FivePointCentralVertexTwo}
\end{align}
The five-point Magnus Compton amplitude, $iN_5$, is obtained by summing over the three possible choices of central vertex $b$. A sum over all $(a\,b\,c)\in S_3$ counts each tree twice, since $(a\,b\,c)$ and $(c\,b\,a)$ give the same contribution. Hence
\begin{align}
    iN_5 = -4(i\lambda)^3\sum_{ (a\,b\,c)\in C_3}c_{ab}c_{bc}\, P_b(D_a, D_{ab})\, , 
\end{align}
where the sum is over all 3 cyclic permutations and
\begin{align}
    P_b(D_a,D_{ab})
     =
    \Bigg[&
        \frac13
        \frac{1}{(D_a+i\vareps)^2}
        \frac{1}{(D_{ab}+i\vareps)^2}
        +
        \frac13
        \frac{1}{(D_a-i\vareps)^2}
        \frac{1}{(D_{ab}-i\vareps)^2}
        \nonumber\\
        &
        +
        \frac16
        \frac{1}{(D_a+i\vareps)^2}
        \frac{1}{(D_{ab}-i\vareps)^2}
        +
        \frac16
        \frac{1}{(D_a-i\vareps)^2}
        \frac{1}{(D_{ab}+i\vareps)^2}
    \Bigg]\,.
\end{align}
Thus the classical five-point result is
built entirely from spanning trees, with each edge producing a
squared retarded or advanced linear propagator and each causal assignment
weighted by the appropriate Murua coefficient. This agrees with the
propagator structure of the Murua-dressed worldline result
\cite{Gonzo:2026yha}.


\section{\texorpdfstring{$(n+2)$}{(n+2)}-point Magnus Compton amplitudes}\label{sec:GeneralClassicalLimit}

In this section we derive the general Magnus Compton amplitude and show the cancellation of all hyperclassical terms, together with a class of classical and quantum contributions. 

\subsection{General expression}
We consider the Magnus Compton amplitude for the emission of $n$ massless quanta, focusing on a single diagram with emissions in the order $q_1,\dots, q_n$. Let $\sigma_j=\pm$ specify the retarded $(+)$ or advanced $(-)$ prescription of the $j^{\text{th}}$ propagator. The corresponding momentum-space contribution is
\begin{equation}
\label{eq:gen-expre}
    (i\lambda)^n\prod_{r=1}^{n-1}\frac{i}{(p-q_{1\dots r})^2-m^2+\sigma_r i\vareps}\ ,
\end{equation}
where we introduced $q_{1\dots r}\coloneq q_1+\cdots +q_r$. Expanding the denominator in barred variables gives
\begin{align}
    (p-q_{1\dots r})^2-m^2+\sigma_r i\vareps &= -2\bar{p}\cdot q_{1\dots r}-q_{1\dots r}\cdot q_{r+1\dots n}+\sigma_r i\vareps \nn \\
    &= -D_{1\dots r}-\sum_{i\leq r<j} c_{ij} + \sigma_ri\vareps \ ,
\end{align}
where  $D_{1\dots r}\coloneq \sum_{i=1}^r D_i$, with 
$D_i$ and $c_{ij}$ defined in \eqref{eq:invariants-1} and \eqref{eq:invariants-2}, respectively. Rewriting the retarded and advanced propagators using the Schwinger parametrisation \eqref{eq:Schw}, the contribution to the amplitude is
\begin{align}
    (i\lambda)^n\sigma_1\cdots\sigma_{n-1}\int\! d\tau_{12}\cdots &d\tau_{n-1\,n}\ \theta(-\sigma_1\tau_{12})\cdots \theta(-\sigma_{n-1}\tau_{n-1\,n}) \nn\\
    &\times\exp\left(i\sum_{r=1}^{n-1}D_{1\dots r}\, \tau_{r\,r+1} + i \sum_{r=1}^{n-1}\sum_{i\leq r<j}c_{ij}\, \tau_{r\,r+1}\right)\ .
\end{align}
The exponentials can be simplified by changing the order of summation and using momentum conservation,%
\footnote{Strictly speaking this would be automatically accounted for if we had Schwinger-parameterised $\delta(2\bar{p}\cdot q_{1\dots n})$, but we choose to suppress this in the main text. See Appendix \ref{section: New Eikonal Identity} for a derivation where this integration is kept visible.}
$\sum_{i=1}^n D_i=0$. This results in
\begin{equation}
    (i\lambda)^n\left(\prod_{j=1}^{n-1}\sigma_j\right) \int\!\left[\prod_{i=1}^{n-1} d\tau_{i\, i+1}\,  
    \theta(-\sigma_i\tau_{i\, i+1})\right]\ \,  e^{i\sum_{i=1}^n D_i \tau_i}e^{i\sum_{i<j}c_{ij}\tau_{ij}}\ .
\end{equation}
Now let $\rho=(\rho_1\,\ldots\,\rho_n)$ denote an ordering of the
 emissions of $n$ massless particles. Summing over all emission orderings and causal assignments of $\sigma_j$, the Magnus Compton amplitude for the emission of $n$ massless particles can be cast in the form 
\begin{equation}\label{eq:NMatrixElemSchwinger}
iN_{n+2}
= (i\lambda)^n
\sum_{\rho\in S_n}
\int
\Big(\prod_{j=1}^{n-1}
d\tau_{\rho_j\rho_{j+1}}\Big)\,
\Theta_\rho^{\rm M}\,
e^{i\sum_{i=1}^{n}D_i\tau_i}\,
e^{i\sum_{i<j}
c_{\rho_i\rho_j}\tau_{\rho_i\rho_j}}\, , 
\end{equation}
where
\begin{equation}
\boxed{
\Theta_\rho^{\rm M}
=
\sum_{\vec{\sigma}\in\{\pm1\}^{n-1}}
\omega(\vec\sigma)\, 
(-1)^{k(\vec \sigma)}\, 
\prod_{j=1}^{n-1}
\theta\big(-\sigma_j\tau_{\rho_j\rho_{j+1}}\big)}
\, . 
\label{eq:ThetaMagnusGeneral}
\end{equation}
We recall that 
$\vec{\sigma}= (\sigma_1, \ldots , \sigma_{n-1})$ labels a causal assignment of the propagators,  $\omega(\vec\sigma)$ is the Murua coefficient of the corresponding oriented chain, and we also used 
\begin{align}
    \prod_{j=1}^{n-1}\sigma_j = (-1)^{k(\vec\sigma)}\, , 
\end{align}
where $k(\vec\sigma)$ is  the number of advanced propagators.
The exponent $\sum_{i=1}^nD_i\tau_i$ is a straightforward generalisation of the four-point case. 
Importantly, it is invariant under permutations, a fact we will be using in the following.%
\footnote{\label{foot:Measures}Note also that the measure is invariant under permutations. Package the coordinates (including the one from the suppressed $\delta(2\bar{p}\Cdot q_{1\dots n})$ integral) as $\boldsymbol{\Delta\tau}=(\tau_{12},\tau_{23},\dots,\tau_{n-1\, n},\tau_n)$ and also let $\boldsymbol{\tau}=(\tau_1,\dots,\tau_n)$. These two can be related by $\boldsymbol{\Delta\tau} =M\boldsymbol{\tau}$, where $M_{ij}=\delta_{i j}-\delta_{i\ j-1}$. Note that $M$ is invertible, $(M^{-1})_{ij}=\sum_{\ell=i}^n\delta_{\ell j}$ and $M_{ij}(M^{-1})_{jk}=\delta_{ik}$.
Given some permutation $\sigma$, consider its permutation matrix $P_\sigma$, acting on $\boldsymbol{\tau}$ as $\boldsymbol{\tau}\mapsto P_\sigma\boldsymbol{\tau}$. Then its action on $\boldsymbol{\Delta\tau}$ is $\boldsymbol{\Delta\tau}\mapsto MP_\sigma M^{-1}\boldsymbol{\Delta\tau}$, so the Jacobian of the overall transformation is $J=|\det(MP_\sigma M^{-1})| = |\det P_\sigma| = |\pm 1|=1$.}

It will be useful to rewrite the general expression for $iN_{n+2}$ in \eqref{eq:NMatrixElemSchwinger} using variables that make the expansion in terms of graphs more transparent, writing it as 
\begin{equation}\label{eq:NExpressionQFT}
    iN_{n+2}=(i\lambda)^n\int\!\left(\prod_{j=1}^{n-1}d\tau_{j\, j+1}\right) e^{i\sum_{i=1}^n D_i\tau_i} \ \mathcal{X}_n
    \ ,
\end{equation}
where we used the permutation invariance of both the measure and $e^{i \sum_iD_i\tau_i}$. 
Here we have defined 
\begin{equation}
\label{eq:XnDefinition}
    \mathcal{X}_n
    =
    \sum_{\rho\in S_n}
    \Theta_\rho^{\rm M}\ e^{i\seee_\rho}\, , \quad \seee_\rho= \sum_{i<j}c_{\rho_i\rho_j}\tau_{\rho_i\rho_j}\,.
\end{equation}
This is the key quantity we will be studying. 
Expanding the soft exponential separates the hyperclassical, classical, and quantum orders: 
\begin{equation}
\label{eq:softexp}
    e^{i\seee_\rho}
    =
    1+i\seee_\rho
    -\frac{1}{2}\seee_\rho^2
    -\frac{i}{6}\seee_\rho^3+\cdots.
\end{equation}
Since $c_{ij}\sim \hbar^2$ and $\tau_i\sim\hbar^{-1}$, at order $r$ in the expansion of $e^{i \seee_{\rho}}$
we obtain  a factor of $\hbar^r$. Combined with the $\hbar$ factors from the measure, the overall scaling is $\hbar^{r-(n-1)}$. This means that for 
\begin{equation} \label{eq:classical-scaling-r}
\begin{aligned}
r<n-1 &\quad &&\text{the contributions are hyperclassical},\\
r=n-1 &\quad &&\text{the contributions are classical},\\
r>n-1 &\quad &&\text{the contributions are quantum}. 
\end{aligned}
\end{equation}
In the following, it will be convenient to rewrite  the exponent $\seee_\rho$ in \eqref{eq:XnDefinition} above as
\begin{equation}
\label{eq:rewriting}
    \seee_\rho=\sum_{i<j}c_{\rho_i\rho_j}\tau_{\rho_i\rho_j}=\sum_{i<j} s_\rho(i,j)\, c_{ij}\tau_{ij}\,,
\end{equation}
with the sign factors
\begin{equation}
\label{eq:signfactors}
    s_\rho(i,j)
    =
    \begin{cases}
      +1, & \text{if $i$ appears before $j$ in the ordering $\rho$}\,,\\
      -1, & \text{if $j$ appears before $i$ in the ordering $\rho$}\,.
    \end{cases}
\end{equation}
The second equality follows because, for every ordering $\rho$, $\seee_\rho$ contains the same set of monomials $c_{ij}\tau_{ij}$, with $i<j$; the dependence on $\rho$ enters only through their signs, according to whether the relative ordering of $i$ and $j$ is preserved or reversed. These sign factors will play an important role in the cancellations to come; we present a particularly impactful consequence in the following subsection.

\subsection{Cancellations due to parity}
Consider a diagram together with its partner obtained by reversing both the emission ordering and all causal arrows. Their sum is proportional to
\begin{align}
e^{i\sum_{i<j}c_{ij}\tau_{ij}}+(-1)^{n-1}\,e^{-i\sum_{i<j}c_{ij}\tau_{ij}}\, , \end{align}
from the symmetry of $c_{ij}$ and the antisymmetry of $\tau_{ij}$, with the
factor $(-1)^{n-1}$ arising from the change in the product $\prod_{j=1}^{n-1}\sigma_j$
under the flip of all $n-1$ arrows.  Expanding and combining the soft exponentials, the $m^{\rm th}$-order term is accompanied by the factor
$1+(-1)^{n-1}(-1)^{m}=1+(-1)^{m+n-1}$, and therefore cancels pairwise
whenever $m+n$ is \emph{even}.  For example, the four-point leading 
hyperclassical term has $n=2$ and $m=0$, and the five-point subleading
hyperclassical term has $n=3$ and $m=1$; consistently, the classical
terms, with $m=n-1$ and hence $m+n=2n-1$ odd, survive.  For the
hyperclassical orders with $m+n$ odd (the leading-order term at five
points, $n=3$, $m=0$,  and the subleading term at six points, $n=4$,
$m=1$), the cancellation relies 
on the Murua coefficients, as we will show in Section~\ref{sec:generalcancellationviashuffles}. The pattern of parity cancellations can be seen visually in Table \ref{tab:paritycancellations}.
\begin{table}[]
    \centering
    \[\begin{array}{c|c|c|c|c|c|} \cline{2-6} & 4\text{-point} & 5\text{-point} & 6\text{-point} & 7\text{-point} & 8\text{-point} \\ \hline \multicolumn{1}{|c|}{\text{Leading $\hbar$ }} & \cmark & \xmark & \cmark & \xmark & \cmark \\[4pt] \multicolumn{1}{|c|}{\text{NL $\hbar$}} & \shadecell & \cmark & \xmark & \cmark & \xmark \\[4pt] \multicolumn{1}{|c|}{\text{NNL $\hbar$}} & \cmark & \shadecell & \cmark & \xmark & \cmark \\[4pt] \multicolumn{1}{|c|}{\text{N}^3\text{L $\hbar$}} & Q & \cmark & \shadecell & \cmark & \xmark \\[4pt] \multicolumn{1}{|c|}{\text{N}^4\text{L $\hbar$}} & \cmark & Q & \cmark & \shadecell & \cmark \\ \hline \end{array} \]
\caption{Table summarising the parity cancellations of Magnus Compton amplitudes. Check marks (\cmark) indicate cases in which the parity cancellation occurs (regardless of whether it is hyperclassical or quantum), while crosses ($\xmark$) indicate cases in which the hyperclassical cancellations are more subtle. $Q$ represents genuine non-zero quantum contributions. Shaded boxes represent classical contributions.}
    \label{tab:paritycancellations}
\end{table}

Note that these cancellations also occur for quantum terms; no $\hbar$ expansion is needed. Specifically, the only non-vanishing terms are those of $\cO(\hbar^{2k})$. This reflects an invariance under   
\begin{equation}
    \hbar\to -\hbar\, , 
\end{equation} 
which exchanges retarded and advanced prescriptions. Reinstating factors of $\hbar$, the propagators are given by 
\begin{equation}
    \frac{i}{-2 \bar{p}\Cdot \bar{q}_i + i \sigma \, \vareps \, {\rm sgn}(\hbar) }  \, ,  
\end{equation}
where $q = \hbar \bar{q}$, that is, $\bar{q}$ is the ($\hbar$-independent) wavevector. 

\subsection{Cancellation  of hyperclassical/disconnected terms via shuffles}
\label{sec:generalcancellationviashuffles}
The remaining cancellations must therefore occur through a different mechanism, which we now discuss. First, we show how the various terms in the expansion can be understood as graphs over the time coordinates $\{\tau_1,\dots,\tau_n\}$. This provides a clean separation between hyperclassical, classical, and quantum orders in terms of graph connectedness. We then show that disconnected graphs can be written as sums over shuffles of permutations of vertices. This reduces the cancellation of hyperclassical terms, and certain classical terms, to algebraic identities between sums of permutations, which we prove in the subsequent section using the Malvenuto-Reutenauer Hopf algebra of permutations.

Expanding the soft exponential $e^{i\seee_\rho}$ in \eqref{eq:XnDefinition} according to \eqref{eq:softexp} generates monomials labelled by collections of non-negative integers $a_{ij}$. We associate each monomial with an undirected multigraph%
\footnote{A multigraph  is a graph where repeated edges (bubbles) are allowed to appear.} $\hat g$ on the $n$ emission vertices, where $a_{ij}(\hat g)$ denotes the number of edges between vertices $i$ and $j$. We then write
\begin{equation}
    \mu(\hat g)=\prod_{i<j}(c_{ij}\tau_{ij})^{a_{ij}(\hat g)}\, , 
\end{equation}
effectively making the identification 
\begin{align}
    (c_{ij}\tau_{ij})^{a_{ij}} &\leftrightarrow \text{edge between }i \text{ and } j \text{ with multiplicity } a_{ij}\ .
\end{align}
In principle, every possible such product can appear, with each $0\leq a_{ij}(\hat g)<\infty$. 
There is a one-to-one correspondence between a graph $\hat g$ and the monomial $\mu(\hat g)$ it represents. For example, 
\begin{align}
e^{c_{12}\tau_{12}+c_{13}\tau_{13}+c_{23}\tau_{23}} &= 1 + c_{12}\tau_{12} + \cdots + c_{12}\tau_{12} c_{13}\tau_{13}+ \cdots + \frac{1}{2!}(c_{23}\tau_{23})^2 + \cdots \nn\\
&=
    \begin{tikzpicture}[baseline={([yshift=-1.95ex]current bounding box.center)}, thick,font=\small]
    \node[draw,circle,fill=black,inner sep=1.8pt,label=above:$1$] (v1) at (0,0) {};
    \node[draw,circle,fill=black,inner sep=1.8pt,label=above:$2$] (v2) at (0.5,0) {};
    \node[draw,circle,fill=black,inner sep=1.8pt,label=above:$3$] (v3) at (1,0) {};
    \end{tikzpicture}
    +
    \begin{tikzpicture}[baseline={([yshift=-1.95ex]current bounding box.center)}, thick,font=\small]
    \node[draw,circle,fill=black,inner sep=1.8pt,label=above:$1$] (v1) at (0,0) {};
    \node[draw,circle,fill=black,inner sep=1.8pt,label=above:$2$] (v2) at (0.5,0) {};
    \node[draw,circle,fill=black,inner sep=1.8pt,label=above:$3$] (v3) at (1,0) {};
    \draw (v1) -- (v2);
    \end{tikzpicture}
    +\cdots+
    \begin{tikzpicture}[baseline={([yshift=-1.05ex]current bounding box.center)}, thick,font=\small]
    \node[draw,circle,fill=black,inner sep=1.8pt,label=above:$1$] (v1) at (0,0) {};
    \node[draw,circle,fill=black,inner sep=1.8pt,label=above:$2$] (v2) at (0.5,0) {};
    \node[draw,circle,fill=black,inner sep=1.8pt,label=above:$3$] (v3) at (1,0) {};
    \draw (v1) -- (v2);
    \draw[out=-60,in=-120] (v1) to (v3);
    \end{tikzpicture}
    +\cdots+\frac{1}{2!}\
    \begin{tikzpicture}[baseline={([yshift=-1.55ex]current bounding box.center)}, thick,font=\small]
    \node[draw,circle,fill=black,inner sep=1.8pt,label=above:$1$] (v1) at (0,0) {};
    \node[draw,circle,fill=black,inner sep=1.8pt,label=above:$2$] (v2) at (0.5,0) {};
    \node[draw,circle,fill=black,inner sep=1.8pt,label=above:$3$] (v3) at (1,0) {};
    \draw[out=60,in=120] (v2) to (v3);
    \draw[out=-60,in=-120] (v2) to (v3);
    \end{tikzpicture}
    +\cdots
\end{align}
With this map, we can express
\begin{equation}
    \mathcal{X}_{n}=\sum_{\hat g}\mathcal{X}_n(\hat g)\mu(\hat g)\ ,
\end{equation}
with the sum running over all possible graphs on $n$ vertices, with any edge multiplicities. The coefficient associated with the multigraph $\hat g$ is 
\begin{equation}
    \mathcal{X}_n(\hat g)=\frac{i^{|E_{\hat g}|}}{\prod_{i<j}a_{ij}(\hat g)!}\sum_{\rho\in S_n}\Theta_\rho^{\rm M} \prod_{i<j}s_{\rho}(i,j)^{a_{ij}(\hat g)}\ .
    \label{eq:monomial-coefficient}
\end{equation}
Here  $|E_{\hat g}|{=}\sum_{i<j}a_{ij}(\hat g)$ is the total number of edges of the graph, and we recall that $s_\rho(i,j)$ was defined in \eqref{eq:signfactors}.

\paragraph{Disconnected diagrams}
All graphs contributing to $\mathcal{X}_n$ can be split into two categories, connected and disconnected. We will now see that all of the disconnected diagrams (and a subset of connected ones) vanish. Importantly, this will be the mechanism underlying the cancellation of \textit{all} hyperclassical terms.

Any disconnected multigraph $\hat g^{\rm disc}$ can be written in terms of connected components
$C_1,\ldots,C_k$ and corresponding vertex sets $v_1,\dots, v_k$ such that   $v_1\cup\cdots \cup v_k = \{1,\dots,n\}$. Since there are no edges joining different components,
every factor $s_\rho(i,j)^{a_{ij}(\hat g)}$ involves two vertices $i$ and $j$
belonging to the same component. Therefore, for the disconnected graph $\hat g^{\rm disc}$ the overall sign of each term in the sum splits into independent signs
\begin{equation}
    s_\rho(\hat g^{\rm disc})=s_{\rho}(C_1)\cdots s_{\rho}(C_k) = \left(\prod_{i\to j\in C_1}s_\rho(i,j)^{a_{ij}(C_1)}\right)\cdots \left(\prod_{i\to j\in C_k}s_\rho(i,j)^{a_{ij}(C_k)}\right)\ .
\end{equation}
This quantity depends only on the relative ordering of the vertices within each component, and is independent of the relative orderings between the vertices of different components. Therefore, fixing the internal ordering $\lambda_i$ of $v_i$ in each component
$C_i$, we have that $s_\rho(C_i)=s_{\lambda_i}(C_i)$.

In light of this, it is natural to split the sum over $\rho\in S_n$ into multiple smaller sums preserving the sign of each $s_{\lambda_i}(C_i)$. Crucially, 
the permutations that do so are precisely the shuffles
\begin{equation}
    \rho\in\lambda_1\shuffle \cdots \shuffle \lambda_k\,.
\end{equation} 
This reorganisation is crucial: the shuffle product is the operation underlying the
Hopf-algebraic structure introduced in Section~\ref{Section Hopf Algebras}.

The coefficient of a disconnected multigraph then decomposes into a
linear combination of shuffle sums of dressed theta functions:
\begin{equation}
    \mathcal{X}_n (\hat g^{\rm disc})=\frac{i^{|E_{\hat g}|}}{\prod_{i<j}a_{ij}(\hat g)!}\sum_{\lambda_1\dots\lambda_k}s_{\lambda_1}(C_1)\cdots s_{\lambda_k}(C_k)\sum_{\rho\in\lambda_1\shuffle\cdots\shuffle\lambda_k}
    \Theta_\rho^{\rm M},
    \label{eq:disconnectedmonomialcoefficient}
\end{equation}
where the outer sum runs over all choices of the internal component orderings.
In fact, we claim that
\begin{equation}
\label{eq:GeneralShuffleFormula}
\boxed{
    \sum_{\rho\in\rho_1\shuffle \cdots \shuffle \rho_k} \Theta_\rho^{\rm M} = 0
}\, 
\end{equation}
for $k\geq 2$, assuming each $\rho_i$ is non-empty. This identity, which we will prove in  Section~\ref{Hyperclassical Cancellations Proof Section}, implies that 
\begin{equation}
\label{eq:discvan}
    \mathcal{X}_n (\hat g^{\rm disc}) = 0 \, ,  
\end{equation}
so that  disconnected diagrams do not contribute at all.  

As a simple example, consider the case with  three  vertices and the monomial
\begin{equation}
    \mu(\hat g)=c_{12}\tau_{12}\,.
\end{equation}
Its graph consists of the connected component $\{1,2\}$ and the isolated
vertex $\{3\}$,
\begin{align}
    \hat g=
    \begin{tikzpicture}[baseline={([yshift=-1.8ex]current bounding box.center)}, thick]
    \node[draw,circle,fill=black,inner sep=1.8pt,label=above:$\tau_1$] (v1) at (0,0) {};
    \node[draw,circle,fill=black,inner sep=1.8pt,label=above:$\tau_2$] (v2) at (1.2,0) {};
    \node[draw,circle,fill=black,inner sep=1.8pt,label=above:$\tau_3$] (vn) at (2.4,0) {};
    \draw (v1) -- (v2);
\end{tikzpicture}\ 
\end{align}
The coefficient in $\mathcal{X}_3$ corresponding to this monomial is
\begin{equation}
    \mathcal{X}_3(\hat g)=i\sum_{\rho\in S_3}\Theta^{\rm M}_\rho s_{\rho}(1,2)\ .
\end{equation}
If the internal ordering of the first component is $(1\,2)$,
the possible shuffles with the isolated vertex, which leave $s_\rho (1,2)$ unchanged,  are
\begin{equation}
\label{eq:s1}
    (1\,2)\shuffle (3)=(1\,2\,3)+(1\,3\,2)+(3\,1\,2)\,.
\end{equation}
If instead the internal ordering is $(2\,1)$, the possible shuffles are
\begin{equation}
\label{eq:s2}
    (2\,1)\shuffle(3)=(2\,1\,3)+(2\,3\,1)+(3\,2\,1)\,.
\end{equation}
Since $s_\rho(1,2)=+1$ for every ordering in
$(1\,2)\shuffle(3)$ and $s_\rho(1,2)=-1$ for every ordering in
$(2\,1)\shuffle(3)$, $\mathcal X_3(\hat g)$ can be simplified as
\begin{equation}
    \mathcal X_3(\hat g)
    =
    i
    \left[
        \sum_{\rho\in(1\,2)\shuffle(3)}
        \Theta_\rho^{\rm M}
        -
        \sum_{\rho\in(2\,1)\shuffle(3)}
        \Theta_\rho^{\rm M}
    \right].
\end{equation}
By \eqref{eq:GeneralShuffleFormula}, however, each shuffle sum vanishes separately:
\begin{equation}
    \sum_{\rho\in(1\,2)\shuffle(3)}
    \Theta_\rho^{\rm M}
    =
    \Theta_{123}^{\rm M}
    +\Theta_{132}^{\rm M}
    +\Theta_{312}^{\rm M}
    =0\,,
\end{equation}
and
\begin{equation}
    \sum_{\rho\in (2\,1)\shuffle(3)}
    \Theta_\rho^{\rm M}
    =
    \Theta_{213}^{\rm M}
    +\Theta_{231}^{\rm M}
    +\Theta_{321}^{\rm M}
    =0\,,
\end{equation}
hence $\mathcal X_3(\hat g)=0$.
Thus the shuffle decomposition makes the vanishing of this disconnected multigraph explicit.

\paragraph{Edge reduction}
Since $s_\rho(j,k)=\pm1$, we have that $s_\rho(j,k)^{a_{jk}}=\pm1$. Even edge multiplicities drop out of the ordering sign: $s_\rho(j,k)^{2p}=1$, so the cancellation works as if there were no edge between $j$ and $k$. Odd multiplicities reduce to a single factor: $s_\rho(j,k)^{2p+1}=s_\rho(j,k)$, so the cancellation is identical to the graph where there is only one edge between $j$ and $k$. Diagrammatically, this equality can be represented as follows for the case of $a_{jk}=2$:
\begin{equation}
\begin{tikzpicture}[baseline={([yshift=-0.8ex]current bounding box.center)}, thick]
    \draw[fill=blue!20!white] (-0.5,0) circle(0.5cm) node{$T_1$};
    \draw[fill=blue!20!white] (3.5,0) circle(0.5cm)node{$T_2$};
    \node[draw,circle,fill=black,inner sep=1.8pt] (v1) at (0,0) {};
    \node[draw,circle,fill=black,inner sep=1.8pt] (v2) at (1,0) {};
    \node[draw,circle,fill=black,inner sep=1.8pt] (v3) at (2,0) {};
    \node[draw,circle,fill=black,inner sep=1.8pt] (v4) at (3,0) {};
    \draw (v1) -- (v2);
    \draw[out=60, in=120] (v2) to (v3);
    \draw[out=-60,in=-120] (v2) to (v3);
    \draw (v3) -- (v4);
    \draw ($(v1)+(0.1,0.4)$) node{$i$};
    \draw ($(v2)+(0,0.4)$) node{$j$};
    \draw ($(v3)+(0.1,0.4)$) node{$k$};
    \draw ($(v4)+(-0.1,0.4)$) node{$l$};
\end{tikzpicture}
\longrightarrow
\begin{tikzpicture}[baseline={([yshift=-0.8ex]current bounding box.center)}, thick]
    \draw[fill=blue!20!white] (-0.5,0) circle(0.5cm) node{$T_1$};
    \draw[fill=blue!20!white] (3.5,0) circle(0.5cm)node{$T_2$};
    \node[draw,circle,fill=black,inner sep=1.8pt] (v1) at (0,0) {};
    \node[draw,circle,fill=black,inner sep=1.8pt] (v2) at (1,0) {};
    \node[draw,circle,fill=black,inner sep=1.8pt] (v3) at (2,0) {};
    \node[draw,circle,fill=black,inner sep=1.8pt] (v4) at (3,0) {};
    \draw (v1) -- (v2);
    \draw (v3) -- (v4);
    \draw ($(v1)+(0.1,0.4)$) node{$i$};
    \draw ($(v2)+(0,0.4)$) node{$j$};
    \draw ($(v3)+(0.1,0.4)$) node{$k$};
    \draw ($(v4)+(-0.1,0.4)$) node{$l$};
\end{tikzpicture}\ ,
\end{equation}
representing coefficients of the form
\begin{equation}
    \cdots s_\rho(i,j)s_\rho(j,k)^2s_\rho(k,l)\cdots = \cdots s_\rho(i,j)s_\rho(k,l)\cdots .
\end{equation}
We refer to this process as edge reduction. As an example for three vertices, the graph representing the repeated-edge monomial
\begin{equation}
    \mu(\hat g)=c_{12}^{2}\tau_{12}^{2}
\end{equation}
is represented as
\begin{align}
     \begin{tikzpicture}[baseline={([yshift=-1ex]current bounding box.center)}, thick]
    \node[draw,circle,fill=black,inner sep=1.8pt,label=above:$\tau_1$] (v1) at (0,0) {};
    \node[draw,circle,fill=black,inner sep=1.8pt,label=above:$\tau_2$] (v2) at (1.2,0) {};
    \node[draw,circle,fill=black,inner sep=1.8pt,label=above:$\tau_3$] (vn) at (2.4,0) {};
    \draw (v1) to[out=60,in=120] (v2);
    \draw (v1) to[out=-60,in=-120] (v2);
\end{tikzpicture}\ .
\end{align}
In this case
$s_\rho(1,2)^2=1$, so the two internal orderings enter with the same sign:
\begin{equation}
    \sum_{\rho\in(1\,2)\shuffle(3)}
    \Theta_\rho^{\rm M}
    +
    \sum_{\rho\in(2\,1)\shuffle(3)}
    \Theta_\rho^{\rm M} = \sum_{\rho \in S_3} \Theta^{\rm M}_\rho = 0\ ,
\end{equation}
 showing the same cancellation mechanism as for the following graph
\begin{align}
    \begin{tikzpicture}[baseline={([yshift=-1.8ex]current bounding box.center)}, thick]
    \node[draw,circle,fill=black,inner sep=1.8pt,label=above:$\tau_1$] (v1) at (0,0) {};
    \node[draw,circle,fill=black,inner sep=1.8pt,label=above:$\tau_2$] (v2) at (1.2,0) {};
    \node[draw,circle,fill=black,inner sep=1.8pt,label=above:$\tau_3$] (vn) at (2.4,0) {};
\end{tikzpicture}\ .
\label{fig:3dots}
\end{align}
Therefore, 
\begin{center}
\fbox{\parbox{0.85\textwidth}{\centering
{A contribution can be non-zero only if its edge-reduced graph is~connected.}}}%
\end{center}

\paragraph{Hyperclassical terms}
As argued in \eqref{eq:classical-scaling-r}, hyperclassical terms are associated with graphs with a total number of edges  $|E_{\hat g}|=r<n-1$, meaning they are necessarily disconnected. Therefore, they can be written in the form \eqref{eq:disconnectedmonomialcoefficient} and thus vanish due to \eqref{eq:GeneralShuffleFormula}. In the simplest case, when $r=0$ and the diagram is fully disconnected, \eqref{eq:GeneralShuffleFormula} reduces to
\begin{equation}
    \sum_{\rho\in(1)\shuffle\cdots \shuffle (n)}\Theta_{\rho}^{\mathrm M} = \sum_{\rho\in S_n} \Theta_{\rho}^{\mathrm M}=0\ ,
\end{equation}
which is exactly what appears in the leading hyperclassical term 
\begin{equation}
    (i\lambda)^n\int\!\left(\prod_{j=1}^{n-1}d\tau_{j\, j+1}\right) e^{i\sum_{i=1}^n D_i\tau_i} \sum_{\rho\in S_n}\Theta_{\rho}^{\mathrm M}\ .
\end{equation}
An example of a hyperclassical graph at four vertices is 
\begin{align}
    \begin{tikzpicture}[baseline={([yshift=-1.8ex]current bounding box.center)}, thick]
    \node[draw,circle,fill=black,inner sep=1.8pt,label=above:$\tau_1$] (v1) at (0,0) {};
    \node[draw,circle,fill=black,inner sep=1.8pt,label=above:$\tau_2$] (v2) at (1.2,0) {};
    \node[draw,circle,fill=black,inner sep=1.8pt,label=above:$\tau_3$] (v3) at (2.4,0) {};
    \node[draw,circle,fill=black,inner sep=1.8pt,label=above:$\tau_4$] (vn) at (3.6,0) {};
    \draw (v1) -- (v2);
\end{tikzpicture}\,.
\end{align}
\paragraph{Classical terms}
At classical order, $r=n-1$. This is the first order in the expansion where a connected graph can appear. Repeated edges can nevertheless produce disconnected graphs even at classical order. For example,
\begin{equation}
    \begin{tikzpicture}[baseline={([yshift=-0.5ex]current bounding box.center)}, thick]
    \node[draw,circle,fill=black,inner sep=1.8pt,label=above:$\tau_1$] (v1) at (0,0) {};
    \node[draw,circle,fill=black,inner sep=1.8pt,label=above:$\tau_2$] (v2) at (1.2,0) {};
    \node[draw,circle,fill=black,inner sep=1.8pt,label=above:$\tau_3$] (v3) at (2.4,0) {};
    \node[draw,circle,fill=black,inner sep=1.8pt,label=above:$\tau_4$] (v4) at (3.6,0) {};
    \draw (v1)--(v2);
    \draw[out=60,in=120] (v3) to (v4);
    \draw[out=-60,in=-120] (v3) to (v4);
    \end{tikzpicture}\ .
\end{equation}
The monomials corresponding to these graphs would represent linear propagators raised to cubic or higher powers after back-integration. Any such graph would never correspond to a spanning tree, meaning that these also vanish.

The remaining graphs are spanning trees, which have only one component, so no shuffle decomposition exists and \eqref{eq:GeneralShuffleFormula} does not apply. One example of such a graph is
\begin{equation}
    \begin{tikzpicture}[baseline={([yshift=-0.5ex]current bounding box.center)}, thick]
    \node[draw,circle,fill=black,inner sep=1.8pt,label=below:$\tau_1$] (v1) at (0,0) {};
    \node[draw,circle,fill=black,inner sep=1.8pt,label=below:$\tau_2$] (v2) at (1.2,0) {};
    \node[draw,circle,fill=black,inner sep=1.8pt,label=below:$\tau_3$] (v3) at (2.4,0) {};
    \node[draw,circle,fill=black,inner sep=1.8pt,label=below:$\tau_4$] (v4) at (3.6,0) {};
    \draw (v1)--(v2);
    \draw[out=30,in=150] (v1) to (v3);
    \draw[out=60,in=120] (v1) to (v4);
    \end{tikzpicture}\ .
\end{equation}
In Appendix~\ref{spanningtreesnotshufflessection} we argue why these terms are non-zero, and Section~\ref{Section new Murua for general} evaluates these explicitly.

\paragraph{Quantum terms}
At orders $r>n-1$ in the expansion of the soft exponential $e^{i\seee_\rho}$, corresponding to quantum terms, similar considerations apply. In particular, an edge of even multiplicity $2p$ can be disregarded, whereas an edge of odd multiplicity $2p+1$ can be reduced to a single edge. The only non-zero contributions can come from graphs which are   connected, even after reducing the multiplicity of each edge $\text{mod }2$. For example, the contributions from the graph
\begin{equation}
    \begin{tikzpicture}[baseline={([yshift=-0.5ex]current bounding box.center)}, thick]
    \node[draw,circle,fill=black,inner sep=1.8pt,label=below:$\tau_1$] (v1) at (0,0) {};
    \node[draw,circle,fill=black,inner sep=1.8pt,label=below:$\tau_2$] (v2) at (1.2,0) {};
    \node[draw,circle,fill=black,inner sep=1.8pt,label=below:$\tau_3$] (v3) at (2.4,0) {};
    \node[draw,circle,fill=black,inner sep=1.8pt,label=below:$\tau_4$] (v4) at (3.6,0) {};
    \draw[out=60,in=120] (v1) to (v2);
    \draw[out=-60,in=-120] (v1) to (v2);
    \draw (v2) to (v3);
    \draw[out=30,in=150] (v2) to (v4);
    \end{tikzpicture}\ 
\end{equation}
are zero after edge reduction. In contrast, the contributions from the graph
\begin{equation}
    \begin{tikzpicture}[baseline={([yshift=-0.5ex]current bounding box.center)}, thick]
    \node[draw,circle,fill=black,inner sep=1.8pt,label=below:$\tau_1$] (v1) at (0,0) {};
    \node[draw,circle,fill=black,inner sep=1.8pt,label=below:$\tau_2$] (v2) at (1.2,0) {};
    \node[draw,circle,fill=black,inner sep=1.8pt,label=below:$\tau_3$] (v3) at (2.4,0) {};
    \node[draw,circle,fill=black,inner sep=1.8pt,label=below:$\tau_4$] (v4) at (3.6,0) {};
    \draw (v1) to (v2);
    \draw[out=60,in=120] (v1) to (v2);
    \draw[out=-60,in=-120] (v1) to (v2);
    \draw (v2) to (v3);
    \draw[out=30,in=150] (v2) to (v4);
    \end{tikzpicture}\ 
\end{equation}
are non-zero after edge reduction.

Summarising, the shuffle identity
\eqref{eq:GeneralShuffleFormula} has three immediate consequences:
\begin{itemize}
    \item[{\bf 1.}]
    all hyperclassical contributions vanish, since their associated graphs
    are necessarily disconnected;
    \item[{\bf 2.}]
    disconnected contributions appearing at classical and quantum orders
    vanish by the same mechanism;
    \item[{\bf 3.}] connected contributions which, after edge reduction, become disconnected, also vanish.
\end{itemize}
Thus, the cancellations of these apparently different sectors are controlled by a single combinatorial structure: the shuffle of the orderings of
disconnected components. In the next section, we will show that these cancellations have a natural algebraic origin, which will also provide the framework for understanding the surviving spanning-tree contributions.

\section{Hopf-algebraic structure of the classical limit}
\label{Section Hopf Algebras}

The results of the previous section leave us with three related questions:
\begin{enumerate}
    \item Why do all disconnected contributions to the Magnus amplitude cancel before the proper-time integrations are performed?
    \item Why do all spanning-tree contributions, which form the classical sector, survive?
    \item The QFT Magnus diagrams are built entirely from  {\it chain} topologies and therefore involve only Murua coefficients of directed chains, whereas the corresponding WQFT description contains {\it arbitrary} directed tree topologies. How can the latter coefficients already be encoded in the former?
\end{enumerate}
Section~\ref{sec:generalcancellationviashuffles} reduces the first two questions to the ordering structure of disconnected and connected graphs. There we showed that hyperclassical terms and classical contributions  with linearised propagators raised to cubic or higher powers (after back-integration)  
 can be represented by disconnected multigraphs, and that the coefficient of each disconnected monomial can be written as a sum of dressed theta functions $\Theta_\rho^{\rm M}$ over shuffles of the orderings of its connected components, cf. \eqref{eq:disconnectedmonomialcoefficient}. We claimed in \eqref{eq:GeneralShuffleFormula}, but did not prove, that every such shuffle sum vanishes identically. By contrast, spanning trees do not admit such a decomposition into independent components and are expected to survive.

This distinction has a natural algebraic origin. Total orderings of the proper times can be represented by words, while the sums over disconnected components become shuffle products of words. 
Remarkably, the sum over causal assignments appearing in the dressed theta functions is governed by a familiar object in the algebra of words known as the adjoint first Eulerian idempotent $e_1^\ast$. A property of the Hopf algebra known as Friedrichs' criterion then implies that $e_1^\ast$ annihilates shuffle products, providing a general proof of the cancellation of hyperclassical and disconnected classical terms. By contrast, spanning trees are not shuffle products and survive the action of $e_1^\ast$, as we will show below.

The same algebraic structure also answers the third question. By describing the proper-time regions of directed trees through their linear extensions, we will show that the action of the same Eulerian projector, whose coefficients are the Murua coefficients of directed chains, also reconstructs the Murua coefficients of arbitrary directed tree topologies.

The word algebra introduced below therefore provides a common description of the causal ordering, the connected classical contributions, and the matching of Magnus amplitudes to WQFT.

 To make the link between the classical expansion of Section~\ref{sec:GeneralClassicalLimit} and Hopf algebras explicit, we focus on  the dressed theta function $\Theta^{\rm M}_{1\dots n}$. Recall  from \eqref{eq:ThetaMagnusGeneral} that it is given by      
 \begin{equation}
 \label{eq:MuruaThetaReminder}
 \Theta_\rho^{\rm M}
=
\sum_{\vec{\sigma}\in\{\pm1\}^{n-1}}
\omega(\vec\sigma)
(-1)^{k(\vec\sigma)}
\prod_{j=1}^{n-1}
\theta\big(-\sigma_j\tau_{\rho_j\rho_{j+1}}\big)\, ,
\end{equation}
where $\sigma_i$ keeps track of whether the $i^{\rm{th}}$ propagator is retarded ($+1$) or advanced ($-1$),  $k(\vec\sigma)$ counts the number of advanced propagators, and $\omega (\vec\sigma)$ is the Murua coefficient for a chain given in \eqref{eq:muruaclosed}.
Our aim  now is to rewrite \eqref{eq:MuruaThetaReminder} as a weighted sum over proper-time orderings labelled by permutations. The key observation is that, although $\vec{\sigma}$ specifies the full assignment of retarded and advanced propagators, its coefficient depends only on the number $k(\vec{\sigma})$ of advanced propagators, and not on their positions. It is therefore useful to reorganise the sum over causal assignments in terms of the fully ordered proper-time regions compatible with each assignment. These regions are labelled by permutations, whose relevant combinatorial data are their descent sets. We therefore begin by recalling the notion of a descent.

 Given a permutation  $\rho\in S_n$, its descent set and descent number are defined by  \begin{align} 
D(\rho) &\coloneq \bigl\{ i\in\{1,\ldots,n-1\}: \rho(i)>\rho(i+1) \bigr\}\,, \label{eq:PermDescentSet} \\ d(\rho) &\coloneq |D(\rho)|\,. \label{eq:PermDescentNumber} \end{align} 
Thus, a descent occurs at position $i$ whenever the entry of the permutation decreases from position $i$ to position $i+1$. For example, for some permutations in $S_4$, 
\begin{equation} 
\begin{aligned} 
D((1\,2\,3\,4))&=\varnothing\,, &\qquad d((1\,2\,3\,4))&=0\,, \\ 
D((2\,3\,4\,1))&=\{3\}\,, & d((2\,3\,4\,1))&=1\,, \\ 
D((3\,4\,2\,1))&=\{2,3\}\,, & d((3\,4\,2\,1))&=2\,, \\ 
D((4\,3\,2\,1))&=\{1,2,3\}\,, & d((4\,3\,2\,1))&=3\,. 
\end{aligned} 
\label{eq:S4-descent-examples}
\end{equation} 
The complementary set
\begin{equation} 
A(\rho) = \{1,\ldots,n-1\}\setminus D(\rho) \label{eq:PermAscentSet} \end{equation} 
is the ascent set of $\rho$. 
We can then  return to the causal assignments in \eqref{eq:MuruaThetaReminder}. Each sign vector $\vec{\sigma}=(\sigma_1,\ldots,\sigma_{n-1})$ determines a subset \begin{equation} D_{\vec{\sigma}} \coloneq \bigl\{ i\in\{1,\ldots,n-1\}: \sigma_i=-1 \bigr\}\,, \qquad A_{\vec{\sigma}} \coloneq \{1,\ldots,n-1\}\setminus D_{\vec{\sigma}}\,. \label{eq:CausalDescentAscentSets} 
\end{equation} For each adjacent pair $(i,i+1)$, the sign $\sigma_i$ selects the orientation of the corresponding theta function: $\sigma_i=-1$ gives $\theta_{i\,i+1}$, whereas $\sigma_i=+1$ gives $\theta_{i+1\,i}$. Moreover, 
\begin{equation} 
k(\vec{\sigma}) = |D_{\vec{\sigma}}|\,, \qquad\quad  \prod_{i=1}^{n-1}\sigma_i = (-1)^{|D_{\vec{\sigma}}|}\,. \label{eq:CausalDescentCardinality} 
\end{equation} 
For example, 
\begin{align} 
\vec{\sigma}=(1,-1,1) &\longrightarrow D_{\vec{\sigma}}=\{2\}\,, \qquad \quad \ \ \,  A_{\vec{\sigma}}=\{1,3\}\,, \label{eq:CausalDescentExampleOne} \\ \vec{\sigma}=(-1,-1,-1,1) &\longrightarrow D_{\vec{\sigma}}=\{1,2,3\}\,, \qquad A_{\vec{\sigma}}=\{4\}\,. \label{eq:CausalDescentExampleTwo} 
\end{align} 
There is a one-to-one correspondence between causal assignments $\vec{\sigma}\in\{\pm1\}^{n-1}$ and subsets $D_{\vec{\sigma}}\subseteq\{1,\ldots,n-1\}$. We may therefore rewrite the sum over causal assignments in \eqref{eq:MuruaThetaReminder} as a sum over subsets $D\subseteq\{1,\ldots,n-1\}$: 
\begin{equation} 
\Theta^{\rm M}_{1\cdots n} = \sum_{D\subseteq\{1,\ldots,n-1\}} \frac{(-1)^{|D|}} {n\binom{n-1}{|D|}} \left( \prod_{i\in D}\theta_{i\,i+1} \right) \left( \prod_{j\notin D}\theta_{j+1\,j} \right). \label{eq:ThetaMagnusSubsetSum} \end{equation} 
The reason for using the notation $D$ becomes clear when each theta-function product in \eqref{eq:ThetaMagnusSubsetSum} is decomposed into fully ordered proper-time regions. 
For a fixed subset $D$, these regions are labelled by the permutations $\rho\in S_n$ whose descent set satisfies $D(\rho)=D$.
Using the theta-function identity derived in \cite{Brandhuber:2025igz}, we have \begin{equation} \left( \prod_{i\in D}\theta_{i\,i+1} \right) \left( \prod_{j\notin D}\theta_{j+1\,j} \right) = \sum_{\substack{\rho\in S_n\\D(\rho)=D}} \bigl( \theta_{n\,n-1}\cdots\theta_{32}\theta_{21} \bigr)\rho^{-1}, \label{eq:ThetaDescentDecomposition} \end{equation} 
where we use the right-action convention for the permutation $\rho^{-1}$, which implies $\bigl( \theta_{n\,n-1}\cdots\theta_{21} \bigr)\rho^{-1} = \theta_{\rho^{-1}(n)\,\rho^{-1}(n-1)}\cdots\theta_{\rho^{-1}(2)\rho^{-1}(1)}$ (see below  \eqref{eq:PermRightAction} for more details). Thus, for a fixed causal assignment, the product of theta functions does not generally specify a unique ordering of all proper times. Instead, \eqref{eq:ThetaDescentDecomposition} decomposes the corresponding region into fully ordered proper-time regions, labelled by the permutations $\rho\in S_n$ satisfying $D(\rho)=D$.  Here the advanced propagators select the positions in $D$, while the same positions become descents in each compatible permutation.
Substituting this decomposition into \eqref{eq:ThetaMagnusSubsetSum} and summing over all subsets $D$, each permutation appears exactly once, since every $\rho\in S_n$ has a unique descent set.
Crucially, for each permutation appearing in the decomposition of 
the causal assignment~$\vec\sigma$, we have 
\begin{equation} k(\vec\sigma)  = |D(\rho)| = d(\rho). \label{eq:advanced-propagators-descents} \end{equation} 
Thus, the number $d(\rho)$ of descents of $\rho$ is precisely the number $k(\vec \sigma)$ of advanced propagators in the corresponding causal assignment. This identification is the key step that turns the causal sum into an Eulerian-idempotent sum.
We therefore obtain
\begin{align} 
\begin{split}
\Theta^{\rm M}_{1\cdots n} &= \sum_{\rho\in S_n} \frac{(-1)^{d(\rho)}} {n\binom{n-1}{d(\rho)}} \theta_{\rho^{-1}(n)\,\rho^{-1}(n-1)} \cdots \theta_{\rho^{-1}(2)\,\rho^{-1}(1)} \\ &= \bigl( \theta_{n\,n-1}\cdots\theta_{32}\theta_{21} \bigr) \left( \sum_{\rho\in S_n} \frac{(-1)^{d(\rho)}} {n\binom{n-1}{d(\rho)}}\, \rho^{-1} \right). \label{eq:DressedThetaFinalForm} 
\end{split}
\end{align} 
In the second line, we have expressed each fully ordered proper-time region as a relabelling of the reference region $\theta_{n\,n-1}\cdots\theta_{32}\theta_{21}$.

The weighted permutation sum in parentheses in \eqref{eq:DressedThetaFinalForm} is a familiar object in the algebra of words and permutations: it is precisely the adjoint first Eulerian idempotent $e_1^\ast$, defined explicitly in \eqref{eq:AdjointEulerian2}, whose coefficients are the signed Murua coefficients of directed chains. Its right action on words defines the adjoint first Eulerian projector $\pi_1^\ast$, introduced in \eqref{eq:AdjointEulerian1}. Consequently, identities among sums of dressed theta functions $\Theta_\rho^{\rm M}$ can be translated into identities among words acted on by $e_1^\ast$. In particular, the cancellations encountered in Section~\ref{sec:GeneralClassicalLimit} can be understood as consequences of the algebraic properties of this idempotent, rather than as separate identities among theta functions. This is what motivates the Hopf-algebraic viewpoint developed in the following.

\subsection{Word algebra and the adjoint Eulerian projector}
\label{World algebra and the adjoint Section}
We briefly review the algebraic ingredients required to prove in full generality the cancellation of hyperclassical and disconnected classical terms, following \cite{reutenauer2003free}. The same framework will also describe directed trees through their linear extensions and yield a new formula for the Murua coefficients of arbitrary topologies using only the known Murua coefficients of chain topologies.

Let $A$ be a set, known as the alphabet. Its elements, known as letters, can be concatenated to produce words. Here, letters label the emission vertices, and a word records a total ordering of their proper times. Thus, for example, the word $231$ represents the ordered region
$\tau_2<\tau_3<\tau_1$.

Now let $\mathbb{R}\langle A\rangle$ denote the vector space spanned by words made out of letters in $A$, with coefficients in $\mathbb{R}$. For example, if $A=\{a,b,c\}$, some possible words $w$ would be $abc,bca,aab,ccc$ and so on, and some possible elements of $\mathbb{R}\langle A\rangle$ are
\begin{align}
    P_1 &= \frac{1}{2}abc-\frac{1}{4}acb-\frac{1}{4}bac-\frac{1}{4}bca-\frac{1}{4}cab+\frac{1}{2}cba \\
    P_2 &= abc + bac + bca
    \label{eq:P2}\\
    P_3 &= abc - acb \ .
\end{align}
We can naturally define a Lie bracket on words: if $u,v\in\mathbb{R}\langle A\rangle$, we define 
\begin{align}
    [u,v] = uv-vu\, ,
\end{align}
where the `product' refers to concatenation. By linearity we can extend this to any element in $\mathbb{R}\langle A\rangle$. 

There is a subspace of $\mathbb{R}\langle A\rangle$, denoted by $\mathscr{L}(A)$, which is the smallest subspace containing $A$ which is closed under the Lie bracket. Its elements are called Lie polynomials and include every element of $A$, together with anything that can be built from them via Lie brackets or their sums. For example, $P_1$ is a Lie polynomial, as $P_1 = \frac{1}{4}[a,[b,c]]+\frac{1}{4}[c,[b,a]]$, whereas $P_2$ and $P_3$ are not.

We can then introduce a second product $\shuffle$ on $\mathbb{R}\langle A\rangle$, called the shuffle product. It is defined as for permutations (only it is more general as repeated letters are allowed), for instance if $u=u'a$ and $v=v'b$ are words (with $a$ and $b$ their final letters),
\begin{equation}
    u\shuffle v= (u'\shuffle v)a + (u\shuffle v')b\ .
\end{equation}
Again this can be extended to any element of $\mathbb{R}\langle A\rangle$ by linearity. This is precisely the operation encountered in Section~\ref{sec:generalcancellationviashuffles},  now applied to words as opposed to permutations: if two disconnected components have fixed internal orderings $u$ and $v$, then $u\shuffle v$ sums all possible ways of interleaving their vertices while preserving those internal orderings. Among the examples above, the polynomial $P_2$ introduced in \eqref{eq:P2} is the shuffle product $P_2=a\shuffle bc$, whereas neither $P_1$ nor $P_3$ is a shuffle product.

Finally, we can equip this vector space with the scalar product for
which the words form an orthonormal basis,
\begin{equation}\label{eq:InnerProdWords}
    \langle u,v\rangle=\delta_{u,v}\, .
\end{equation}
Lie polynomials satisfy the important property, called Friedrichs' criterion,  that for any $\mathcal{P}\in\mathscr{L}(A)$,
\begin{equation}
\label{eq:FriedrichsCrit}
    \langle \mathcal{P},u\shuffle v\rangle = 0
\end{equation}
for any $u$ and $v$ (non-empty), that is,  Lie polynomials are orthogonal to shuffle products with respect to the scalar product \eqref{eq:InnerProdWords}. Importantly, this orthogonality will provide the cancellation mechanism needed below:
the disconnected contributions found in Section~\ref{sec:generalcancellationviashuffles} are shuffles, whereas the Eulerian projector maps onto Lie polynomials. 

With this, we can verify that $P_2$ and $P_3$ are not Lie polynomials: using $a\shuffle bc=abc+bac+bca$ and the orthonormality of the word basis, only identical words contribute, hence  
\begin{align} \left\langle P_2,a\shuffle bc\right\rangle &= \left\langle abc,abc\right\rangle +\left\langle bac,bac\right\rangle +\left\langle bca,bca\right\rangle \nonumber\\ &=1+1+1=3\, ,  \end{align} 
and similarly, 
\begin{align} \left\langle P_3,a\shuffle bc\right\rangle  &= \left\langle abc,abc\right\rangle =1\, .  \end{align}
By contrast,  
\begin{align}
    \langle P_1,a\shuffle bc\rangle = \frac{1}{2}-\frac{1}{4}-\frac{1}{4}=0\, , 
\end{align}
and similarly for all other shuffles, 
hence $P_1$ is a Lie polynomial. 

In what follows, we will only be interested in words of a special type: these are made from letters in the alphabet $\{1,\dots,n\}$ with the letters appearing once each. In other words, we will focus on words $w_\rho=\rho(1)\rho(2)\cdots\rho(n)$ for $\rho\in S_n$. The vector space spanned by such words is denoted by $E_n$, and such polynomials are called multilinear. These basis elements are in a one-to-one correspondence with permutations, and we will often use them interchangeably when it is convenient. For example,
\begin{equation}
    2314 -\frac{1}{2}\,2341 \leftrightarrow (2\,3\,1\,4)-\frac{1}{2}(2\,3\,4\,1)
    \label{eq:using-words-perms-interchangeably}
\end{equation}

\paragraph{Action of permutations on words}
For a word $u=a_1\cdots a_n$ and a permutation
$\rho\in S_n$, we use the right action%
\footnote{The right action is more natural, since
$(u\rho_1)\rho_2=(a_{\rho_1(1)}\cdots a_{\rho_1(n)})\rho_2$, so then applying the second permutation $(u\rho_1)\rho_2=a_{\rho_1(\rho_2(1))}\cdots a_{\rho_1(\rho_2(n))} = u(\rho_1\circ\rho_2)$, causing the order in the permutation product to agree with how the successive permutations are written. In contrast, working with left actions the orders would be reversed relative to one another,  that is $\rho_1 (\rho_2 u) = (\rho_2\circ \rho_1)u$.
}
\begin{equation}\label{eq:PermRightAction}
    u\rho\coloneq a_{\rho(1)}\cdots a_{\rho(n)}\, .
\end{equation}
For example, for the word $u=abc$,
\begin{equation}
    u (3\,2\,1) = cba\ ,\qquad u(2\,3\,1) = bca\,, \qquad u(3\,1\,2) = cab\ .
\end{equation}
With respect to the canonical scalar product, this action satisfies%
\footnote{\label{foot:1}This is guaranteed because $\langle u\rho,v\rho\rangle=\langle u,v\rangle$ for any $u,v$. 
}
\begin{equation}
    \langle u,v\rho\rangle
    =
    \langle u\rho^{-1},v\rangle .
    \label{eq:PermutationAdjointScalarProduct}
\end{equation}
Using the same examples as above, $\rho^{-1}$ is
\begin{align}
    &u(3\,2\,1)^{-1} = u(3\,2\,1) = cba\\
    &u(2\,3\,1)^{-1}=u(3\,1\,2)=cab\, , 
\end{align}
since $(3\,2\,1)(3\,2\,1)=(1\,2\,3)$, so that $(3\,2\,1)^{-1}=(3\,2\,1)$ and $(2\,3\,1)(3\,1\,2)=(1\,2\,3)$ so that $(2\,3\,1)^{-1}=(3\,1\,2)$.

\paragraph{Canonical Projections}
The final concept we need to introduce is the first canonical projection $\pi_1$,
also known in the literature as the first Eulerian idempotent  
\cite{SOLOMON1968363},
\begin{equation}
    \pi_1 \colon \mathbb{R}\langle A\rangle \to \mathscr{L}(A)\ , \qquad
    \pi_1(P) \coloneq  P\, e_1\ .
\end{equation}
It is a linear map from the
vector space of words,  of which $P$ is one,   onto the subspace of Lie polynomials. When acting on a word $w$ of length $n$, it is realised by%
\footnote{Here $e_1$ is an element of $\mathbb{R}S_n$, which can be thought of as formal linear combinations of permutations with coefficients in $\mathbb{R}$. Elements inherit the product from both $\mathbb{R}$ and $S_n$, so that for $r_{1,2}\in\mathbb{R}$ and $\sigma,\sigma'\in S_n$, we have that $(r_1 \sigma)(r_2\sigma')= r_1 r_2 (\sigma\circ\sigma')$, indicating the composition of permutations by $\circ$ for clarity.}
\begin{equation}
    \pi_1(w) = \sum_{\rho\in S_n}\frac{(-1)^{d(\rho)}} {n\binom{n-1}{d(\rho)}}\, w\rho
\end{equation}
that is $e_1$ is the following linear combination of permutations: 
\begin{equation}
\label{eq:e1-def}
    e_1 \coloneq \sum_{\rho\in S_n} \frac{(-1)^{d(\rho)}}{n \binom{n-1}{d(\rho)} }\,\rho\ .
\end{equation}
Note that 
the magnitude of the coefficient multiplying the permutation $\rho$ in  \eqref{eq:e1-def} is precisely the corresponding  Murua coefficient of chains \eqref{eq:muruaclosed} with $d(\rho)$ flipped arrows,  with  the factor $(-1)^{d(\rho)}$ arising from the product of  causal signs. 
Crucially, the role of  $\pi_1$  is to project the full word space onto the subspace of Lie polynomials. 
 For example, for a word of length two, 
\begin{equation} 
e_1=\frac{1}{2}\bigl((1\,2)-(2\,1)\bigr), 
\end{equation} 
 hence 
\begin{equation} 
\pi_1(ab) = ab\,e_1 = \frac{1}{2}(ab-ba) = \frac{1}{2}[a,b]. 
\end{equation} 
Thus $\pi_1$ removes the symmetric shuffle component $\frac{1}{2}(ab+ba)=\frac{1}{2}(a\shuffle b)$ and retains the antisymmetric Lie component.
For words of length three,%
\footnote{In $S_3$, the identity permutation has no descents, the permutations $(1\,3\,2)$, $(2\,1\,3)$, $(2\,3\,1)$ and $(3\,1\,2)$ have one descent, and the reversed permutation $(3\,2\,1)$ has two descents. Their coefficients in $e_1$ are therefore $1/3$, $-1/6$ and $1/3$, respectively.}
\begin{equation} 
e_1 = \frac{1}{3}(1\,2\,3) -\frac{1}{6} \bigl((1\,3\,2)+(2\,1\,3)+(2\,3\,1)+(3\,1\,2)\bigr) +\frac{1}{3}(3\,2\,1)\, , 
\end{equation} 
and acting on the word $abc$ we get 
\begin{align} \pi_1(abc) &= abc\,e_1 \nonumber\\ &= \frac{1}{3}abc -\frac{1}{6} \bigl(acb+bac+bca+cab\bigr) +\frac{1}{3}cba \nonumber\\ &= \frac{1}{6} \Bigl( [a,[b,c]] +[[a,b],c] \Bigr). 
\end{align}
The structure of the first Eulerian idempotent closely resembles the combination of permutations appearing in \eqref{eq:DressedThetaFinalForm}, except that $\rho$ is replaced by $\rho^{-1}$. To make this correspondence precise, we consider the adjoint of $\pi_1$, denoted by $\pi_1^\ast$, defined using the inner product \eqref{eq:PermutationAdjointScalarProduct}.  Explicitly,
\begin{equation}
\pi_1^\ast(P) = P e_1^\ast\ ,
\label{eq:AdjointEulerian1}
\end{equation} where the adjoint first Eulerian idempotent is
\begin{equation}
\boxed{
e_1^\ast = \sum_{\rho\in S_n} \frac{(-1)^{d(\rho)}} {n\binom{n-1}{d(\rho)}}\, \rho^{-1}}\, .
\label{eq:AdjointEulerian2} 
\end{equation} 
Comparing \eqref{eq:AdjointEulerian2} with
\eqref{eq:DressedThetaFinalForm}, we therefore arrive at the central relation 
\begin{equation}
    \boxed{
    \Theta^{\rm M}_{1\cdots n}
    =
    \bigl(
        \theta_{n\,n-1}\cdots\theta_{32}\theta_{21}
    \bigr)e_1^* 
    }\, , 
    \label{eq:DressedThetaEulerian}
\end{equation} 
where the permutations appearing in $e_1^*$ act from the right by
relabelling the indices of the ordered theta-function region. Equation  
\eqref{eq:DressedThetaEulerian} gives the desired dictionary: the Murua-weighted causal sum in the Schwinger representation is the adjoint first Eulerian idempotent acting on the corresponding word.

\subsection{Application to hyperclassical cancellations}
\label{Hyperclassical Cancellations Proof Section}
We can now apply these algebraic results to
Section~\ref{sec:GeneralClassicalLimit}. Recall from
\eqref{eq:GeneralShuffleFormula} that the cancellation of hyperclassical
and disconnected classical contributions is equivalent  to proving
\begin{equation}
    \sum_{\rho\in u\shuffle v}\Theta^{\rm M}_{\rho}=0 \,.
    \label{eq:ShuffleSumV2}
\end{equation}
It is sufficient to establish the identity for two non-empty words $u$ and
$v$, since the corresponding result for a shuffle of any number of
connected components follows iteratively. We will now prove
\eqref{eq:ShuffleSumV2} in full generality using the algebraic properties
of the first Eulerian projector. 

From \eqref{eq:DressedThetaFinalForm}, the dressed theta function
$\Theta_{1\cdots n}^{\rm M}$ can be written in the basis of totally ordered proper-time regions.  Let
\begin{equation}
    R_\rho
    \coloneq
    \theta_{\rho_2\rho_1}
    \theta_{\rho_3\rho_2}
    \cdots
    \theta_{\rho_n\rho_{n-1}}\,,
    \qquad
    \rho\in S_n\,,
    \label{eq:RProperTimeRegion}
\end{equation}
so that $R_\rho$ represents the region
\begin{equation}
    \tau_{\rho_1}<\tau_{\rho_2}<\cdots<\tau_{\rho_n}\,.
\end{equation}
Let $\mathscr R_n$ denote the vector space spanned by these $n!$ totally ordered regions, and recall that  $E_n$ is the multilinear degree-$n$ subspace of the word algebra.  We then  introduce the linear isomorphism 
\begin{equation}
    \label{eq:T-linear-map}
    \mathcal T:
    \mathscr R_n
    \longrightarrow
    E_n\,,
    \qquad
    \mathcal T(R_\rho)=\rho\,.
\end{equation} 
Thus $\mathcal T$ simply replaces a totally ordered proper-time region
by the word carrying the same ordering.
Using the definition of $\Theta_{1\cdots n}^{\rm M}$ in \eqref{eq:DressedThetaFinalForm}, together with
\eqref{eq:AdjointEulerian1}, we  obtain
\begin{equation}
\label{eq:Ttheta}
    \mathcal T\!\left(\Theta_{1\cdots n}^{\rm M}\right)
     =
    1\cdots n  \  e_1^* = \pi_1^*(1\cdots n) \,.
\end{equation}
More generally, relabelling the reference ordering by
$\rho\in S_n$ gives
\begin{equation}
    \boxed{
    \mathcal T\!\left(\Theta_\rho^{\rm M}\right)
    =
     \rho\,  e_1^*
     =
    \pi_1^*(\rho)\,.
    }
    \label{eq:ThetaAdjointEulerian}
\end{equation}
Here $1\cdots n$ is regarded as the reference word, or equivalently the identity permutation.
In this sense, the adjoint Eulerian projector is precisely the
word-algebra representation of the dressed causal weights.
For example, when $n=2$, 
\begin{equation}
    e_1^*
    =
    \frac12((1\,2)-(2\,1))\,,
\end{equation} 
while
\begin{equation}
    \Theta_{12}^{\rm M}
    =
    \frac12(R_{12}-R_{21})
    =
    \frac12(\theta_{21}-\theta_{12})\,.
\end{equation}
Hence
\begin{equation}
    \mathcal T(\Theta_{12}^{\rm M})
    =
    \frac12(12-21)
     =
    12\,e_1^*\,  = \pi_1^*(12),
\end{equation}
and similarly
\begin{equation}
    \mathcal T(\Theta_{21}^{\rm M})
    =
    \frac12(21-12)
    =
    21\,e_1^*\, = \pi_1^*(21).
\end{equation}
This example illustrates the general correspondence \eqref{eq:ThetaAdjointEulerian}: under the map $\mathcal T$, the dressed causal weights are represented by the action of the adjoint Eulerian projector on words. We can now use this correspondence to prove \eqref{eq:ShuffleSumV2}.
We first show that the adjoint Eulerian projector annihilates every non-trivial shuffle product; we then use the map $\mathcal T$ to translate this algebraic identity back into the corresponding cancellation of dressed theta functions.  Consider any element  $X\in\mathbb R\langle A\rangle$. Since $\pi_1$ projects onto the space of Lie polynomials $\mathscr L(A)$, Friedrichs' criterion \eqref{eq:FriedrichsCrit} implies that \begin{align} \left\langle X,\pi_1^\ast(u\shuffle v) \right\rangle &= \left\langle \pi_1(X),u\shuffle v \right\rangle = 0 \end{align} for any two non-empty words $u$ and $v$. Since the canonical scalar product is non-degenerate, it follows that 
\begin{equation} \boxed{ \pi_1^\ast(u\shuffle v) = (u\shuffle v)e_1^\ast = 0\,. }
\label{eq:Eulerian-annihilates-shuffle} 
\end{equation} 
We now translate this identity back to the proper-time representation.
Applying $\mathcal T$ to the shuffle sum and using \eqref{eq:ThetaAdjointEulerian}, we then obtain 
\begin{align} \mathcal T\!\left( \sum_{\rho\in u\shuffle v} \Theta_\rho^{\rm M} \right) &= \sum_{\rho\in u\shuffle v} \pi_1^\ast(\rho) \nonumber\\ &= \pi_1^\ast(u\shuffle v) = 0\,. \label{eq:shuffle-theta-under-T} \end{align} Since $\mathcal T$ is an isomorphism, \begin{equation} \boxed{ \sum_{\rho\in u\shuffle v} \Theta_\rho^{\rm M} = 0 } \,,\label{eq:proof-GeneralShuffleFormula} \end{equation} which proves the claim made in \eqref{eq:GeneralShuffleFormula}.

Now recall from \eqref{eq:monomial-coefficient} that $\mathcal X_n(\hat g)$ denotes the coefficient multiplying the monomial $\mu(\hat g)$ in the soft expansion of  the Schwinger integrand $\mathcal X_n$. If $\hat g$ is disconnected, then \eqref{eq:disconnectedmonomialcoefficient} expresses $\mathcal X_n(\hat g)$ as a linear combination of shuffle sums over the orderings of the connected components of $\hat g$. Each such shuffle sum vanishes by \eqref{eq:proof-GeneralShuffleFormula}. It follows that
\begin{equation}
\boxed{ \mathcal X_n(\hat g^{\rm disc})=0\,, } 
\end{equation} 
where $\hat{g}^{\rm disc}$ denotes any disconnected multigraph.
In particular, all hyperclassical and disconnected classical contributions vanish. 
The cancellation therefore occurs entirely at the level of causal orderings, before any proper-time integration. A disconnected graph leaves the relative ordering of its connected components unconstrained, and summing over these interleavings produces precisely a shuffle product. The adjoint Eulerian projector annihilates this shuffle, so the contribution vanishes before integration.

In conclusion, at classical order, the only  surviving graphs are the spanning trees. In 
Appendix~\ref{spanningtreesnotshufflessection}, we 
prove  that the word polynomial associated with a spanning tree is not annihilated by the adjoint Eulerian projector.


\section{New formula for Murua coefficients}
\label{Section new Murua for general}
From the field theory  perspective, the Magnus Compton amplitude is built entirely from chain topologies, with retarded and advanced propagators assigned in
all possible ways. The only Murua coefficients that appear  are therefore those of directed chains. However, as we have seen, additional graph topologies emerge upon taking the classical limit. These arise from expanding $e^{i\seee_\rho}$: in our diagrammatic representation, the massless leg insertions form the vertices, while each factor $c_{ij}\tau_{ij}$ adds an edge between the vertices $i$ and $j$. Products of such factors therefore generate topologies beyond the chain.

The corresponding WQFT diagrams are more general \cite{Kim:2024svw}: arbitrary tree topologies occur, and every directed tree is weighted by its own Murua coefficient. WQFT Magnus diagrams follow analogous rules to the QFT Magnus diagrams: all possible topologies are drawn, each propagator can be retarded or advanced, and each directed diagram is weighted by the Murua coefficient \cite{Gonzo:2026yha}.
Therefore, if the classical limit of the QFT calculation agrees with WQFT, where are the Murua coefficients
of these non-chain topologies encoded on the QFT side? We will show that they are already contained in the chain coefficients through the adjoint Eulerian projector. The bridge between the two descriptions is provided by the so-called linear extensions of a directed tree, which we introduce below.

First, recall that the WQFT action corresponding to the QFT model considered in this paper  is%
\footnote{Note that the factor of $\lambda/m^2$ has been chosen so that the mass dimension of $\lambda_{\rm{WQFT}}$ equals that of $\lambda_{\rm{QFT}}$.}
\begin{equation}
    S=\int\!d^4x\ \frac{1}{2}\left(\partial^\mu h\right)\left(\partial_\mu h\right)-\frac{m}{2}\int\! d\tau\left(\dot{x}^2(\tau)-\frac{\lambda}{m^2} h(x(\tau))\right)\ .
\end{equation}
The corresponding Feynman rules are
\begin{align}
    \begin{tikzpicture}[baseline={([yshift=-1.2ex]current bounding box.center)},font=\small]
    \draw (0,0) -- (1.5,0);\draw[fill=black] (0,0) circle (1pt)  node[left]{$\mu$};
    \draw[fill=black] (1.5,0) circle (1pt) node[right]{$\nu$};
    \draw[-Latex] (0.2, 0.15) -- (1.3, 0.15) node[midway, above]{$\omega$};
    \end{tikzpicture}
    &= -\frac{i}{m} \frac{\eta^{\mu\nu}}{\omega^2}\\
	\begin{tikzpicture}[baseline={([yshift=-1.6ex]current bounding box.center)}]
		\draw[dotted] (-0.6,0) -- (0,0);
		\draw[vector] (0,-1) -- (0,0);
		\draw (0,0) .. controls (0.2, 0.55) and (0.75, 0.9) .. (1, 1) node[right]{\small $z^{\mu_n}(-\omega_n)$};
		\draw (0,0) .. controls (0.3, 0.4) and (0.5, 0.5) .. (1,0.6) node[right]{\small $z^{\mu_{n-1}}(-\omega_{n-1})$};
		\draw (0,0) -- (1,0) node[right]{\small $z^{\mu_1}(-\omega_1)$};
		\node at (1, 0.4){$\vdots$};
		\draw[-Latex] (0.15,-0.1) -- (0.15,-0.95) node[midway, right]{\small $q$};
		\draw[fill=black] (0,0) circle (2pt);
	\end{tikzpicture}
	&= i^{n+1} \frac{\lambda}{2m}\, e^{iq\cdot b} \hat{\delta}\left(q \cdot v+\sum_{j=1}^n \omega_{j}\right) q^{\mu_1}\cdots q^{\mu_n} \ .
\end{align}
For a given tree-level graph topology $\hat g$, the tree-level Compton process is
\begin{equation}
    \left.iT^{\rm{WQFT}}_{n}\right|_{\hat g}=
    \begin{tikzpicture}[baseline={([yshift=-0.8ex]current bounding box.center)}]
    \draw[dotted] (0,0) -- +(-1.25,0);
    \draw[dotted] (0,0) -- +(1.25,0);
    \draw[vector] (0,0) -- +(-130:1) node[shift=(-130:0.3)]{$q_1$};
    \node at (-90:0.8) {$\dots$};
    \draw[vector] (0,0) -- +(-50:1) node[shift=(-50:0.3)]{$q_n$};
    \draw[fill=white] (0,0) circle(0.6cm and 0.3cm) node{$\hat g$};
    \end{tikzpicture}
    =
    i^{2n-1}\left(\frac{\lambda}{2m^2}\right)^n m\, \hat\delta(q_{1\dots n}\cdot v) \prod_{i\leftrightarrow j\in \hat g} \frac{q_i\cdot q_j}{\omega_{ij}^2}\ ,
\end{equation}
where $\omega_{ij}=v\cdot q_I$ for $I\subset\{1,\dots, n\}$ is the (graph-dependent) energy flowing through the edge $i{\to} j$. We have set $b=0$.  To calculate the matrix element of $N$ (as opposed to $T$), we should sum over combinations of retarded and advanced propagators, weighted by the corresponding Murua coefficients \cite{Gonzo:2026yha}. 
Therefore
\begin{equation}
    \left.iN^{\rm{WQFT}}_n\right|_{\hat g}= i^{2n-1}\left(\frac{\lambda}{2m^2}\right)^n m\, \hat\delta(q_{1\dots n}\cdot v) \sum_{g\in A(\hat g)} \omega(g)\prod_{i\leftrightarrow j\in \hat g} \frac{q_i\cdot q_j}{(\omega_{ij}\pm i\varepsilon)^2}\, ,
\end{equation}
with the sign of each $i\varepsilon$ determined by the arrow structure in $g$. Using the same Schwinger parameters as in the QFT calculation and reintroducing $c_{ij}=q_i\cdot q_j$ gives
\begin{align}
    \left.iN^{\rm{WQFT}}_n\right|_{\hat g}&= i^{2n-1}2^{n-1}\lambda^n\, \hat\delta(2m\, q_{1\dots n}\cdot v) \sum_{g\in A(\hat g)}\omega(g)\int\!\prod_{i\to j \in g}\left(d\tau_{ij}\ c_{ij}\tau_{ij} e^{-i \tau_{ij}2m\, \omega_{ij}}\theta_{ji}\right)\nn\\
    &= (i\lambda)^n\, \hat{\delta}(2m\, q_{1\dots n}\cdot v) \int \prod_{i\to j \in g_0}\left( d\tau_{ij}\ c_{ij}\tau_{ij}e^{-i \tau_{ij}2m\, \omega_{ij}}\right) \mathcal{X}_n^{\text{WQFT}}(\hat g)\ ,
\end{align}
where $g_0$ is some reference orientation%
\footnote{
Note that the choice of $g_0$ does not affect the full expression. If the reference orientation is changed, any sign flips due to the product of $\tau_{ij}$ would be compensated by an overall sign in \eqref{WQFT integrand}.
}
of $\hat g$, chosen to combine all terms in the sum under a common integral.
The coefficient associated with a given undirected topology $\hat g$ in the Schwinger parameter integral is
\begin{equation}
    \mathcal{X}^{\rm{WQFT}}_n(\hat g)=2^{n-1} i^{n-1}\sum_{g\in A(\hat g)}(-1)^{s(g_0, g)}\omega(g)\left(\prod_{(i\to j)\in g}\theta_{ji}\right) \,,
    \label{WQFT integrand}
\end{equation}
where $s(g_0,g)$ counts the number of flips of edges required to go from $g_0$ to $g$. Equation \eqref{WQFT integrand} generalises the chain expressions in \eqref{eq:ThetaMagnusGeneral} and \eqref{eq:monomial-coefficient} to general topologies. Note, however, that in the worldline description there is no sum over permutations $\rho$ of the external massless particles. For example, 
\begin{equation}\label{eq:XWQFTExample}
    \mathcal{X}^{\rm{WQFT}}_4\left(
    \begin{tikzpicture}[baseline={([yshift=-0.5ex]current bounding box.center)}, thick,font=\small]
    \node[draw,circle,fill=black,inner sep=1.8pt,label=below:$\tau_1$] (v1) at (0,0) {};
    \node[draw,circle,fill=black,inner sep=1.8pt,label=below:$\tau_2$] (v2) at (0.5,0) {};
    \node[draw,circle,fill=black,inner sep=1.8pt,label=below:$\tau_3$] (v3) at (1,0) {};
    \node[draw,circle,fill=black,inner sep=1.8pt,label=below:$\tau_4$] (v4) at (1.5,0) {};
    \draw (v1)--(v2);
    \draw[out=30,in=150] (v1) to (v3);
    \draw[out=60,in=120] (v1) to (v4);
    \end{tikzpicture}
    \right) = 2^3 i^3 \sum_{\sigma_{1,2,3}\in\{\pm1\}} (-1)^{k(\vec\sigma)}\omega(\vec\sigma)\theta(-\sigma_1\tau_{12})\theta(-\sigma_2\tau_{13})\theta(-\sigma_3\tau_{14})\,.
\end{equation}
We therefore see Murua coefficients associated with graphs beyond chains. From the field theory  perspective, however, only chain Murua coefficients appear explicitly. The coefficients of these more general topologies must therefore emerge from combinations of chain coefficients upon taking the classical limit. This will require us to compare the $N$-matrix for both QFT and WQFT. However, it will be more convenient to compare $\mathcal{X}$ instead. To that end, we need to ensure the remainder of the integrand matches. Starting with the measure, we want to ensure that the transformation
\begin{equation}
    d\boldsymbol{\Delta\tau}_{g_0}=\prod_{i\to j\in g_0}d\tau_{ij}\ d\tau_n \longrightarrow d\tau_{12}\cdots d\tau_{n-1\, n} d\tau_n = d\boldsymbol{\Delta\tau}
\end{equation}
has unit Jacobian. Firstly, we need to introduce the incidence matrix of the graph $g$. It is a matrix such that
\begin{equation}
    Q(g)_{ij}=
    \begin{cases}
    1 &\text{if edge } e_j \text{ originates from } i\ ,  \\
    -1 &\text{if edge } e_j \text{ terminates at } i\ , \\
    0 &\text{otherwise }\ .
    \end{cases}
\end{equation}
It can be shown that
\begin{equation}
    \boldsymbol{\Delta\tau}_g=\begin{pmatrix}
        &&Q(g)^{\top}&& \\
        0&0&\dots&0&1
    \end{pmatrix}
    \begin{pmatrix}
    \tau_1\\ \tau_2\\ \vdots \\ \tau_n
    \end{pmatrix}
    = \begin{pmatrix}
        &&Q(g)^{\top}&& \\
        0&0&\dots&0&1
    \end{pmatrix} M^{-1}\boldsymbol{\Delta\tau}\ .
\end{equation}
The matrix $M$ was introduced in Footnote~\ref{foot:Measures}, and has determinant 1.  Additionally, using properties of the incidence matrix \cite{Bapat2010}, it can be shown that the Jacobian for the overall transformation is 1. Continuing with the example graph in \eqref{eq:XWQFTExample}, its incidence matrix is
\begin{equation}
    Q\left(
    \begin{tikzpicture}[baseline={([yshift=-0.5ex]current bounding box.center)}, thick,font=\small]
    \node[draw,circle,fill=black,inner sep=1.8pt,label=below:$\tau_1$] (v1) at (0,0) {};
    \node[draw,circle,fill=black,inner sep=1.8pt,label=below:$\tau_2$] (v2) at (0.5,0) {};
    \node[draw,circle,fill=black,inner sep=1.8pt,label=below:$\tau_3$] (v3) at (1,0) {};
    \node[draw,circle,fill=black,inner sep=1.8pt,label=below:$\tau_4$] (v4) at (1.5,0) {};
    \draw (v1)--(v2);
    \draw[out=30,in=150] (v1) to (v3);
    \draw[out=60,in=120] (v1) to (v4);
    \end{tikzpicture}
    \right)=
    \begin{pmatrix}
        1  & 1  & 1\\
        -1 & 0  & 0\\
        0  & -1 & 0 \\
        0  & 0  & -1
    \end{pmatrix}\ .
\end{equation}
Indeed, its associated measure $d\tau_{12}d\tau_{13}d\tau_{14}d\tau_4$ is related to the standard measure $d\tau_{12}d\tau_{23}d\tau_{34}d\tau_4$ via
\begin{equation}
    \begin{pmatrix}
        \tau_{12}\\ \tau_{13} \\ \tau_{14} \\\tau_4
    \end{pmatrix}
    =
    \begin{pmatrix}
        1 & -1 & 0  & 0  \\
        1 & 0  & -1 & 0  \\
        1 & 0  & 0  & -1 \\
        0 & 0  & 0  & 1
    \end{pmatrix}
    \begin{pmatrix}
        1 & 1 & 1 & 1\\
        0 & 1 & 1 & 1\\
        0 & 0 & 1 & 1\\
        0 & 0 & 0 & 1\\
    \end{pmatrix}
    \begin{pmatrix}
        \tau_{12} \\ \tau_{23} \\ \tau_{34} \\ \tau_4
    \end{pmatrix}\ ,
\end{equation}
and a quick calculation confirms the matrix has determinant 1. What remains is to confirm that the exponential factors match too. Recall that the $\omega_{ij}$ factors are solved such that they obey energy conservation. This means that at each vertex the total energy going in must equal the energy of the outgoing massless quanta:
\begin{equation}
    \sum_{j=1}^n\omega_{ji}=q_i\cdot v\ ,
\end{equation}
for each vertex $i$, where $\omega_{ji}=0$ if $i$ and $j$ are not connected. Therefore, using the antisymmetry of $\tau_{ij}$ and $\omega_{ij}$,
\begin{equation}
    -2im\sum_{i\to j\in g_0}\tau_{ij}\omega_{ij}= 2im\sum_{i=1}^n\tau_i \sum_{j=1}^n\omega_{ji} = i\sum_{i=1}^n\tau_i\ 2m q_i\cdot v=i\sum_{i=1}^n\tau_i D_i\ ,
\end{equation}
which is precisely the exponent appearing in \eqref{eq:NExpressionQFT}, recalling that we previously defined $D_i\coloneq 2q_i\cdot \bar{p}=2mq_i\cdot v$. Overall, the WQFT expression is
\begin{equation}
    \left.iN^{\rm{WQFT}}_n\right|_{\hat g} = (i\lambda)^n\, \hat{\delta}(2m\, q_{1\dots n}\cdot v) \int\! d\tau_{12}\cdots d\tau_{n-1\, n}\ e^{i\sum_{i=1}^n D_i\tau_i}\prod_{i\to j \in g_0}\left( c_{ij}\tau_{ij}\right) \mathcal{X}_n^{\text{WQFT}}(\hat g)
\end{equation}

\subsection{Linear extensions and proper-time regions}

We now show that comparing the QFT and WQFT expressions gives a simple new formula for the Murua coefficient of an arbitrary directed tree in terms of chain Murua coefficients. The appropriate combinatorial language is that of {\it linear extensions.}

Let $g$ be a directed tree on the vertex set $\{1,\ldots,n\}$. Its oriented edges impose causal constraints on the relative proper times of the vertices and hence define a partial order. A  \emph{linear extension} of this partial order is a total ordering of the vertices that is compatible with all such constraints. 

We denote by $\operatorname{LE}(g)$ the set of \emph{linear extensions} of $g$. Each $f=f_1\cdots f_n\in\operatorname{LE}(g)$ is a permutation of the vertex labels, written as a word, and corresponds to the totally ordered proper-time region 
\begin{equation} 
\tau_{f_1}<\tau_{f_2}<\cdots<\tau_{f_n}\,. \end{equation} 
We associate with $g$ the word polynomial 
\begin{equation} 
\mathcal L_g \coloneq \sum_{f\in\operatorname{LE}(g)} f \,. 
\label{eq:linear-extension-polynomial} 
\end{equation} 
The number of terms in this polynomial is equal to the number of linear extensions of the partial order, $\phi(g)$.

For example, consider the directed tree represented by a three-vertex star,
\begin{equation}
g=
    \begin{tikzpicture}[baseline={([yshift=-.8ex]current bounding box.center)}, thick, scale=1]
    \draw[causArrow] (0,0) -- (0.5,0.4);
    \draw[causArrow] (0,0) -- (0.5,-0.4);

    \draw[fill=white] (0,0) circle(3pt);
    \draw[fill=white] (0.5,0.4) circle(3pt);
    \draw[fill=white] (0.5,-0.4) circle(3pt);

    \node[xshift=-0.4cm] at (0,0) {$\tau_1$};
    \node[yshift=0.4cm] at (0.5,0.4) {$\tau_2$};
    \node[yshift=-0.4cm] at (0.5,-0.4) {$\tau_3$};
\end{tikzpicture}
\end{equation}
This ordering requires $\tau_1<\tau_2$ and $\tau_1<\tau_3$, but imposes no
relative ordering between $\tau_2$ and $\tau_3$. Hence
\begin{equation}
    \operatorname{LE}(g)=\{123,132\},
    \qquad
    \mathcal L_g=123+132 .
\end{equation}
Similarly, for the four-vertex star
\begin{equation}
g=
    \begin{tikzpicture}[
    baseline={([yshift=-.8ex]current bounding box.center)},
    thick,
    scale=1
]
    \draw[causArrow] (0,0) -- (0.7,-0.5);
    \draw[causArrow] (0,0) -- (0.7,0);
    \draw[causArrow] (0,0) -- (0.7,0.5);

    \draw[fill=white] (0,0) circle(3pt);
    \draw[fill=white] (0.7,-0.5) circle(3pt);
    \draw[fill=white] (0.7,0) circle(3pt);
    \draw[fill=white] (0.7,0.5) circle(3pt);

    \node[xshift=-0.4cm]  at (0,0)       {$\tau_1$};
    \node[yshift=0.4cm] at (0.7,0.5)   {$\tau_2$};
    \node[xshift=0.4cm] at (0.7,0)     {$\tau_3$};
    \node[yshift=-0.4cm] at (0.7,-0.5)  {$\tau_4$};
\end{tikzpicture}
\end{equation}
the vertex $1$ must appear first, while the other three vertices can appear
in any order. Thus $\mathcal{L}_g$ is the sum of $3!$ terms, 
\begin{align}
    \mathcal L_g
    &=
    1234+1243+1324+1342+1423+1432 .
    \label{eq:four-star-linear-extensions}
\end{align}
If instead one edge is reversed,
\begin{equation}
g=
    \begin{tikzpicture}[
    baseline={([yshift=-.8ex]current bounding box.center)},
    thick,
    scale=1
]
    \draw[causArrow] (-0.5,0) -- (0,0);
    \draw[causArrow] (0,0) -- (0.5,0.4);
    \draw[causArrow] (0,0) -- (0.5,-0.4);

    \draw[fill=white] (-0.5,0) circle(3pt);
    \draw[fill=white] (0,0) circle(3pt);
    \draw[fill=white] (0.5,0.4) circle(3pt);
    \draw[fill=white] (0.5,-0.4) circle(3pt);

    \node[xshift=-0.4cm] at (-0.5,0){$\tau_2$};
    \node[shift={(-0.1cm,0.4cm)}] at (0,0) {$\tau_1$};
    \node[yshift=0.4cm] at (0.5,0.4) {$\tau_3$};
    \node[yshift=-0.4cm] at (0.5,-0.4) {$\tau_4$};


\end{tikzpicture}
\end{equation}
we must have $2$ before $1$, and $1$ before both $3$ and $4$. Therefore
\begin{equation}
    \mathcal L_g
    =
    2134+2143 .
    \label{eq:four-star-one-flip-linear-extensions}
\end{equation}
In contrast, the totally directed chain
\begin{equation}
g=
    \begin{tikzpicture}[baseline={([yshift=-2ex]current bounding box.center)}, thick,scale=1]
        \draw[massive, causArrow] (0,0) -- (0.6, 0);
        \draw[massive, causArrow] (0.6, 0) -- (1.2, 0);
        \draw[massive, causArrow] (1.2, 0) -- (1.8, 0);
        \draw[fill=white] (0,0) circle(3pt);
        \draw[fill=white] (0.6,0) circle(3pt);
        \draw[fill=white] (1.2,0) circle(3pt);
        \draw[fill=white] (1.8,0) circle(3pt);
        \node[yshift=0.4cm]  at (0,0)       {$\tau_1$};
        \node[yshift=0.4cm] at (0.6,0)   {$\tau_2$};
        \node[yshift=0.4cm] at (1.2,0)     {$\tau_3$};
        \node[yshift=0.4cm] at (1.8,0)  {$\tau_4$};
        \end{tikzpicture}
\end{equation}
has a unique linear extension,
\begin{equation}
    \mathcal L_g=1234 .
\end{equation}
There is a corresponding proper-time interpretation. Let
\begin{equation}
    \Theta_g
    \coloneq
    \prod_{(i\to j)\in g}\theta_{ji}
\end{equation}
denote the proper-time region associated with the directed tree $g$, with the same arrow and $\theta$-function conventions as above. This region is the union of all totally ordered regions compatible with the arrows of $g$, namely those labelled by the linear extensions of $g$. Therefore
\begin{equation}
    \Theta_g
    =
    \sum_{f\in\operatorname{LE}(g)}R_f \,,
    \label{eq:tree-region-linear-extensions}
\end{equation}
where $R_f$ denotes the totally ordered proper-time region associated with the word $f$, defined above \eqref{eq:RProperTimeRegion}. Under the map from proper-time regions to words introduced
above \eqref{eq:T-linear-map},  this is simply
\begin{equation}\label{eq:ThetaGLinearExtensions}
    \mathcal{T}\!\left(\Theta_g\right)
    =
    \mathcal L_g \,.
\end{equation}
For example, the region associated with the three-vertex star above
decomposes into the two totally ordered regions
\begin{equation}
    \Theta_g
    =
    R_{123}+R_{132}\,,
\end{equation}
corresponding to
$\mathcal L_g=123+132$.

A useful consequence is that the sets of linear extensions associated
with two different orientations of the same underlying tree are disjoint. Consider two orientations $g,g'$ of the same graph $\hat g$.
If $g\neq g'$, then they must differ by the orientation of at least one
edge. Suppose this happens for the edge connecting $i$ and $j$. Then
\begin{equation}
    i\longrightarrow j\in g\,,
    \qquad
    j\longrightarrow i\in g'\,.
\end{equation}
Therefore, every word in $\operatorname{LE}(g)$ has $i$ before $j$, whereas every word
in $\operatorname{LE}(g')$ has $j$ before $i$. Hence
\begin{equation}
    \operatorname{LE}(g)
    \cap
    \operatorname{LE}(g')
    =
    \varnothing \,.
    \label{eq:linear-extension-blocks-disjoint}
\end{equation}
Conversely, every total ordering of the vertices induces a unique
orientation on each edge of the fixed underlying tree $\hat g$. We
therefore have the disjoint decomposition
\begin{equation}
    S_n
    =
    \bigsqcup_{g\in A(\hat g)}
    \operatorname{LE}(g)\,.
    \label{eq:orientation-partition}
\end{equation}

\subsection{From the sign-weighted word sum to linear extensions}

Recall that the coefficient of a given multigraph $\hat g$ in the classical expansion is given by $\mathcal{X}_n(\hat g)$ in \eqref{eq:monomial-coefficient}. When $\hat g$ corresponds to a spanning tree,
\begin{equation}
    \mathcal{X}_n(\hat g^{\text{spanning}}) =i^{n-1}\sum_{\rho\in S_n} \Theta_\rho^{\rm M} \prod_{(i\to j)\in g_0} s_{\rho}(i,j) = i^{n-1}\sum_{\rho\in S_n} \Theta_\rho^{\rm M}\epsilon_{g_0}(\rho)\ ,
    \label{eq:qftXwithg0}
\end{equation}
since for $\hat g$ to be a spanning tree, it cannot have any repeated factors (or edges in the corresponding graph $\hat g$), and we have introduced
\begin{equation}
    \epsilon_{g_0}(\rho)
    \coloneq
    \prod_{(i\to j)\in g_0}s_\rho(i,j)\,.
    \label{eq:epsilon-product-of-signs}
\end{equation}
Note that in \eqref{eq:monomial-coefficient} we assumed $g_0$ is the graph where $i < j$ for each $(i\to j)$. In this case we keep $g_0$ arbitrary for consistency with the WQFT result. We now compare \eqref{eq:qftXwithg0} to the coefficient arising from WQFT,
\begin{equation}
    \mathcal{X}^{\rm{WQFT}}_n(\hat g)=2^{n-1} i^{n-1}\sum_{g\in A(\hat g)}(-1)^{s(g_0, g)}\omega(g)\,\Theta_g \,.
\end{equation}
Having isolated the classical contribution in the QFT calculation, we now match it to the corresponding WQFT expression. This requires
\begin{equation}
    \boxed{\mathcal{X}_n(\hat g^{\text{spanning}})=\mathcal{X}^{\rm{WQFT}}_n(\hat g)}\ .
\end{equation}
It is convenient to compare them after applying the map $\mathcal T$ of \eqref{eq:T-linear-map}:
\begin{align}
\mathcal{T}\left(
    \mathcal{X}_n(\hat g^{\text{spanning}}) 
\right) = 
    i^{n-1}\sum_{\rho\in S_n}
    \epsilon_{g_0}(\rho)\,
    \mathcal{T}(\Theta_\rho^{\rm M})
&=
i^{n-1}\sum_{\rho\in S_n}
\epsilon_{g_0}(\rho)\,
\pi_1^*(\rho) \\ \nonumber &=
i^{n-1}\pi_1^*
\left(
    F_{g_0}
\right)\,,
\label{eq:tree-coefficient-word-map}
\end{align}
where we have defined the sign-weighted word polynomial
\begin{equation}
    F_{g_0}
    \coloneq
    \sum_{\rho\in S_n}
    \epsilon_{g_0}(\rho)\,\rho\,,
    \label{eq:reference-sign-word-sum}
\end{equation}
recalling that $g_0\in A(\hat g)$ is the chosen reference orientation. Therefore, under the map $\mathcal{T}$ introduced in the previous section, $\pi_1^*(F_{g_0})$ is precisely the word-valued counterpart of the coefficient
of the spanning-tree monomial in \eqref{eq:monomial-coefficient}, $\mathcal{X}_n(\hat g^{\text{spanning}})$, up to a factor of $i^{n-1}$.

On the other hand, applying $\mathcal{T}$ to the expression from WQFT gives
\begin{align}
    \mathcal{T}\left(
    \mathcal{X}_n^{\text{WQFT}}(\hat{g})\right) 
 &=  2^{n-1} i^{n-1}\sum_{g\in A(\hat g)}(-1)^{s(g_0, g)}\omega(g)\,\mathcal{T}\left(\Theta_g \right) \\ \nonumber 
 &= 2^{n-1} i^{n-1}\sum_{g\in A(\hat g)}(-1)^{s(g_0, g)}\omega(g)\,\mathcal{L}_g
\end{align}
using \eqref{eq:ThetaGLinearExtensions}.
Matching the QFT and WQFT expressions therefore requires
\begin{equation}
    \pi^*_1(F_{g_0})=2^{n-1}\sum_{g\in A(\hat g)}(-1)^{s(g_0, g)}\omega(g)\,\mathcal{L}_g\ .
    \label{eq:initial-QFT-WQFT-equality}
\end{equation}
To simplify the left-hand side, introduce
\begin{equation}
    r_\rho(i,j)
    =
    \begin{cases}
        1, & \text{if $i$ appears before $j$ in $\rho$}\,,\\
        0, & \text{if $i$ appears after $j$ in $\rho$}\,,
    \end{cases}
\end{equation}
so that
\begin{equation}
    s_\rho(i,j)=2r_\rho(i,j)-1 \,.
\end{equation}
The sign-weighted polynomial may then be written as
\begin{equation}
    F_{g_0}
    =
    \sum_{\rho\in S_n}
    \left[
        \prod_{(i\to j)\in g_0}
        \bigl(2r_\rho(i,j)-1\bigr)
    \right]\rho \,.
    \label{eq:F-r-indicators}
\end{equation}
Expanding the product gives a sum over subsets of the $n-1$ edges of
$\hat g$. The contribution in which the factor $2r_\rho(i,j)$ is selected
from every term in the product is
\begin{equation}
    2^{n-1}
    \sum_{\rho\in S_n}
    \left[
        \prod_{(i\to j)\in g_0}
        r_\rho(i,j)
    \right]\rho \,.
\end{equation}
The product $\prod_{(i\to j)\in g_0}r_\rho(i,j)$ equals one if and only if the word $\rho$ respects every arrow of $g_0$, i.e. if and only if
$\rho\in\operatorname{LE}(g_0)$. Hence this term is
\begin{equation}
    2^{n-1}\mathcal L_{g_0}\,.
\end{equation}
Every other term in the expansion of
\eqref{eq:F-r-indicators} contains ordering conditions associated with only
a proper subset of the edges of $\hat g$. Since $\hat g$ is a tree, removing at least one edge produces a disconnected graph. If its connected
components are $g_1,\ldots,g_k$, then the corresponding compatible word
sum is the shuffle
\begin{equation}
    \mathcal L_{g_1}
    \shuffle\cdots\shuffle
    \mathcal L_{g_k}\,,
    \qquad
    k\geq2\,.
\end{equation}
We therefore obtain
\begin{equation}
    \boxed{F_{g_0}
    =
    2^{n-1}\mathcal L_{g_0}
    +
    \bigl(\text{shuffle terms}\bigr)\,.}
    \label{eq:F-linear-extensions-plus-shuffles}
\end{equation}
As
shown in Section~\ref{Hyperclassical Cancellations Proof Section}, the adjoint Eulerian projector annihilates every shuffle. Applying $\pi_1^*$ to
\eqref{eq:F-linear-extensions-plus-shuffles} therefore gives
\begin{equation}
    \pi_1^*
    \left(
        F_{g_0}
    \right)
    =
    2^{n-1}
    \pi_1^*
    \left(
        \mathcal L_{g_0}
    \right)\,.
    \label{eq:F-linear-extension-projection}
\end{equation}
After an application of the map $\mathcal{T}^{-1}$, this identity can be equivalently seen as
\begin{equation}
    \sum_{\rho\in S_n}
    \epsilon_{g_0}(\rho)\,
    \Theta_\rho^{\rm M}
    =
    2^{n-1}
    \sum_{u\in\operatorname{LE}(g_0)}
    \Theta_u^{\rm M}\, .
    \label{eq:preview-linear-extension-identity}
\end{equation}
Combining \eqref{eq:initial-QFT-WQFT-equality} with \eqref{eq:F-linear-extension-projection} gives
\begin{equation}
    \boxed{
    \pi_1^*
    \left(
        \mathcal L_{g_0}
    \right)
    =
    \sum_{g\in A(\hat g)}
    (-1)^{s(g_0,g)}
    \omega(g)\,
    \mathcal L_g \,,
    }
    \label{eq:Eulerian-linear-extension-Murua}
\end{equation}
or, equivalently, in terms of proper-time regions,
\begin{equation}
  \sum_{f\in\operatorname{LE}(g_0)}
  \Theta_f^{\rm M}
  =
  \sum_{g\in A(\hat{g})}
  (-1)^{s(g_0,g)}
  \omega(g)\,\Theta_g\,.
  \label{eq:propagator-tree-murua-identity}
\end{equation}
Equation \eqref{eq:Eulerian-linear-extension-Murua} provides the link between the Eulerian-projector structure of the classical QFT Magnus amplitudes and the Murua coefficients associated with general directed trees in WQFT. Its significance is that no generic Murua coefficients appear explicitly on the QFT side: all coefficients entering $\pi_1^*$ are those of directed chains. Nevertheless, after resolving the partial ordering of $g_0$ into its linear extensions, the same expression reorganises into the WQFT sum involving Murua coefficients for all directed versions of the underlying tree. Thus, the Murua coefficients of more general directed trees are encoded in the chain coefficients of the QFT description together with the combinatorics of linear extensions.

Recall from \eqref{eq:orientation-partition} that the sets of linear extensions associated with two different orientations $g,g'\in A(\hat g)$ of the same undirected graph $\hat g$ are disjoint. Hence, for any $\gamma \in \operatorname{LE}(g)$,
\begin{equation}
    \langle \gamma,\mathcal{L}_{g'}\rangle=\delta_{gg'}\ ,
\end{equation}
allowing us to project out the term associated with $g$ from the right-hand side of \eqref{eq:Eulerian-linear-extension-Murua}. Picking an arbitrary $\gamma\in \LE(g)$,
\begin{equation}
    \langle \gamma\,,\pi_1^*(\mathcal L_{g_0})\rangle
    =
    (-1)^{s(g_0,g)}
    \omega(g)\,.
    \label{eq:block-constant-Murua}
\end{equation}
Here, $\langle \gamma,P\rangle$ denotes the coefficient of the
word $\gamma$ in the word polynomial $P$. Thus, up to the orientation sign relative to $g_0$, the Murua coefficient of the directed tree $g$ can be read directly from the coefficient of any one of its linear extensions in
$\pi_1^*(\mathcal L_{g_0})$.

This makes explicit how the Murua coefficients of general directed trees are encoded by the Eulerian projector: although the coefficients entering the projector are those associated with directed chains, its action on the polynomial of linear extensions reorganises them to produce the coefficients of arbitrary directed trees.

\subsection{A formula for arbitrary Murua coefficients}

To extract the coefficient of a given directed tree $g$, we may choose the reference orientation itself to be $g_0=g$. Since
$s(g,g)=0$, equation \eqref{eq:block-constant-Murua} gives
\begin{align}
    \omega(g)
    &=
    \langle\gamma\,,
    \pi_1^*(\mathcal L_g)\rangle =
\sum_{f\in\operatorname{LE}(g)}
    \langle\gamma\,,
    \pi_1^*(f)  \rangle
=\sum_{f\in\operatorname{LE}(g)}
    \langle\gamma\,,
    f e_1^* \rangle\, ,\label{eq:omega-g}
\end{align}
using \eqref{eq:linear-extension-polynomial}.
With the right-action convention of Section~\ref{World algebra and the adjoint Section}, the coefficient of
$\gamma$ in $f e_1^*$ is 
\begin{equation}
    \langle\gamma\,,
    f e_1^*\rangle
    =
    \sum_{\rho\in S_n}\frac{(-1)^{d(\rho)}}{n\binom{n-1}{d(\rho)}}\langle\gamma,f\rho^{-1}\rangle
    =
    \frac{
        (-1)^{d(\gamma^{-1}f)}
    }{
        n
        \binom{n-1}{d(\gamma^{-1}f)}
    } \,.
\end{equation}
It follows that the Murua coefficient of an arbitrary directed tree can be written entirely in terms of its linear extensions
\begin{equation}
    \boxed{
    \omega(g)
    =
    \sum_{f\in\operatorname{LE}(g)}
    \frac{
        (-1)^{d(\gamma^{-1}f)}
    }{
        n
        \binom{n-1}{d(\gamma^{-1}f)}
    }\,.
    }
    \label{eq:general-Murua-linear-extensions}
\end{equation}
As a reminder, \eqref{eq:using-words-perms-interchangeably} treats the multilinear words appearing in linear extensions and permutations interchangeably.

The dependence on $\gamma$ in
\eqref{eq:general-Murua-linear-extensions} is only apparent. Since
$\gamma$ may be chosen to be any linear extension of $g$, we may relabel the vertices according to one chosen linear extension so that
\begin{equation}
    \gamma=(1\,2\,\cdots\, n) \,.
\end{equation}
With this choice $\gamma^{-1}=(1\,2\,\cdots\, n)$, and the formula reduces to
\begin{equation}
    \boxed{
    \omega(g)
    =
    \sum_{f\in\operatorname{LE}(g)}
    \frac{
        (-1)^{d(f)}
    }{
        n
        \binom{n-1}{d(f)}
    } \,.
    }
    \label{eq:general-Murua-linear-extensions-identity}
\end{equation}
Thus, the Murua coefficient of any directed tree is determined solely by the descent statistics of its linear extensions. Equivalently, defining $\ell_n^{(d)}$ as a directed chain with $d$ flipped arrows, for which
\begin{equation}
\omega\bigl(\ell_n^{(d)}\bigr)
=
\frac{1}{
n\binom{n-1}{d}
}\,,
\end{equation}
we may write
\begin{equation}
\omega(g)
=
\sum_{f\in\operatorname{LE}(g)}
(-1)^{d(f)}
\omega\bigl(\ell_n^{(d(f))}\bigr)\,.
\label{eq:general-Murua-from-chains}
\end{equation}
This makes the reconstruction from chains explicit. Each linear extension $f$ contributes the Murua coefficient of the chain with $d(f)$ flipped arrows, weighted by the sign $(-1)^{d(f)}$. Summing over all linear extensions therefore reconstructs the Murua coefficient of the general directed tree $g$. Thus, although only chain Murua coefficients enter the Eulerian projector governing the classical QFT Magnus amplitudes, they determine the Murua coefficients of arbitrary directed trees through their linear extensions. Equation \eqref{eq:general-Murua-from-chains} provides a direct combinatorial prescription for computing $\omega(g)$ in terms of the descent statistics of $\operatorname{LE}(g)$.

\paragraph{Examples}

We first return to the three-vertex star
\begin{equation}
g=
    \begin{tikzpicture}[baseline={([yshift=-.8ex]current bounding box.center)}, thick, scale=1]
    \draw[causArrow] (0,0) -- (0.5,0.4);
    \draw[causArrow] (0,0) -- (0.5,-0.4);

    \draw[fill=white] (0,0) circle(3pt);
    \draw[fill=white] (0.5,0.4) circle(3pt);
    \draw[fill=white] (0.5,-0.4) circle(3pt);

    \node[xshift=-0.4cm] at (0,0) {$\tau_1$};
    \node[yshift=0.4cm] at (0.5,0.4) {$\tau_2$};
    \node[yshift=-0.4cm] at (0.5,-0.4) {$\tau_3$};
\end{tikzpicture}
\end{equation}
for which
\begin{equation}
    \mathcal L_g=123+132 \,.
\end{equation}
Taking $\gamma=(1\,2\,3)$, the two relative permutations have descent numbers 
\begin{equation}
    d((1\,2\,3))=0\,,
    \qquad
    d((1\,3\,2))=1 \,.
\end{equation}
Equation \eqref{eq:general-Murua-linear-extensions} therefore gives
\begin{equation}
    \omega\left(
    \begin{tikzpicture}[baseline={([yshift=-.8ex]current bounding box.center)}, thick, scale=1]
    \draw[causArrow] (0,0) -- (0.5,0.4);
    \draw[causArrow] (0,0) -- (0.5,-0.4);
    \draw[fill=white] (0,0) circle(3pt);
    \draw[fill=white] (0.5,0.4) circle(3pt);
    \draw[fill=white] (0.5,-0.4) circle(3pt);
\end{tikzpicture}
    \ \right)
    =
    \frac{1}{3}
    -
    \frac{1}{3\binom{2}{1}}
    =
    \frac13-\frac16
    =
    \frac16 \,,
\end{equation}
in agreement with the known Murua coefficient.

As another example, consider the four-vertex star
\begin{equation}
g=
    \begin{tikzpicture}[
    baseline={([yshift=-.8ex]current bounding box.center)},
    thick,
    scale=1
]
    \draw[causArrow] (0,0) -- (0.7,-0.5);
    \draw[causArrow] (0,0) -- (0.7,0);
    \draw[causArrow] (0,0) -- (0.7,0.5);

    \draw[fill=white] (0,0) circle(3pt);
    \draw[fill=white] (0.7,-0.5) circle(3pt);
    \draw[fill=white] (0.7,0) circle(3pt);
    \draw[fill=white] (0.7,0.5) circle(3pt);

    \node[xshift=-0.4cm]  at (0,0)       {$\tau_1$};
    \node[yshift=0.4cm] at (0.7,0.5)   {$\tau_2$};
    \node[xshift=0.4cm] at (0.7,0)     {$\tau_3$};
    \node[yshift=-0.4cm] at (0.7,-0.5)  {$\tau_4$};
\end{tikzpicture}
\end{equation}
Its linear-extension polynomial is
\begin{equation}
    \mathcal L_g
    =
    1234+1243+1324+1342+1423+1432 \,.
\end{equation}
Choosing $\gamma=(1\,2\,3\,4)$, we find
\begin{equation}
    \omega\left(
    \begin{tikzpicture}[
    baseline={([yshift=-.8ex]current bounding box.center)},
    thick,
    scale=1
]
    \draw[causArrow] (0,0) -- (0.7,-0.5);
    \draw[causArrow] (0,0) -- (0.7,0);
    \draw[causArrow] (0,0) -- (0.7,0.5);
    \draw[fill=white] (0,0) circle(3pt);
    \draw[fill=white] (0.7,-0.5) circle(3pt);
    \draw[fill=white] (0.7,0) circle(3pt);
    \draw[fill=white] (0.7,0.5) circle(3pt);
\end{tikzpicture}
    \ \right)
    =
    \frac14
    -
    4\frac{1}{4\binom31}
    +
    \frac{1}{4\binom32}
    =
    \frac14-\frac13+\frac1{12}
    =
    0 \,,
\end{equation}
which reproduces the vanishing Murua coefficient of the completely
outgoing four-vertex star.

Now reversing one edge,
\begin{equation}
g=
    \begin{tikzpicture}[
    baseline={([yshift=-.8ex]current bounding box.center)},
    thick,
    scale=1
]
    \draw[causArrow] (-0.5,0) -- (0,0);
    \draw[causArrow] (0,0) -- (0.5,0.4);
    \draw[causArrow] (0,0) -- (0.5,-0.4);

    \draw[fill=white] (-0.5,0) circle(3pt);
    \draw[fill=white] (0,0) circle(3pt);
    \draw[fill=white] (0.5,0.4) circle(3pt);
    \draw[fill=white] (0.5,-0.4) circle(3pt);

    \node[xshift=-0.4cm] at (-0.5,0){$\tau_2$};
    \node[shift={(-0.1cm,0.4cm)}] at (0,0) {$\tau_1$};
    \node[yshift=0.4cm] at (0.5,0.4) {$\tau_3$};
    \node[yshift=-0.4cm] at (0.5,-0.4) {$\tau_4$};
\end{tikzpicture}
\end{equation}
In this case
\begin{equation}
    \mathcal L_g
    =
    2134+2143 \,.
\end{equation}
Taking $\gamma=(2\,1\,3\,4)$, we have
\begin{equation}
    d\bigl((2\,1\,3\,4)^{-1}(2\,1\,3\,4)\bigr)=0\,,
    \qquad
    d\bigl((2\,1\,3\,4)^{-1}(2\,1\,4\,3)\bigr)=1\,,
\end{equation}
and therefore
\begin{equation}
    \omega\left(
    \begin{tikzpicture}[
    baseline={([yshift=-.8ex]current bounding box.center)},
    thick,
    scale=1
]
    \draw[causArrow] (-0.5,0) -- (0,0);
    \draw[causArrow] (0,0) -- (0.5,0.4);
    \draw[causArrow] (0,0) -- (0.5,-0.4);
    \draw[fill=white] (-0.5,0) circle(3pt);
    \draw[fill=white] (0,0) circle(3pt);
    \draw[fill=white] (0.5,0.4) circle(3pt);
    \draw[fill=white] (0.5,-0.4) circle(3pt);
\end{tikzpicture}
    \ \right)
    =
    \frac14
    -
    \frac{1}{4\binom31}
    =
    \frac14-\frac1{12}
    =
    \frac16 \,.
\end{equation}
Again this agrees with the known one-flip star coefficient.

It is also useful to see how the familiar chain coefficients arise from
the same formula. For the fully directed chain
\begin{equation}
g=
\begin{tikzpicture}[baseline={([yshift=-1.7ex]current bounding box.center)}, thick]
    \draw[causArrow] (0,0) -- (0.75,0);
    \draw[causArrow=0.6] (0.75,0) -- (1.25,0);
    \draw[causArrow=0.4] (1.95,0) -- (2.45,0);
    \draw[fill=white] (0,0) circle(3pt) node[yshift=0.4cm]{$\tau_1$};
    \draw[fill=white] (0.75,0) circle(3pt) node[yshift=0.4cm]{$\tau_2$};
    \node at (1.6,0) {$\cdots$};
    \draw[fill=white] (2.45,0) circle(3pt) node[yshift=0.4cm]{$\tau_n$};
\end{tikzpicture}
\end{equation}
there is only one linear extension,
\begin{equation}
    \mathcal L_g=12\cdots n \,.
\end{equation}
Taking $\gamma=(1\,2\,\cdots\, n)$ therefore gives immediately
\begin{equation}
    \omega(g)=\frac1n \,,
\end{equation}
which is the $k=0$ case of the chain formula
\begin{equation}
    \omega(\ell_n^{(k)})
    =
    \frac{1}{
        n\binom{n-1}{k}
    } \,.
\end{equation}

\subsection{Loop graphs}
The analysis above can be repeated for higher orders in the classical expansion. From the QFT side, this involves connected graphs containing more than $n-1$ edges, which necessarily means the underlying (undirected) graph $\hat g$ contains loops. Repeating the previous WQFT calculation, now for diagrams containing loops of worldline excitations, gives
\begin{align}
    \left.iN_n^{\rm WQFT}\right|_{\hat g}=\frac{i^{2n-1+L}2^{n-1+L}}{\prod_{i<j}a_{ij}(\hat g)!}\lambda^{n}&\hat{\delta}(2mq_{1\dots n}\cdot v)\sum_{g\in A(\hat g)}\omega(g)\nn\\*
    &\int\! d\tau_{12}\cdots d\tau_{n-1\, n}\ e^{i\sum_{i=1}^nD_i\tau_i}\prod_{i\to j\in g}\left( c_{ij}\tau_{ij}\theta_{ji}\right)\ .
\end{align}
Note that the $L$ loop integrals which were present in the momentum space expression cancelled with $L$ Schwinger parameter integrals, leaving only $n-1$ integrals (compared to the original $n-1+L$ Schwinger integrals, one for each propagator). The remaining terms in the exponent can be placed in standard form as in the tree-level case, and similarly for the measure. This means that, in the case where $\hat g$ is an $L$-loop graph,
\begin{equation}
    \left.iN^{\rm{WQFT}}_n\right|_{\hat g} = (i\lambda)^n \hat\delta(2mq_{1\dots n}\cdot v)\int\! d\tau_{12}\cdots d\tau_{n-1\, n}\ e^{i\sum_{i=1}^nD_i\tau_i}\prod_{i\to j\in g_0}(c_{ij}\tau_{ij}) \mathcal{X}^{\rm{WQFT}}_n(\hat g)\ ,
\end{equation}
as before, where now
\begin{equation}
    \mathcal{X}^{\rm{WQFT}}_n(\hat g) = \frac{2^{n-1+L}i^{n-1+L}}{\prod_{i<j}a_{ij}(\hat g)!}\sum_{g\in A(\hat g)}(-1)^{s(g_0,g)}\omega(g) \prod_{i\to j \in g}\theta_{ji}\ .
\end{equation}
On the QFT side, for a loop graph we have
\begin{equation}
    \mathcal{X}_n(\hat g) = \frac{i^{n-1+L}}{\prod_{i<j}a_{ij}(\hat g)!}\sum_{\rho\in S_n}\Theta_\rho^{\rm M}\epsilon_{g_0}(\rho)\ .
\end{equation}
The sign factor $\epsilon_{g_0}(\rho)$ was defined in \eqref{eq:epsilon-product-of-signs}, only now we allow for repeated edges. We can proceed as we did previously, applying the map $\mathcal{T}$, and taking $\gamma\in\LE(g)$ to project out the Murua coefficient $\omega(g)$, and picking $g_0=g$ to simplify the sign factor. What fails compared to the tree-level case is that in the loop-level case $\pi_1^*(F_g)\not\propto\pi_1^*(\mathcal{L}_g)$, so this simplification cannot be made. Finally, by a smart choice of labelling of $g$, we can pick $\gamma = 12\dots n$. This means that, for loop graphs,
\begin{equation}\label{eq: Murua Loop}
    \boxed{\omega(g)= 2^{-(n-1+L)}\sum_{f\in S_n}\frac{(-1)^{d(f)}}{n\binom{n-1}{d(f)}}\epsilon_g(f)\,.}
\end{equation}
This equation makes manifest the edge reduction rule for ``banana loops'' previously found in \cite{Guo:2026xaw}: a pair of repeated edges can be removed from a graph and this only changes the Murua coefficient by a factor of 4\footnote{Due to differences of convention this factor is $-4$ in \cite{Guo:2026xaw}.}
\begin{equation}
\omega(g_{r+2})
   =\frac{1}{4}\,\omega(g_r),
\label{eq:banana-rule}
\end{equation}
where $g_r$ is a graph $g$ with a specific edge with multiplicity $r$.
This is clear since the sign factors in $\epsilon_g(f)$ of \eqref{eq: Murua Loop} only depend on whether the number of repeated edges is odd or even. The overall factor of 4 that appears when removing a pair of repeated edges simply comes from the change in the number of loops $L$ in the prefactor $2^{-(n-1+L)}$. 

If the number of repeated edges is even, then the edge can be entirely removed. 
If the graph becomes disconnected after this, then the Murua coefficient is zero. For example, these edge-reduction rules give
\begin{equation}
\begin{aligned}
\omega\!\left(
\begin{tikzpicture}[baseline=-0.5ex, font=\small, thick]
    \draw[massive, causArrow] (0,0) to[bend left=35]  (1,0);
    \draw[massive, causArrow] (0,0) -- (1,0);
    \draw[massive, causArrow] (0,0) to[bend right=35] (1,0);
    \draw[fill=white] (0,0) circle(3pt);
    \draw[fill=white] (1,0) circle(3pt);
\end{tikzpicture}
\right)
&=
\frac{1}{4}\,
\omega\!\left(
\begin{tikzpicture}[baseline=-0.5ex, font=\small, thick]
    \draw[massive, causArrow] (0,0) -- (1,0);
    \draw[fill=white] (0,0) circle(3pt);
    \draw[fill=white] (1,0) circle(3pt);
\end{tikzpicture}
\right),
\\[1em]
\omega\!\left(
\begin{tikzpicture}[baseline=-0.5ex, font=\small, thick]
    \draw[massive, causArrow] (0,0) to[bend left=27]  (1,0);
    \draw[massive, causArrow] (0,0) to[bend right=27] (1,0);
    \draw[fill=white] (0,0) circle(3pt);
    \draw[fill=white] (1,0) circle(3pt);
\end{tikzpicture}
\right)
&=0 .
\end{aligned}
\label{eq:banana-reduction}
\end{equation}

\subsection{New identity for Murua coefficients}
From \eqref{eq:initial-QFT-WQFT-equality}, we can also derive a new identity for Murua coefficients. Consider taking the inner product of \eqref{eq:initial-QFT-WQFT-equality} with a sum of every permutation in $S_n$,
\begin{align}
    S=\langle\sum_{\sigma\in S_n}\sigma,\pi_1^*(F_{g_0})\rangle &= 2^{n-1+L}\sum_{g\in A(\hat g)}(-1)^{s(g_0,g)}\omega(g)\langle\sum_{\sigma\in S_n}\sigma,\mathcal{L}_g\rangle \nn\\*
    &=2^{n-1+L}\sum_{g\in A(\hat g)}(-1)^{s(g_0,g)}\omega(g) \phi(g)\ ,
\end{align}
where we used orthonormality of the inner product and the fact that the number of terms in $\mathcal{L}_g$ is $\phi(g)$, the number of linear extensions of the graph $g$. On the other hand, using adjointness,
\begin{equation}
    S=\langle\sum_{\sigma\in S_n}\sigma,\pi_1^*(F_{g_0})\rangle = \langle\pi_1\left(\sum_{\sigma\in S_n}\sigma\right),F_{g_0}\rangle\ .
\end{equation}
We will now show that this is identically zero. Consider the simpler object 
\begin{align}
    \langle\pi_1\left(\sum_{\sigma\in S_n}\sigma\right),\sigma'\rangle &= \sum_{\sigma\in S_n}\sum_{\rho\in S_n}\frac{(-1)^{d(\rho)}}{n\binom{n-1}{d(\rho)}}\langle \sigma\rho,\sigma' \rangle\nn \\
    &=\sum_{\rho\in S_n}\frac{(-1)^{d(\rho)}}{n\binom{n-1}{d(\rho)}}\ .
\end{align}
Here we used the definition of the projector $\pi_1$ in the first line. In the second line we used the fact that the inner product is orthonormal, to collapse the sum over $\sigma$. Breaking up the sum over all of $S_n$ into smaller sums with constant descent,
\begin{equation}
    \sum_{\rho\in S_n}\frac{(-1)^{d(\rho)}}{n\binom{n-1}{d(\rho)}} = \sum_{k=0}^{n-1}\sum_{\substack{\rho\in S_n\\d(\rho)=k}} \frac{(-1)^{d(\rho)}}{n\binom{n-1}{d(\rho)}}=\sum_{k=0}^{n-1}\frac{(-1)^{k}A(n,k)}{n\binom{n-1}{k}} = 0\ ,
\end{equation}
where $A(n,k)$, known as the Eulerian number, counts the number of permutations of $n$ elements with $k$ descents. The final equality is shown in Appendix \ref{app:EulerianId}. Putting it all together, we find that
\begin{equation}
    \boxed{
    \sum_{g\in A(\hat g)}
    (-1)^{s(g_0,g)}
    \phi(g)\,\omega(g)
    =
    0 \,.
    }
    \label{eq:magic-identity-from-linear-extensions}
\end{equation}

\subsection{Bootstrapping Murua coefficients}
The identity \eqref{eq:magic-identity-from-linear-extensions}, combined with the edge contraction rule \cite{Kim:2024svw,Guo:2026xaw}
\begin{align}\label{eq:edgecontraction}
\omega\left(\begin{tikzpicture}[baseline={([yshift=-.5ex]current bounding box.center)},thick]
        \draw[causArrow] (0,0) -- (1,0);
        \draw[noncausArrow] (1,0) -- (1.75,0.5);%
        \draw[noncausArrow] (1,0) -- (1.75,0.0);%
        \draw[noncausArrow] (1,0) -- (1.75,-0.5);%
        \draw[noncausArrow] (0,0) -- (-.75,0.5);%
        \draw[noncausArrow] (0,0) -- (-.75,0.0);%
        \draw[noncausArrow] (0,0) -- (-.75,-0.5);%
        \draw[fill=white] (0,0) circle(3pt);
        \draw[fill=white] (1,0) circle(3pt);
        \draw[fill=white,color=white] (-.75,0) ellipse (10pt and 20pt);
        \draw[fill=white,color=white] (1.75,0) ellipse (10pt and 20pt);
        \draw[pattern=north east lines] (-.75,0) ellipse (10pt and 20pt);
        \draw[pattern=north east lines] (1.75,0) ellipse (10pt and 20pt);
        \filldraw[fill=white,color=white] (-.75,0) circle (7pt);
        \filldraw[fill=white,color=white] (1.75,0) circle (7pt);
        \node (middle) at (-.75,0) {A}; 
        \node (middle) at (1.75,0) {B}; 
\end{tikzpicture}\right)+
\omega\left(\begin{tikzpicture}[baseline={([yshift=-.5ex]current bounding box.center)},thick]
        \draw[causArrow] (1,0) -- (0,0);
        \draw[noncausArrow] (1,0) -- (1.75,0.5);%
        \draw[noncausArrow] (1,0) -- (1.75,0.0);%
        \draw[noncausArrow] (1,0) -- (1.75,-0.5);%
        \draw[noncausArrow] (0,0) -- (-.75,0.5);%
        \draw[noncausArrow] (0,0) -- (-.75,0.0);%
        \draw[noncausArrow] (0,0) -- (-.75,-0.5);%
        \draw[fill=white] (0,0) circle(3pt);
        \draw[fill=white] (1,0) circle(3pt);
        \draw[fill=white,color=white] (-.75,0) ellipse (10pt and 20pt);
        \draw[fill=white,color=white] (1.75,0) ellipse (10pt and 20pt);
        \draw[pattern=north east lines] (-.75,0) ellipse (10pt and 20pt);
        \draw[pattern=north east lines] (1.75,0) ellipse (10pt and 20pt);
        \filldraw[fill=white,color=white] (-.75,0) circle (7pt);
        \filldraw[fill=white,color=white] (1.75,0) circle (7pt);
        \node (middle) at (-.75,0) {A}; 
        \node (middle) at (1.75,0) {B}; 
\end{tikzpicture}\right)=
\omega\left(\begin{tikzpicture}[baseline={([yshift=-.5ex]current bounding box.center)},thick]
        \draw[noncausArrow] (0,0) -- (0.75,0.5);%
        \draw[noncausArrow] (0,0) -- (0.75,0.0);%
        \draw[noncausArrow] (0,0) -- (0.75,-0.5);%
        \draw[noncausArrow] (0,0) -- (-.75,0.5);%
        \draw[noncausArrow] (0,0) -- (-.75,0.0);%
        \draw[noncausArrow] (0,0) -- (-.75,-0.5);%
        \draw[fill=white] (0,0) circle(3pt);
        \draw[fill=white,color=white] (-.75,0) ellipse (10pt and 20pt);
        \draw[fill=white,color=white] (0.75,0) ellipse (10pt and 20pt);
        \draw[pattern=north east lines] (-.75,0) ellipse (10pt and 20pt);
        \draw[pattern=north east lines] (0.75,0) ellipse (10pt and 20pt);
        \filldraw[fill=white,color=white] (-.75,0) circle (7pt);
        \filldraw[fill=white,color=white] (0.75,0) circle (7pt);
        \node (middle) at (-.75,0) {A}; 
        \node (middle) at (0.75,0) {B}; 
\end{tikzpicture}\right)\,
\end{align}
and the fact that the Murua coefficient of a single vertex is one:
\begin{equation}
    \omega\left(\begin{tikzpicture}[font=\small, thick,baseline={([yshift=-0.6ex]current bounding box.center)}]
        \draw[fill=white] (0,0) circle(3pt);
    \end{tikzpicture}\right)=1\,,
\end{equation}
is enough to \textit{fix all tree-level Murua coefficients} recursively. In addition, at loop-level we can also fix all coefficients simply by including the banana reduction rule
\eqref{eq:banana-rule}, with the rules above. 

As an example, at tree-level the first Murua coefficients can be found using
\begin{equation}
    \omega\left(\begin{tikzpicture}[font=\small, thick]
        \draw[causArrow] (0,0) -- (1,0);
        \draw[fill=white] (0,0) circle(3pt);
        \draw[fill=white] (1,0) circle(3pt);
    \end{tikzpicture}\right)+\omega\left(\begin{tikzpicture}[font=\small, thick]
        \draw[causArrow] (1,0) -- (0,0);
        \draw[fill=white] (0,0) circle(3pt);
        \draw[fill=white] (1,0) circle(3pt);
    \end{tikzpicture}\right)= \omega\left(\begin{tikzpicture}[font=\small, thick]
        \draw[fill=white] (0,0) circle(3pt);
    \end{tikzpicture}\right) \implies  \omega\left(\begin{tikzpicture}[font=\small, thick]
        \draw[causArrow] (0,0) -- (1,0);
        \draw[fill=white] (0,0) circle(3pt);
        \draw[fill=white] (1,0) circle(3pt);
    \end{tikzpicture}\right)=\frac{1}{2}
\end{equation}
and so the identity \eqref{eq:magic-identity-from-linear-extensions} plays no role.

At three vertices there are three unique diagrams
\begin{equation}
    \begin{tikzpicture}[font=\small, thick]
        \draw[massive, causArrow] (0,0) -- (1, 0);
        \draw[massive, causArrow] (1, 0) -- (2, 0);
        \draw[fill=white] (0,0) circle(3pt);
        \draw[fill=white] (1,0) circle(3pt);
        \draw[fill=white] (2,0) circle(3pt);
    \end{tikzpicture}\quad     \begin{tikzpicture}[font=\small, thick]
        \draw[massive, causArrow] (0,0) -- (1, 0);
        \draw[massive, causArrowR] (1, 0) -- (2, 0);
        \draw[fill=white] (0,0) circle(3pt);
        \draw[fill=white] (1,0) circle(3pt);
        \draw[fill=white] (2,0) circle(3pt);
    \end{tikzpicture}\quad 
    \begin{tikzpicture}[font=\small, thick]
        \draw[massive, causArrowR] (0,0) -- (1, 0);
        \draw[massive, causArrow] (1, 0) -- (2, 0);
        \draw[fill=white] (0,0) circle(3pt);
        \draw[fill=white] (1,0) circle(3pt);
        \draw[fill=white] (2,0) circle(3pt);
    \end{tikzpicture}\,,
\end{equation}
However, the contraction rule only gives two independent equations
\begin{equation}\label{eq: exampleBootstrap1}
\begin{split}
    &\omega\left(\begin{tikzpicture}[font=\small, thick]
        \draw[massive, causArrow] (0,0) -- (1, 0);
        \draw[massive, causArrow] (1, 0) -- (2, 0);
        \draw[fill=white] (0,0) circle(3pt);
        \draw[fill=white] (1,0) circle(3pt);
        \draw[fill=white] (2,0) circle(3pt);
    \end{tikzpicture}\right)+\omega\left(\begin{tikzpicture}[font=\small, thick]
        \draw[massive, causArrow] (0,0) -- (1, 0);
        \draw[massive, causArrowR] (1, 0) -- (2, 0);
        \draw[fill=white] (0,0) circle(3pt);
        \draw[fill=white] (1,0) circle(3pt);
        \draw[fill=white] (2,0) circle(3pt);
    \end{tikzpicture}\right)= \frac{1}{2}\,,\\
    &\omega\left(\begin{tikzpicture}[font=\small, thick]
        \draw[massive, causArrow] (0,0) -- (1, 0);
        \draw[massive, causArrow] (1, 0) -- (2, 0);
        \draw[fill=white] (0,0) circle(3pt);
        \draw[fill=white] (1,0) circle(3pt);
        \draw[fill=white] (2,0) circle(3pt);
    \end{tikzpicture}\right)+\omega\left(\begin{tikzpicture}[font=\small, thick]
        \draw[massive, causArrowR] (0,0) -- (1, 0);
        \draw[massive, causArrow] (1, 0) -- (2, 0);
        \draw[fill=white] (0,0) circle(3pt);
        \draw[fill=white] (1,0) circle(3pt);
        \draw[fill=white] (2,0) circle(3pt);
    \end{tikzpicture}\right)= \frac{1}{2}\,.
    \end{split}
\end{equation}
The final additional constraint comes from \eqref{eq:magic-identity-from-linear-extensions}
\begin{equation}\label{eq: exampleBootstrap2}
    \omega\left(\begin{tikzpicture}[font=\small, thick]
        \draw[massive, causArrow] (0,0) -- (1, 0);
        \draw[massive, causArrow] (1, 0) -- (2, 0);
        \draw[fill=white] (0,0) circle(3pt);
        \draw[fill=white] (1,0) circle(3pt);
        \draw[fill=white] (2,0) circle(3pt);
    \end{tikzpicture}\right)- 
    2\, \omega\left(\begin{tikzpicture}[font=\small, thick]
        \draw[massive, causArrow] (0,0) -- (1, 0);
        \draw[massive, causArrowR] (1, 0) -- (2, 0);
        \draw[fill=white] (0,0) circle(3pt);
        \draw[fill=white] (1,0) circle(3pt);
        \draw[fill=white] (2,0) circle(3pt);
    \end{tikzpicture}\right)-2\, \omega\left(\begin{tikzpicture}[font=\small, thick]
        \draw[massive, causArrowR] (0,0) -- (1, 0);
        \draw[massive, causArrow] (1, 0) -- (2, 0);
        \draw[fill=white] (0,0) circle(3pt);
        \draw[fill=white] (1,0) circle(3pt);
        \draw[fill=white] (2,0) circle(3pt);
    \end{tikzpicture}\right)+\omega\left(\begin{tikzpicture}[font=\small, thick]
        \draw[massive, causArrowR] (0,0) -- (1, 0);
        \draw[massive, causArrowR] (1, 0) -- (2, 0);
        \draw[fill=white] (0,0) circle(3pt);
        \draw[fill=white] (1,0) circle(3pt);
        \draw[fill=white] (2,0) circle(3pt);
    \end{tikzpicture}\right)=0\,.
\end{equation}
Combining the above equations gives the expected Murua coefficients
\begin{equation}
   \omega\left( \begin{tikzpicture}[font=\small, thick]
        \draw[massive, causArrow] (0,0) -- (1, 0);
        \draw[massive, causArrow] (1, 0) -- (2, 0);
        \draw[fill=white] (0,0) circle(3pt);
        \draw[fill=white] (1,0) circle(3pt);
        \draw[fill=white] (2,0) circle(3pt);
    \end{tikzpicture}\right)=\frac{1}{3}\,,\quad     \omega\left(\begin{tikzpicture}[font=\small, thick]
        \draw[massive, causArrow] (0,0) -- (1, 0);
        \draw[massive, causArrowR] (1, 0) -- (2, 0);
        \draw[fill=white] (0,0) circle(3pt);
        \draw[fill=white] (1,0) circle(3pt);
        \draw[fill=white] (2,0) circle(3pt);
    \end{tikzpicture}\right)=
    \omega\left(\begin{tikzpicture}[font=\small, thick]
        \draw[massive, causArrowR] (0,0) -- (1, 0);
        \draw[massive, causArrow] (1, 0) -- (2, 0);
        \draw[fill=white] (0,0) circle(3pt);
        \draw[fill=white] (1,0) circle(3pt);
        \draw[fill=white] (2,0) circle(3pt);
    \end{tikzpicture}\right)=\frac{1}{6}\,.
\end{equation}
Note that the contraction rule and the identity \eqref{eq:magic-identity-from-linear-extensions} themselves imply that the Murua coefficient remains unchanged if the directions of all the arrows in a graph are swapped. 

For larger tree the construction works in the same way, and a general proof is given in Appendix~\ref{app: recursionProof}.


\subsection{Relation to Wilson line exponentiation}
\label{sec:wilson-webs}

A closely related physical and combinatorial structure to what we have studied appears in the exponentiation of correlators of multiple Wilson lines. In the soft approximation, one can represent a hard parton with four-velocity $v_i^\mu$  by a semi-infinite Wilson line extending from the hard interaction vertex along the classical trajectory $x_i^\mu(s) =sv_i^\mu$,  
\begin{equation} 
\Phi_{v_i} = \mathcal P \exp\left[ i\int_0^\infty ds\, v_i^\mu A_\mu^a\big(x_i(s)\big)T_i^a \right]\, , \end{equation} 
where $\mathcal P$ orders the colour generators along the trajectory, and $T_i^a$ are the generators in the representation of  parton $i$.
The two settings share the same basic ordering structure; our Magnus Compton amplitudes describe the successive emission of soft massless particles from a hard massive line.  In the soft limit of gauge theory, hard energetic particles are similarly replaced by Wilson lines, from which soft gluons are emitted at ordered positions along the corresponding trajectories. In both settings, soft emissions are ordered along a hard line and summed over their possible orderings.  In the Wilson line case the ordering acts non-trivially on colour generators, whereas in our scalar model the relevant ordering information is instead carried by the causal proper-time regions and their retarded or advanced prescriptions. Wilson line exponentiation therefore provides a natural comparison for the ordering and connectedness structures found here.

Correlators of several Wilson
lines exponentiate, 
\begin{equation}
    Z = \langle \prod_{i=1}^L \Phi_i\rangle = e^W\,,
\end{equation}
and their exponent is organised in terms of sets of
diagrams known as webs~\cite{Gardi:2013ita,Dukes:2013gea,Dukes:2013wa}.
For a web $\mathcal W$, whose diagrams $D$ are related by permutations of
the gluon attachments along the Wilson lines, the contribution to the exponent takes the form
\begin{equation}
    W_{\mathcal W}
    =
    \sum_{D,D'\in\mathcal W}
    F(D)\,R_{DD'}\,C(D')\,,
\end{equation}
where $F(D)$ and $C(D)$ are respectively the kinematic and colour factors
of the diagrams and $R$ is the corresponding web-mixing matrix.  A
fundamental property of these matrices is idempotence,
\begin{equation}
    R^2=R\,,
\end{equation}
so that they act as projectors onto the combinations of diagrams which
enter the exponent.  
Furthermore, the exponentiated colour factors
selected in this way are fully connected~\cite{Gardi:2013ita}. Thus,
although the objects on which the projection acts are different, there
is a clear structural parallel with the Magnus amplitudes studied here:
in both cases exponentiation is accompanied by an idempotent projection
which removes reducible combinations and retains a connected sector.

The parallel extends to the poset description of web-mixing matrices developed in~\cite{Dukes:2013gea,Dukes:2013wa}.
Certain web diagrams $D$ naturally define a partially ordered set $P_D$,
whose elements are the irreducible subdiagrams of $D$.  The compatible
total orderings are the linear extensions $L(P_D)$ of this poset.  If
$p=|P_D|$, the diagonal element of the web-mixing matrix is (see Eq.~3 of \cite{Dukes:2013wa})
\begin{equation}
    R_{DD}
    =
    \sum_{\pi\in L(P_D)}
    \frac{(-1)^{d(\pi)}}
    {p\binom{p-1}{d(\pi)}}\,.
    \label{eq:web-poset}
\end{equation}
Here $d(\pi)$ denotes the number of descents of the corresponding linear
extension, after choosing a natural labelling of the poset.

Equation~\eqref{eq:web-poset} is very similar to the result \eqref{eq:general-Murua-linear-extensions-identity} for the Murua coefficient of an arbitrary directed tree, which we rewrite here for convenience: 
\begin{equation}
    \omega(g)
    =
    \sum_{f\in\operatorname{LE}(g)}
    \frac{(-1)^{d(f)}}
    {n\binom{n-1}{d(f)}}\,.
    \label{eq:murua-web-comparison}
\end{equation}
Indeed, the two quantities are obtained by applying precisely the same
descent-weighted functional to the linear extensions of a poset.  Defining
for a poset $P$ on $p$ elements
\begin{equation}
    \mathcal E(P)
    \coloneq
    \sum_{\pi\in\operatorname{LE}(P)}
    \frac{(-1)^{d(\pi)}}
    {p\binom{p-1}{d(\pi)}}\,,
\end{equation}
we may write the two results simply as
\begin{equation}
    R_{DD}=\mathcal E(P_D)\,,
    \qquad
    \omega(g)=\mathcal E(P_g)\,,
\end{equation}
where $P_g$ is the partial order defined by the arrows of the directed
tree $g$.

There is also a direct connection at the level of the individual
permutation coefficients.  For the class of webs denoted
$(1,1,\ldots,1,n)$ in~\cite{Dukes:2013gea}, the diagrams are themselves
labelled by permutations, and the entries of the web-mixing matrix depend
only on the ascent or descent number of the relative permutation.  Up to
the corresponding ascent/descent convention, their possible values are
\begin{equation}
    \frac{(-1)^d}
    {n\binom{n-1}{d}}\,,
\end{equation}
which are exactly the coefficients of the first Eulerian idempotent 
appearing in our dressed proper-time functions.  Thus the similarity is
not restricted to the final sum over linear extensions: the same weights already occur in the projector which
implements Wilson line exponentiation.

The notion of connectedness is realised somewhat differently in the two
settings.  For multiple Wilson lines, the web-mixing projection selects
fully connected \emph{colour} factors in the exponent.  In the scalar
Magnus amplitudes considered here there is no colour algebra;
instead, the adjoint Eulerian projector annihilates the shuffle products
associated with disconnected proper-time graphs, leaving at classical
order connected \emph{kinematic} spanning trees.  Nevertheless, in both
cases connectedness emerges only after projecting the ordered
contributions which build the exponent.

This parallel is particularly suggestive because the effective
multi-gluon vertices used in proofs of non-Abelian exponentiation have
colour factors built from nested commutators~\cite{Gardi:2013ita}, and 
the Magnus exponent is itself organised precisely in terms of nested commutators (see for example Section~2 of \cite{Brandhuber:2025igz}).
In the present case we have identified the relevant projector explicitly
with the first Eulerian projector,
\begin{equation}
    \pi_1=\log_\star(\operatorname{id}_H)\,,
\end{equation}
which implements the map from the group-like Dyson series to its primitive Magnus logarithm.  It would be interesting to understand
whether the appearance of the same descent-weighted linear-extension
formula in web-mixing matrices reflects a common
Hopf-algebraic origin of the two projection mechanisms.

\section{Conclusions and outlook} 
In this work we have uncovered a Hopf-algebraic framework governing the classical limit of tree-level Magnus Compton amplitudes. In the Schwinger proper-time representation, disconnected contributions organise into shuffle products and are annihilated by the adjoint first Eulerian projector. This proves the cancellation of all hyperclassical contributions at the integrand level, and more generally, of all disconnected contributions at classical and higher orders. At the classical order, this leaves precisely the spanning trees expected from WQFT. 

The same framework establishes a direct correspondence between the directed chains of the QFT Magnus expansion and the arbitrary directed trees appearing in WQFT. By decomposing the proper-time region of a directed tree into its linear extensions, we reconstruct its Murua coefficient entirely from the coefficients of directed chains, as summarised in \eqref{eq:general-Murua-linear-extensions}. The QFT chain sector therefore already contains the complete tree-level Murua data required by the worldline description.

Several directions deserve further investigation. A natural next step is to extend the Schwinger and Hopf-algebraic framework to loop-level Magnus amplitudes, where it would be interesting to determine whether disconnected and iterative contributions continue to be removed at the integrand level. It would also be important to generalise the analysis to theories with momentum-dependent interactions, spin and colour, including scalar QED, QED, QCD and gravity, and to understand how contact interactions modify the graph and shuffle structures found here. Finally, the appearance of the same descent-weighted linear-extension formula in web-mixing matrices suggests a deeper relation with the exponentiation of multiple Wilson lines. Clarifying whether the two constructions share a common Hopf-algebraic mechanism for isolating connected contributions would be particularly interesting. We hope to come back to some of these questions in the near future.

\section*{Acknowledgements}

We would like to thank Jung-Wook Kim for interesting discussions.  This work was
supported by the Science and Technology Facilities Council (STFC) Consolidated
Grant ST/X00063X/1 \textit{``Amplitudes, Strings \& Duality''}.    The work of GRB is
supported by the U.K. Royal Society through Grant URF\textbackslash R1\textbackslash20109. SM and PVM are supported by STFC quota studentships.
No new data were generated or analysed during this study.

\newpage

\appendix
\section{Notation and conventions}
\label{app:notation}
For convenience, we collect here the notation used repeatedly throughout the paper.

\paragraph{Kinematics and proper times.}
We consider the emission of $n$ massless momenta $q_i$ from a massive scalar line,
with
\begin{equation}
    D_i \coloneq 2\, \bar p\!\cdot q_i\,,
    \qquad
    c_{ij} \coloneq q_i\!\cdot q_j\,,
    \qquad
    \sum_{i=1}^n D_i=0\,,
\end{equation}
and assign a proper time $\tau_i$ to each emission vertex. We define
\begin{equation}
    \tau_{ij}\coloneq\tau_i-\tau_j\,,
    \qquad
    \theta_{ij}\coloneq\theta(\tau_{ij})\,.
\end{equation}
Propagators are represented in the 
Schwinger form as
\begin{equation}
\label{eq:Schw-bis}
\frac{i}{A+i\sigma\vareps} = \sigma\int\!d\tau \ \theta(-\sigma\tau) \, e^{-iA\tau}\, , 
\end{equation}
 with $\sigma=+1$ corresponding to the retarded and $\sigma=-1$ to the advanced prescription.

\paragraph{Orderings.}
A permutation
\begin{equation}
    \rho=(\rho_1\,\cdots\,\rho_n)\in S_n
\end{equation}
labels the total proper-time ordering
$\tau_{\rho_1}<\cdots<\tau_{\rho_n}$. We define
\begin{equation}
    s_\rho(i,j)=
    \begin{cases}
        +1, & i \text{ appears before } j \text{ in }\rho\,,\\
        -1, & j \text{ appears before } i \text{ in }\rho\,,
    \end{cases}
\end{equation}
and denote the descent number of $\rho$ by
\begin{equation}
    d(\rho)
    \coloneq
    \big|\{i:\rho(i)>\rho(i+1)\}\big|\,.
\end{equation}
The Murua-weighted dressed proper-time function associated with the ordering
$\rho$ is denoted by $\Theta^{\rm M}_\rho$.

\paragraph{Words and Eulerian projectors.}
Multilinear words on the alphabet $\{1,\ldots,n\}$ are identified with
permutations in $S_n$. Their canonical pairing is
\begin{equation}
    \langle u,v\rangle=\delta_{u,v}\,,
\end{equation}
and the shuffle product is denoted by $\shuffle$.
Permutations act from the right on words. The first Eulerian projector and
its adjoint are denoted by
\begin{equation}
    \pi_1(P)=P e_1\,,
    \qquad
    \pi_1^*(P)=P e_1^*\,,
\end{equation}
with $\pi_1=\log_\star(\mathrm{id}_H)$.

\paragraph{Directed trees.}
We denote an undirected tree by $\hat g$ and the set of its orientations by
$A(\hat g)$. For a directed tree $g$,
\begin{equation}
    \operatorname{LE}(g)
\end{equation}
is the set of linear extensions of the partial order defined by its arrows, and
\begin{equation}
    \mathcal L_g
    \coloneq
    \sum_{f\in \operatorname{LE}(g)} f\,,
    \qquad
    \phi(g)\coloneq|\operatorname{LE}(g)|\,.
\end{equation}
Its Murua coefficient is denoted by $\omega(g)$. For
$g_0,g\in A(\hat g)$,
$s(g_0,g)$ counts the number of edges whose orientations differ.

\section{Murua coefficients for  chains}
\label{app:Murua-chain}

In this section, we prove that for a chain with $n$ vertices (and hence $n-1$ arrows), with all but $k$ arrows pointing in the same direction, which we denote as $\ell_n^{(k)}$, the corresponding Murua coefficient is 
\begin{equation}
\label{eq:muruaclosed-bis}
  \omega(\ell_n^{(k)})
  =
  \frac{1}{n\binom{n-1}{k}}\, , 
\end{equation}
independently of the position of the reversed arrows. 

To do so, we first recall that for the chain with $n$ vertices and all arrows pointing
in the same direction (see Example 8 of \cite{Murua_2006}),
\begin{align}\label{eq:MuruaChain0Flip}
\omega\!\left(
\begin{tikzpicture}[baseline={([yshift=-1.2ex]current bounding box.center)}, thick,font=\small]
    \draw[causArrow] (0,0) -- (0.75,0);
    \draw[causArrow=0.6] (0.75,0) -- (1.25,0);
    \draw[causArrow=0.4] (1.95,0) -- (2.45,0);
    \draw[fill=white] (0,0) circle(3pt) node[yshift=0.4cm]{$1$};
    \draw[fill=white] (0.75,0) circle(3pt) node[yshift=0.4cm]{$2$};
    \node at (1.6,0) {$\cdots$};
    \draw[fill=white] (2.45,0) circle(3pt) node[yshift=0.4cm]{$n$};
\end{tikzpicture}
\right)
    = \frac{1}{n}\ .
\end{align}
We denote by
\begin{align}
\ell_n =
\begin{tikzpicture}[baseline={([yshift=-1.7ex]current bounding box.center)}, thick,font=\small]
    \draw[causArrow] (0,0) -- (0.75,0);
    \draw[causArrow=0.6] (0.75,0) -- (1.25,0);
    \draw[causArrow=0.4] (1.95,0) -- (2.45,0);
    \draw[fill=white] (0,0) circle(3pt) node[yshift=0.4cm]{$1$};
    \draw[fill=white] (0.75,0) circle(3pt) node[yshift=0.4cm]{$2$};
    \node at (1.6,0) {$\cdots$};
    \draw[fill=white] (2.45,0) circle(3pt) node[yshift=0.4cm]{$n$};
\end{tikzpicture}
\end{align}
the chain with $n$ vertices and all $n-1$ arrows pointing to the right. More generally, as we said earlier, let $\ell_n^{(k)}$ denote the same chain but with
$k$ arrows flipped, so that $k$ arrows point to the left and the remaining
$n-1-k$ arrows point to the right. Notice that there are many different possibilities, but as we will see the Murua coefficient will only depend on the number of flips and not their positions.

Pick any edge of $\ell_n$, and consider the two trees obtained by keeping
that edge unflipped or flipping it. The contraction rule gives%
\footnote{See for example Section~6.2 of \cite{Brandhuber:2025igz} for a brief review of the main properties of Murua coefficients, of which the contraction rule is one of the most useful in practical applications.}
\begin{align}
    \omega\left(
    \begin{tikzpicture}[baseline={([yshift=-1.2ex]current bounding box.center)}, thick,font=\small]
    \draw[causArrow=0.6] (0,0) -- (0.5,0);
    \draw[causArrow,blue] (1.3,0) -- (2.05,0);
    \draw[causArrow=0.4] (2.85,0) -- (3.35,0);
    \draw[fill=white] (0,0) circle(3pt) node[yshift=0.4cm]{$1$};
    \node at (0.85,0) {$\cdots$};
    \draw[fill=white] (1.3,0) circle(3pt);
    \draw[fill=white] (2.05,0) circle(3pt);
    \node at (2.5,0) {$\cdots$};
    \draw[fill=white] (3.35,0) circle(3pt) node[yshift=0.4cm]{$n$};
    \end{tikzpicture}
\right)
    +\omega\left(
    \begin{tikzpicture}[baseline={([yshift=-1.2ex]current bounding box.center)}, thick,font=\small]
    \draw[causArrow=0.6] (0,0) -- (0.5,0);
    \draw[causArrow,blue] (2.05,0) -- (1.3,0);
    \draw[causArrow=0.4] (2.85,0) -- (3.35,0);
    \draw[fill=white] (0,0) circle(3pt) node[yshift=0.4cm]{$1$};
    \node at (0.85,0) {$\cdots$};
    \draw[fill=white] (1.3,0) circle(3pt);
    \draw[fill=white] (2.05,0) circle(3pt);
    \node at (2.5,0) {$\cdots$};
    \draw[fill=white] (3.35,0) circle(3pt) node[yshift=0.4cm]{$n$};
    \end{tikzpicture}
\right)
    =\omega\left(
    \begin{tikzpicture}[baseline={([yshift=-1.2ex]current bounding box.center)}, thick,font=\small]
    \draw[causArrow=0.6] (0,0) -- (0.5,0);
    \draw[causArrow=0.4] (1.2,0) -- (1.7,0);
    \draw[fill=white] (0,0) circle(3pt) node[yshift=0.4cm]{$1$};
    \node at (0.85,0) {$\cdots$};
    \draw[fill=white] (1.7,0) circle(3pt) node[yshift=0.4cm]{$n{-}1$};
    \end{tikzpicture}
\right)\ .
\end{align}
Equivalently, we can write
\begin{equation}
    \omega(\ell_n) + \omega(\ell_n^{(1)})
    =
    \omega(\ell_{n-1})\ .
\end{equation}
Notice that the position of the flipped edge in $\ell_n^{(1)}$ does not matter, since contracting any edge always produces $\ell_{n-1}$. Using $\omega(\ell_n)=1/n$, we therefore find
\begin{equation}\label{eq:MuruaChain1Flip}
    \omega(\ell_n^{(1)})
    =
    \frac{1}{n-1} - \frac{1}{n}
    =
    \frac{1}{n(n-1)}
    =
    \frac{1}{n \binom{n-1}{1}}\ .
\end{equation}
Now flip one further edge. Applying the same contraction rule gives
\begin{equation}
    \omega(\ell_n^{(1)}) + \omega(\ell_n^{(2)})
    =
    \omega(\ell_{n-1}^{(1)}).
\end{equation}
Again, the specific locations of the flipped edges are irrelevant. Using the result of \eqref{eq:MuruaChain1Flip}, but now with $n\mapsto n-1$, we have
\begin{equation}
    \omega(\ell_{n-1}^{(1)})
    =
    \frac{1}{(n-1)(n-2)}.
\end{equation}
Hence
\begin{equation}
    \omega(\ell_n^{(2)}) = \frac{1}{(n-1)(n-2)} - \frac{1}{n(n-1)} =
    \frac{2}{n(n-1)(n-2)} =
    \frac{1}{n \binom{n-1}{2}}\ .
\end{equation}
We now prove the general formula by induction. For a fixed $n$, the base case of $k=0$ is given by \eqref{eq:MuruaChain0Flip}. Suppose that
\begin{equation}\label{eq:MuruaChainkFlip}
    \omega(\ell_n^{(k)})
    =
    \frac{1}{n\binom{n-1}{k}}
\end{equation}
for some $0\leq k < n-1$. Consider $\ell_n^{(k)}$ and flip any one of the
remaining right-pointing arrows. The contraction rule gives
\begin{equation}
    \omega(\ell_n^{(k)}) + \omega(\ell_n^{(k+1)})
    =
    \omega(\ell_{n-1}^{(k)})\ .
\end{equation}
Therefore,
\begin{equation}
    \frac{1}{n\binom{n-1}{k}}
    +
    \omega(\ell_n^{(k+1)})
    =
    \frac{1}{(n-1)\binom{n-2}{k}}\ .
\end{equation}
Solving for $\omega(\ell_n^{(k+1)})$, we obtain
\begin{align}
    \omega(\ell_n^{(k+1)})
    &=
    \frac{1}{(n-1)\binom{n-2}{k}}
    -
    \frac{1}{n\binom{n-1}{k}}
    =
    \frac{1}{n\binom{n-1}{k+1}}\ ,
\end{align}
after some quick algebra. This inductive step can be repeated until $k=n-1$, since then there are no further arrows to flip. This case, however, is identical to the $k=0$ case since all the edges face in the same direction, $\ell^{(n-1)}_n=\ell^{(0)}_n$. In particular,
\begin{equation}
    \omega(\ell^{(n-1)}_n)=\omega(\ell^{(0)}_n)=\frac{1}{n}=\frac{1}{n\binom{n-1}{n-1}}\ ,
\end{equation}
so the formula works as expected even for this final case.
Therefore \eqref{eq:MuruaChainkFlip} is true for all $n$ and $0\leq k\leq n-1$.

We can now verify that the Murua coefficients we have found satisfy the expected properties:
\begin{enumerate}
    \item[\textbf{1.}] The coefficients satisfy the contraction rule by construction.
    
    \item[\textbf{2.}] Given that the binomial coefficients satisfy the reflection property
    \begin{equation}
    \binom{n-1}{k}
    =
    \binom{n-1}{n-1-k}\,
    \end{equation}
    the coefficients are invariant under $k\mapsto n-1-k$, i.e.\ the reversal of all arrows
    \begin{equation}
    \omega(\ell_n^{(k)}) = \omega(\ell_n^{(n-1-k)})\ .
    \end{equation}

    \item[\textbf{3.}] Finally, the coefficients obey the sum rule:
    \begin{align}
    \sum_{g \in A(\hat{g})} \omega(g) =
    \sum_{k=0}^{n-1} \binom{n-1}{k} \frac{1}{n\binom{n-1}{k}}
    = \sum_{k=0}^{n-1} \frac{1}{n}
    = 1\ ,
    \end{align}
    since there are $\binom{n-1}{k}$ chains with $n$ vertices and $k$ reversed arrows.
\end{enumerate}

Incidentally, we can also repeat this for the $n$-star,
\begin{equation}
    s_n=
    \begin{tikzpicture}[baseline={([yshift=-2.8ex]current bounding box.center)},thick,font=\small]
    \draw[causArrow] (0,0) -- (90:0.8);
    \draw[causArrow] (0,0) -- (40:0.8);
    \draw[causArrow] (0,0) -- (-10:0.8);
    \draw[causArrow] (0,0) -- (140:0.8);
    \draw[causArrow] (0,0) -- (190:0.8);
    \draw[fill=white] (0,0) circle(3pt);
    \draw[fill=white] (90:0.8) circle(3pt) node[yshift=0.4cm]{$1$};
    \draw[fill=white] (40:0.8) circle(3pt) node[shift={(40:0.4)}]{$2$};
    \draw[fill=white] (-10:0.8) circle(3pt) node[shift={(-10:0.4)}]{$3$};
    \draw[fill=white] (140:0.8) circle(3pt) node[shift={(140:0.4)}]{$n$};
    \draw[fill=white] (190:0.8) circle(3pt) node[shift={(190:0.6)}]{$n{-}1$};
    \node at (-40:0.8){.};
    \node at (-50:0.8){.};
    \node at (-60:0.8){.};
    \node at (220:0.8){.};
    \node at (230:0.8){.};
    \node at (240:0.8){.};
    \end{tikzpicture}\ ,
\end{equation}
and defining $s_n^{(k)}$ as before. Then one can show that
\begin{equation}
    \omega(s_n^{(k)})=(-1)^{n+k} \sum_{j=0}^k\binom{k}{j}B_{n-k+j}\ .
\end{equation}
These satisfy the contraction rule, sum rule and reflection rule thanks to the defining property of the Bernoulli numbers
\begin{equation}
    \sum_{k=0}^n\binom{n}{k}B_k=B_n\ .
\end{equation}

\section{Survival of spanning trees}
\label{spanningtreesnotshufflessection}
In Section~\ref{Hyperclassical Cancellations Proof Section}, we showed that the Eulerian projector removes every disconnected contribution. For this mechanism to reproduce the classical limit of the amplitude, the connected graphs must survive the projection. At classical order the expansion contains $n-1$ edges, so after the disconnected contributions have vanished the relevant surviving graphs are precisely the spanning trees, namely connected, cycle-free graphs containing all vertices. 
We now show algebraically that the word polynomial associated with a spanning tree is not annihilated by the adjoint Eulerian projector.

Let $T$ be a spanning tree on $n$ vertices, with a fixed orientation
assigned to each edge. Recall the sign-weighted word sum associated to $T$,
\begin{equation}
    F_T
\coloneq
\sum_{\rho\in S_n}
\epsilon_T(\rho)\,\rho\,,
\qquad
\epsilon_T(\rho)
\coloneq
\prod_{i\to j\in T}
s_\rho(i,j)\, , 
\end{equation}
where
\begin{equation}
    s_\rho(i,j)
    =
    \begin{cases}
        +1, & \text{if $i$ appears before $j$ in $\rho$}\,,\\
        -1, & \text{if $j$ appears before $i$ in $\rho$}\,.
    \end{cases}
\end{equation}
Under the map $\mathcal{T}$ introduced in the previous section,
$\pi_1^*(F_T)$ is precisely the word-valued counterpart of the coefficient
of the spanning-tree monomial in \eqref{eq:monomial-coefficient}. To show that $\pi_1^*(F_T)$ is non-zero, it is sufficient to find a Lie polynomial which has a non-vanishing pairing with $F_T$. Indeed, if $L_T$ is a Lie polynomial, then
\begin{equation}
    \pi_1(L_T) = L_T\,,
\end{equation}
and hence,
\begin{equation}
\langle
        \pi_1^*(F_{T}), L_{T}
    \rangle = \langle F_{T}, \pi_1(L_{T})\rangle=\langle F_{T}, L_{T}\rangle  .
\end{equation}
We will recursively associate a Lie polynomial $L_T$ to the tree $T$ and show that
\begin{equation}
        \langle F_T,L_T\rangle=2^{n-1}\,.
    \label{eq:TreeLiePairing}
\end{equation}
\paragraph{Base case} For $n=1$, let $T=
\tikz{\node[draw,circle,fill=black,inner sep=1.8pt,label=above:$i$] (0,0){}}
$. Then
\begin{equation}
    F_T=i\,,
    \qquad
    L_T=i\,,
\end{equation}
and therefore
\begin{equation}
    \langle F_T,L_T\rangle=1=2^0\,.
\end{equation}

\paragraph{Inductive step} Let $T$ be an oriented tree with $n$ vertices. Choose a leaf $\ell$
connected to a vertex $v$, and remove $\ell$ to obtain the oriented tree $T'$ with $n-1$ vertices.
\begin{equation}
\begin{aligned}
T &=
\begin{tikzpicture}[thick,baseline={([yshift=-1.8ex]current bounding box.center)}]
    \node[draw,circle,fill=black,inner sep=1.8pt,label=above:$\ell$] (vl) at (0,0) {};
    \node[draw,circle,fill=black,inner sep=1.8pt,label=above:$v$] (vv) at (1.2,0) {};
    \node[draw,circle,fill=black,inner sep=1.8pt] (v0) at (2.4,0) {};

    \draw (vl) -- (vv);
    \draw (vv) -- (v0);
    \draw (v0) -- ++(0.4,0);
    \node at (3.25,0) {$\cdots$};
\end{tikzpicture}
\\[1em]
T' &=
\begin{tikzpicture}[thick,baseline={([yshift=-1.8ex]current bounding box.center)}]
    \node[draw,circle,fill=black,inner sep=1.8pt,label=above:$v$] (vl) at (0,0) {};
    \node[draw,circle,fill=black,inner sep=1.8pt] (v0) at (1.2,0) {};

    \draw (vl) -- (v0);
    \draw (v0) -- ++(0.4,0);
    \node at (2.05,0) {$\cdots$};
\end{tikzpicture}
\end{aligned}
\end{equation}
Assume inductively that
\begin{equation}
    \langle F_{T'},L_{T'}\rangle=2^{n-2}\,.
    \label{eq:TreeInductionAssumption}
\end{equation}
The edge between $\ell$ and $v$ can point in either direction. We define a new polynomial
\begin{equation}
    L_T
    \coloneq
    \begin{cases}
        [\ell,L_{T'}]
        & \text{if }\ell\to v\,,
        \\[2mm]
        [L_{T'},\ell]
        & \text{if } v\to\ell\,.
    \end{cases}
    \label{eq:TreeLieRecursiveDefinition}
\end{equation}
We first consider the case $\ell\to v$. The case $v\to\ell$ will work in
exactly the same way.
It follows that
\begin{equation}
    \langle F_T,L_T\rangle
    =
    \langle F_T,\ell L_{T'}\rangle
    -
    \langle F_T,L_{T'}\ell\rangle\,.
    \label{eq:TreePairingDecomposition}
\end{equation}
Consider first a word appearing in $\ell L_{T'}$. It has the form
$\ell u$, where $u$ is a word on the vertices of $T'$. Since
$\ell$ appears at the beginning, it occurs before every vertex of
$T'$, and in particular before $v$. As the new edge is oriented
$\ell\to v$,
\begin{equation}
    s_{\ell u}(\ell,v)=+1\,.
\end{equation}
All remaining edge signs are precisely those associated with the word $u$ in $T'$.
Since this applies to every word in $T'$, we have
\begin{equation}
    \epsilon_T(\ell u)
    =
    \epsilon_{T'}(u)\,,
\end{equation}
and hence
\begin{equation}
    \langle F_T,\ell L_{T'}\rangle
    =
    \langle \ell F_{T'},\ell L_{T'}\rangle
    =
    \langle F_{T'},L_{T'}\rangle
    =
    2^{n-2}\,,
    \label{eq:LeafFirstPairing}
\end{equation}
where the first equality holds since any word which does not start with $\ell$ drops out of $F_T$ in the scalar product.
Now consider a word appearing in $L_{T'}\ell$. It has the form
$u\ell$, so that $\ell$ appears after every vertex of $T'$, and
in particular after $v$. Therefore,
\begin{equation}
    s_{u\ell}(\ell,v)=-1\,.
\end{equation}
The signs associated with all edges of $T'$ are unchanged, so
\begin{equation}
    \epsilon_T(u\ell)
    =
    -\epsilon_{T'}(u)\,.
\end{equation}
It follows that
\begin{equation}
    \langle F_T,L_{T'}\ell\rangle
    =
    -\langle F_{T'}\ell,L_{T'}\ell\rangle
    =
    -\langle F_{T'},L_{T'}\rangle
    =
    -2^{n-2}\,.
    \label{eq:LeafLastPairing}
\end{equation}
Substituting \eqref{eq:LeafFirstPairing} and
\eqref{eq:LeafLastPairing} into
\eqref{eq:TreePairingDecomposition}, we obtain
\begin{equation}
\label{eq: non-zeroliecomponent}
    \langle F_T,L_T\rangle
    =
    \langle F_{T'},L_{T'}\rangle
    +
    \langle F_{T'},L_{T'}\rangle
    =
    2\,2^{n-2}
    =
    2^{n-1}\, .
\end{equation}
If instead $v\to\ell$, then
\begin{equation}
    L_T=[L_{T'},\ell]
    =
    L_{T'}\ell-\ell L_{T'}\,.
\end{equation}
For a word $u\ell$, the vertex $v$ appears before $\ell$, so
\begin{equation}
    \epsilon_T(u\ell)=\epsilon_{T'}(u)\,.
\end{equation}
For a word $\ell u$, the vertex $\ell$ appears before $v$, so
\begin{equation}
    \epsilon_T(\ell u)=-\epsilon_{T'}(u)\,.
\end{equation}
Therefore
\begin{equation}
    \langle F_T,L_T\rangle
    =
    \langle F_T,L_{T'}\ell\rangle
    -
    \langle F_T,\ell L_{T'}\rangle=
    \langle F_{T'},L_{T'}\rangle -
    \left(-\langle F_{T'},L_{T'}\rangle\right)= 2^{n-1}\,.
\end{equation}
Hence the result holds for either orientation of the edge between
$\ell$ and $v$.
Since $L_T$ is a Lie polynomial,
\begin{equation}
    \pi_1(L_T)=L_T \,.
\end{equation}
We therefore find
\begin{equation}
    \left\langle
        \pi_1^*(F_T),L_T
    \right\rangle = 
    \left\langle
        F_T,\pi_1(L_T)
    \right\rangle = \left\langle
        F_T,L_T
    \right\rangle
    =
    2^{n-1}\neq0 \,.
\end{equation}
Consequently,
\begin{equation}
    \pi_1^*(F_T)\neq0 \,.
\end{equation}
Thus, the sign-weighted word sum associated with a spanning tree survives the adjoint Eulerian projection, in contrast to the shuffle products associated with disconnected graphs. Together with Section~\ref{Hyperclassical Cancellations Proof Section}, this shows that the adjoint Eulerian projector extracts the connected sector of the classical expansion: shuffle products associated with disconnected graphs are annihilated, while
the spanning-tree contributions survive.


\section{New eikonal identity}
\label{section: New Eikonal Identity}
 In this appendix we extend the eikonal identity for Feynman $i \eps$ prescriptions of \cite{Saotome:2012vy} 
\begin{equation}
    \delta(\omega_1+\cdots +\omega_n) \sum_{\text{perms } \omega} \frac{1}{\omega_1+i\vareps}\cdots \frac{1}{\omega_1+\cdots \omega_{n-1}+i\vareps} = (-2\pi i)^{n-1}\delta(\omega_1)\cdots \delta(\omega_n)\ , 
\end{equation}
to more general cases.
In view of applications to Magnus diagrams, we are interested in the situation where there are different signs of the $i \vareps$ prescription of the $n-1$ propagators. We will use similar representations to those in \cite{Saotome:2012vy}, but adapted to allow for a choice of the sign of the $i\vareps$ via $\sigma=\pm$:
\begin{align}
    \delta(\omega_1+\cdots +\omega_n) &= \frac{1}{2\pi}\int\! dt_n\, e^{i(\omega_1+\cdots+\omega_n) t_n}\ , \\
    \frac{1}{\omega+i\sigma \vareps} &=  - i\sigma \int\! d\tau \, \theta(-\sigma\tau) e^{-i(\omega+i\sigma\vareps)\tau}\ .
\end{align}
Notice that the combination of the theta function and the $i\vareps$ in the exponent ensures that the integral converges for either sign of $\sigma$,
\begin{align}
    \sigma=+1 \quad &\implies \int_{-\infty}^0\!d\tau\, e^{-i\omega\tau}e^{\vareps\tau} \longrightarrow \text{convergent}\ , \\
    \sigma=-1 \quad &\implies \int_{0}^\infty\! d\tau\, e^{-i\omega\tau}e^{-\vareps\tau} \ \, \longrightarrow \text{convergent}\ . 
\end{align}
Now consider
\begin{align}
    \text{Eik}(\vec{\sigma}) &\coloneq \delta(\omega_1+\cdots+\omega_n) \sum_{\text{perms } \omega} \frac{1}{\omega_1+i\sigma_1\vareps}\cdots \frac{1}{\omega_1+\cdots +\omega_{n-1}+i\sigma_{n-1}\vareps} \\
    &= \frac{(-i)^{n-1}\sigma}{2\pi}\sum_{\text{perms }\omega}\int\!dt_n \,  e^{i(\omega_1+\cdots+\omega_{n})t_n} \ \prod_{k=1}^{n-1} d\tau_k\  \theta(-\sigma_k\tau_k) \, e^{-i(\sum_{j=1}^{k}\omega_j+i\sigma_k\vareps)\tau_k}  \ ,
\end{align}
 having defined $\vec{\sigma}=(\sigma_1,\ldots,\sigma_{n-1})$ and $\sigma=\sigma_1 \cdots \sigma_{n-1} = \pm 1$. We will suppress the $i\vareps$'s in the exponent as they are merely to ensure convergence, which we will take as guaranteed from now on. Performing a change of basis $\tau_k = t_{k+1}-t_k$,
\begin{align}
    \text{Eik}(\vec{\sigma}) = \frac{(-i)^{n-1}\sigma}{2\pi} \sum_{\text{perms }\omega} \int\! 
    \left(\prod_{j=1}^{n-1} dt_j\ \theta(-\sigma_j(t_{j+1}-t_j)) e^{-i\omega_j(\sum_{k=j}^{n-1} \tau_k-t_n)}\right)dt_n\ e^{i\omega_n t_n} \ .
\end{align}
We have changed the order of summation in the exponent with the aim of making the telescopic cancellation of terms manifest, using $\sum_{k=j}^{n-1} \tau_k = t_n - t_j$.
This means that
\begin{equation}
    \text{Eik}(\vec{\sigma}) =\frac{(-i)^{n-1}\sigma}{2\pi} \int\! dt_n \prod_{i=1}^{n-1}dt_i\ \theta(-\sigma_i(t_{i+1}-t_i)) \,\sum_{\text{perms }\omega}\prod_{j=1}^{n}e^{i \omega_j t_j}\ .
\end{equation}
If the $\theta$ functions were missing, the integrals would trivially evaluate to a product of $\delta(\omega_j)$. Let us consider the simplified case where every $\sigma_j=+1$. If we average over all labellings of the $t$'s, by symmetry
\begin{equation}
\frac{1}{n!}\sum_{\text{perms }t} \theta_{12}\cdots \theta_{n-1\, n} = \frac{1}{n!}\ ,
\end{equation}
meaning that 
\begin{equation}
    \text{Eik}(+\cdots+) = \frac{(-i)^{n-1}}{2\pi\,n!} \int\! dt_1\cdots dt_n\,
    \sum_{\text{perms }\omega}\prod_{j=1}^{n}e^{i \omega_j t_j} = (-2\pi i)^{n-1} \delta(\omega_1)\cdots\delta(\omega_n)\ ,
\end{equation}
matching the original result. Notice that we were able to symmetrise over the variables $t$ without altering the exponentials thanks to the symmetrisation over $\omega$. If the signs are arbitrary, the cancellation is more subtle. Reusing the notation in \cite{Brandhuber:2025igz},
\begin{equation}
\begin{tikzpicture}[font=\small, thick]
        \draw[thetaLine] (0,0) -- (1,0);
        \draw[fill=white] (0,0) circle(3pt);
        \draw[fill=white] (1,0) circle(3pt);
        \node at (0,0.4){$j$};
        \node at (1,0.4){$i$};
    \end{tikzpicture}
    = \theta_{ij}\ ,
\end{equation}
we can represent the product of $\theta$-functions as a directed tree. We can use the diagrammatic identities presented there to decompose the tree as a sum of trees with more edges, essentially resolving the time ordering between nodes which are not connected. For example,
\begin{align}
\begin{tikzpicture}[baseline={([yshift=-.5ex]current bounding box.center)}, thick,font=\footnotesize]
    \draw[thetaLine] (0,0) -- (0.8,0);
    \draw[thetaLine] (0,0) -- (0.4,-0.5);
    \draw[thetaLine] (0.4,-0.5) -- (0.8,-0.5);
    \draw[fill=white] (0,0) circle(3pt) node[yshift=0.4cm,black]{$2$};
    \draw[fill=white] (0.8,0) circle(3pt) node[yshift=0.4cm,black]{$1$};
    \draw[fill=white] (0.4,-0.5) circle(3pt) node[yshift=-0.4cm,black]{$3$};
    \draw[fill=white] (0.8,-0.5) circle(3pt) node[yshift=-0.4cm,black]{$4$};
\end{tikzpicture}
&=
\begin{tikzpicture}[baseline={([yshift=-.5ex]current bounding box.center)}, thick, font=\footnotesize]
    \draw[thetaLine] (0.8,-0.5) -- (1.2,0);
    \draw[thetaLine] (0.4,-0.5) -- (0.8,-0.5);
    \draw[thetaLine] (0,0) -- (1.2,0);
    \draw[thetaLine] (0,0) -- (0.4,-0.5);
    \draw[fill=white] (0.4,-0.5) circle(3pt) node[yshift=-0.4cm,black]{$3$};
    \draw[fill=white] (0.8,-0.5) circle(3pt) node[yshift=-0.4cm,black]{$4$};
    \draw[fill=white] (0,0) circle(3pt) node[yshift=0.4cm,black]{$2$};
    \draw[fill=white] (1.2,0) circle(3pt) node[yshift=0.4cm,black]{$1$};
\end{tikzpicture}
+
\begin{tikzpicture}[baseline={([yshift=-.5ex]current bounding box.center)}, thick, font=\footnotesize]
    \draw[thetaLine] (0,0) -- (0.4,-0.5);
    \draw[thetaLine] (0.4,-0.5) -- (0.8,0);
    \draw[thetaLine] (0,0) -- (0.8,0);
    \draw[thetaLine] (0.4,-0.5) -- (1.2,-0.5);
    \draw[thetaLine] (0.8,0) -- (1.2,-0.5);
    \draw[fill=white] (0,0) circle(3pt) node[yshift=0.4cm,black]{$2$};
    \draw[fill=white] (0.8,0) circle(3pt) node[yshift=0.4cm,black]{$1$};
    \draw[fill=white] (0.4,-0.5) circle(3pt) node[yshift=-0.4cm,black]{$3$};
    \draw[fill=white] (1.2,-0.5) circle(3pt) node[yshift=-0.4cm,black]{$4$};
\end{tikzpicture}
+
\begin{tikzpicture}[baseline={([yshift=-.5ex]current bounding box.center)}, thick, font=\footnotesize]
    \draw[thetaLine] (0.4,0) -- (0.8,-0.5);
    \draw[thetaLine] (0,0) -- (0.4,0);
    \draw[thetaLine] (0.8,-0.5) -- (1.2,-0.5);
    \draw[thetaLine] (0,0) -- (0.8,-0.5);
    \draw[fill=white] (0,0) circle(3pt) node[yshift=0.4cm,black]{$2$};
    \draw[fill=white] (0.4,0) circle(3pt) node[yshift=0.4cm,black]{$1$};
    \draw[fill=white] (0.8,-0.5) circle(3pt) node[yshift=-0.4cm,black]{$3$};
    \draw[fill=white] (1.2,-0.5) circle(3pt) node[yshift=-0.4cm,black]{$4$};
\end{tikzpicture} \nn \\
&=
\begin{tikzpicture}[baseline={([yshift=-.5ex]current bounding box.center)}, thick, font=\footnotesize]
    \draw[thetaLine] (0,0)node[yshift=-0.4cm]{} -- (0.5,0);
    \draw[thetaLine] (0.5,0) -- (1,0);
    \draw[thetaLine] (1,0) -- (1.5,0);
    \draw[fill=white] (0,0) circle(3pt) node[yshift=0.4cm,black]{$2$};
    \draw[fill=white] (0.5,0) circle(3pt) node[yshift=0.4cm,black]{$3$};
    \draw[fill=white] (1,0) circle(3pt) node[yshift=0.4cm,black]{$4$};
    \draw[fill=white] (1.5,0) circle(3pt) node[yshift=0.4cm,black]{$1$};
\end{tikzpicture}
+
\begin{tikzpicture}[baseline={([yshift=-.5ex]current bounding box.center)}, thick, font=\footnotesize]
    \draw[thetaLine] (0,0)node[yshift=-0.4cm]{} -- (0.5,0);
    \draw[thetaLine] (0.5,0) -- (1,0);
    \draw[thetaLine] (1,0) -- (1.5,0);
    \draw[fill=white] (0,0) circle(3pt) node[yshift=0.4cm,black]{$2$};
    \draw[fill=white] (0.5,0) circle(3pt) node[yshift=0.4cm,black]{$3$};
    \draw[fill=white] (1,0) circle(3pt) node[yshift=0.4cm,black]{$1$};
    \draw[fill=white] (1.5,0) circle(3pt) node[yshift=0.4cm,black]{$4$};
\end{tikzpicture}
+
\begin{tikzpicture}[baseline={([yshift=-.5ex]current bounding box.center)}, thick, font=\footnotesize]
    \draw[thetaLine] (0,0)node[yshift=-0.4cm]{} -- (0.5,0);
    \draw[thetaLine] (0.5,0) -- (1,0);
    \draw[thetaLine] (1,0) -- (1.5,0);
    \draw[fill=white] (0,0) circle(3pt) node[yshift=0.4cm,black]{$2$};
    \draw[fill=white] (0.5,0) circle(3pt) node[yshift=0.4cm,black]{$1$};
    \draw[fill=white] (1,0) circle(3pt) node[yshift=0.4cm,black]{$3$};
    \draw[fill=white] (1.5,0) circle(3pt) node[yshift=0.4cm,black]{$4$};
\end{tikzpicture} \nn\\
&\longrightarrow 3\times\
\begin{tikzpicture}[baseline={([yshift=-.5ex]current bounding box.center)}, thick]
    \draw[thetaLine] (0,0) -- (0.5,0);
    \draw[thetaLine] (0.5,0) -- (1,0);
    \draw[thetaLine] (1,0) -- (1.5,0);
    \draw[fill=white] (0,0) circle(3pt);
    \draw[fill=white] (0.5,0) circle(3pt);
    \draw[fill=white] (1,0) circle(3pt);
    \draw[fill=white] (1.5,0) circle(3pt);
\end{tikzpicture}
\end{align}
where the final step is performed by relabelling integration variables to some standard ordering. The main simplification this procedure provides occurs because any additional $\theta$-line beyond the three required to specify a total ordering on the vertices is redundant. Therefore any directed tree can be written as a sum of chains, which are then relabelled to a common ordering. The number of terms appearing in the sum is given by $\phi(g)$, the number of linear extensions of the tree $g$.
Thus, writing the $\theta$-functions as a tree $\tau$, we can effectively take
\begin{equation}
    \theta(-\sigma_1(t_2-t_1))\cdots \theta(-\sigma_{n-1}(t_n-t_{n-1})) \rightarrow \tau \rightarrow \phi(\tau)\,
\underbrace{
    \begin{tikzpicture}[baseline={([yshift=-.5ex]current bounding box.center)},font=\small, thick]
    \draw[thetaLine] (0,0) -- (0.5,0);
    \draw[thin, red] (0.5, 0) -- ++(0.35,0);
    \node at (1.25,0){$\cdots$};
    \draw[thin, red] (2,0) -- ++(-0.35,0);
    \draw[fill=white] (0,0) circle(3pt);
    \draw[fill=white] (0.5,0) circle(3pt);
    \draw[fill=white] (2,0) circle(3pt);
\end{tikzpicture}
}_{n}
\rightarrow \phi(g) \theta_{12}\cdots \theta_{n-1\, n} \ .
\end{equation}
This reduces the problem to the original all-plus case, up to the tree-dependent factor $\phi(g)$. All in all, the following relation holds:
\begin{align}\label{eq:MagicEikonalIdGeneralised}
    \text{Eik}(\vec{\sigma}) &\coloneq \delta(\omega_1+\cdots+\omega_n) \sum_{\text{perms } \omega} \frac{1}{\omega_1+i\sigma_1\vareps}\cdots \frac{1}{\omega_1+\cdots +\omega_{n-1}+i\sigma_{n-1}\vareps}\nn \\ &=\sigma\, \phi(g)(-2\pi i)^{n-1} \delta(\omega_1)\cdots \delta(\omega_n)\ .
\end{align}
Notice that $\phi(\begin{tikzpicture}[baseline={([yshift=-.5ex]current bounding box.center)},font=\small, thick]
    \draw[thetaLine] (0,0) -- (0.5,0);
    \draw[thin, red] (0.5, 0) -- ++(0.35,0);
    \node at (1.25,0){$\cdots$};
    \draw[thin, red] (2,0) -- ++(-0.35,0);
    \draw[fill=white] (0,0) circle(3pt);
    \draw[fill=white] (0.5,0) circle(3pt);
    \draw[fill=white] (2,0) circle(3pt);
\end{tikzpicture})= 1$, meaning that \eqref{eq:MagicEikonalIdGeneralised} is consistent with the result in \cite{Saotome:2012vy}.

\section{Dyson, Magnus, and the Eulerian projector}
\label{sec:Dyson-Magnus-Eulerian}
The appearance of the first Eulerian projector in the main text can be understood directly from the relation between the Dyson and Magnus expansions. When interaction insertions are treated as non-commuting letters in a tensor Hopf algebra, the Dyson series is group-like, whereas its logarithm, the Magnus series, is primitive \cite{reutenauer2003free}. The first Eulerian projector implements precisely this group-like-to-primitive map.
\subsection{The Eulerian projector on group-like elements}
Let $H$ be a connected graded tensor Hopf algebra with multiplication $m$, coproduct $\Delta$, unit $u$ and counit $\epsilon$. The convolution product of two maps $f,g:H\to H$ is
 \begin{equation}   
 f\star g\coloneq m\circ(f\otimes g)\circ\Delta \,. 
 \end{equation}
The first Eulerian projector is the convolution logarithm on the identity map
\begin{equation} 
\pi_1\coloneq\log_\star(\mathrm{id}_H) =\sum_{r\geq1}\frac{(-1)^{r-1}}{r}J^{\star r}\,, \qquad J\coloneq\mathrm{id}_H-u\epsilon \,. 
\end{equation}
An element $G\in H$ is defined to be group-like if
\begin{equation}
\Delta(G)=G\otimes G\,, \qquad \epsilon(G)=1\,. 
\end{equation}
For such a group-like element,
\begin{equation}
    J^{\star r}(G)=(G-1)^r \,.
\end{equation}
Hence the action of the projector on $G$ is
\begin{equation}
\pi_1(G) = \sum_{r\geq1}\frac{(-1)^{r-1}}{r}(G-1)^r = \log G\,. 
\end{equation}
Thus, on group-like elements, the convolution logarithm reduces to the ordinary logarithm.
A short calculation reveals that the logarithm of a group-like element is primitive,
\begin{equation}
\Delta(\log G)=(\log G)\otimes1+1\otimes (\log G)\,. 
\end{equation}
In the Hopf algebra of words, primitive elements form the free Lie algebra and are therefore linear combinations of nested commutators, or Lie polynomials. The first Eulerian projector consequently maps group-like series to a primitive, or Lie, element.

\subsection{Dyson to Magnus}
Consider an interaction insertion $\mathcal{H}(t)$ valued in the word algebra, which by definition is primitive:
\begin{equation}
    \Delta \mathcal{H}(t)=\mathcal{H}(t)\otimes1+1\otimes \mathcal{H}(t)\,.
\end{equation} 
The corresponding word-valued Dyson series is
\begin{equation} 
\mathcal{S}(t,t_0) = 1+ \sum_{n\geq1}
\int_{t>t_1>\cdots>t_n>t_0} dt_1\cdots dt_n\, \mathcal{H}(t_1)\cdots \mathcal{H}(t_n)\,, 
\end{equation}
and satisfies
\begin{equation} 
\frac{d}{dt}\mathcal{S}(t,t_0)=\mathcal{H}(t)\mathcal{S}(t,t_0)\,, \qquad \mathcal{S}(t_0,t_0)=1\,. 
\end{equation}
Applying the coproduct and using that it is an algebra morphism gives
\begin{equation} 
\frac{d}{dt}\Delta(\mathcal{S}) = \bigl(\mathcal{H}(t)\otimes1+1\otimes \mathcal{H}(t)\bigr)\Delta(\mathcal{S})\,. 
\end{equation}
We can similarly examine the differential equation that the product $\mathcal{S}\otimes\mathcal{S}$ satisfies. A short calculation reveals that it satisfies the same differential equation as above. Since the initial conditions of both of these are the same, by uniqueness we must have
\begin{equation}
    \Delta(\mathcal{S})=\mathcal{S}\otimes \mathcal{S}\,.
\end{equation}
Therefore, the Dyson series is group-like. It follows immediately that the word-valued Magnus series
\begin{equation} 
\mathcal{N}\coloneq\log \mathcal{S}=\pi_1(\mathcal{S}) 
\end{equation}
is primitive and hence lies in the free Lie algebra. This is the Hopf-algebraic form of the usual passage from Dyson ordered products to Magnus nested commutators, with the group-like Dyson series mapped to the primitive Magnus series. For example,
\begin{equation}
    \mathcal{N}_1=\int dt_1\,\mathcal{H}(t_1)\,,
\end{equation}
while at second order
\begin{equation}
    \mathcal{N}_2 = \frac12 \int_{t_1>t_2} dt_1dt_2\, [\mathcal{H}(t_1),\mathcal{H}(t_2)]\,.
\end{equation}
Viewing the word-valued insertions as interaction Hamiltonians,
\begin{equation}
    \mathcal{H}(t)\longmapsto -\frac{i}{\hbar}H_I(t)\,,
\end{equation}
recovers
\begin{equation}
    \mathcal{S} \longmapsto \hat S\,, \qquad  \mathcal{N}\longmapsto \log \hat S = \frac{i}{\hbar}\hat N\,.
\end{equation}
\subsection{Connection to hyperclassical cancellations}
At fixed multiplicity $n$, the multilinear words of length $n$ containing $n$ distinct interaction insertions are naturally identified with permutations in $S_n$. The first Eulerian idempotent is then represented by
\begin{equation}
    e_1 = \sum_{\rho\in S_n} \frac{(-1)^{d(\rho)}} {n\binom{n-1}{d(\rho)}}\,\rho \,,
\end{equation}
where $d(\rho)$ is the descent number of $\rho$. With respect to the canonical inner product of words, its adjoint is
\begin{equation}
    e_1^* = \sum_{\rho\in S_n} \frac{(-1)^{d(\rho)}} {n\binom{n-1}{d(\rho)}}\,\rho^{-1}\,.
\end{equation}
Pairing the relation $\mathcal{N}=\pi_1(\mathcal{S})$ with a word $w$ gives
\begin{equation}
    \langle \mathcal{N},w\rangle = \langle \mathcal{S},\pi_1^*(w)\rangle \,.
\end{equation}
Thus a coefficient in the Magnus series is obtained from an Eulerian-weighted combination of Dyson orderings.
This is precisely the structure encountered in the Schwinger representation of Magnus amplitudes. Under the map $\mathcal{T}$ from totally ordered proper-time regions to multilinear words introduced in \eqref{eq:T-linear-map},
\begin{equation}
    \mathcal{T}(\Theta_w^{\rm M})=\pi_1^*(w)\,.
\end{equation} 
The Eulerian coefficients appearing in the dressed proper-time functions are therefore not an accidental match: they are precisely the coefficients of the Eulerian projector at fixed multiplicity, implementing the logarithmic map from the (group-like) Dyson series to the (primitive) Magnus series.
The same viewpoint also explains the cancellation of disconnected contributions. Products of independent ordered regions generate shuffle products, while the adjoint Eulerian projector annihilates shuffles,
\begin{equation}
    \pi_1^*(u\shuffle v)=0\,.
\end{equation}
The logarithmic map from group-like to primitive elements, which we have shown can be thought of as the map from $\hat S$ to $\hat N$, is also responsible for the cancellation of disconnected graphs, as well as the new formula for Murua coefficients derived in Section~\ref{Section new Murua for general}.

\section{Eulerian number identity}
\label{app:EulerianId}
We want to show that for $n\geq 2$
\begin{equation}
\label{eq:GoalEulerianSum}
    \sum_{k=0}^{n-1}(-1)^k\frac{A(n,k)}{n\binom{n-1}{k}}=0\ .
\end{equation}
We begin by recalling the generating function of the Eulerian numbers
\begin{equation}
    1+\sum_{n=1}^\infty \frac{x^n}{n!}\sum_{k=0}^{n-1}A(n,k)t^k = \frac{t-1}{t-e^{(t-1)x}}
\end{equation}
Taking a derivative with respect to $x$ on either side, and evaluating at $x=X(1-u)$ and $t=-u/(1-u)$ shows
\begin{equation}
    \sum_{n=0}^\infty \frac{X^{n}}{n!} \sum_{k=0}^{n}(-1)^{k}u^k(1-u)^{n-k}A(n+1,k) = \frac{e^{-X}}{(u+(1-u)e^{-X})^2}\ .
\end{equation}
Now, we integrate with respect to $u$ between 0 and 1. The left hand side is a simple Beta function integral,
\begin{equation}
    \int_0^1\! du\ u^k (1-u)^{n-k}= \frac{k!(n-k)!}{(n+1)!} = \frac{1}{(n+1)\binom{n}{k}}\ ,
\end{equation}
while the right hand side can be seen as a Feynman parameter integral
\begin{equation}
    \frac{1}{AB}=\int_0^1\! du\ \frac{1}{(Au+(1-u)B)^2}\ .
\end{equation}
Putting it all together,
\begin{equation}
    \sum_{n=0}^\infty \frac{X^n}{n!}\sum_{k=0}^n (-1)^k \frac{A(n+1,k)}{(n+1)\binom{n}{k}} =1\ .
\end{equation}
Comparing coefficients of $X^n$ immediately shows that, for $n>0$,
\begin{equation}
    \sum_{k=0}^n (-1)^k \frac{A(n+1,k)}{(n+1)\binom{n}{k}} = 0\ .
\end{equation}
Setting $n\to n-1$ gives \eqref{eq:GoalEulerianSum}.

\section{Proof of recursive construction of Murua coefficients}\label{app: recursionProof}
Here we prove by induction that the contraction rule combined with the identity \eqref{eq:magic-identity-from-linear-extensions} is enough to fix all Murua coefficients at tree-level recursively. The power of the contraction rule to derive Murua coefficients was already noted in the original work \cite{Kim:2024svw}, where it was combined with  their recursive Murua formula to derive an efficient method for computing tree-level Murua coefficients. Later in \cite{Guo:2026xaw} it was shown that contraction rules in the ``color'' basis were enough to fix all Murua coefficients at any loop order recursively since certain graphs in every topology are zero. An analogous statement was not true in the ``black and white'' basis since these zero graphs do not appear. Instead, in \cite{Guo:2026xaw}, one Murua coefficient was generated for each undirected topology using their full recursive formula before relating this graph to every other one using the contraction rule.
Here we show that in the ``black and white'' basis of retarded and advanced propagators the Murua coefficients can be found from the contraction rule combined with the identity \eqref{eq:magic-identity-from-linear-extensions}. 

Here we will consider only tree graphs. Let $\hat g$ be a fixed undirected tree
with $n$ vertices and $E$ edges, labelled $e_1,\ldots,e_E$. As previously mentioned we define $A(\hat g)$ as the set of all orientations of $\hat g$, of which there are $2^E$.  We will show how the Murua coefficients of all of these graphs can be uniquely derived using the coefficients of smaller graphs. 

If we fix some reference orientation $g_0$ in $A(\hat{g})$, then every other graph can be reached by flipping the edges of $g_0$. Thus, every graph can be uniquely labelled by a vector
\begin{equation}
    \mathbf b=(b_1,\ldots,b_E)\in\{0,1\}^E,
\end{equation}
where $b_i=0$ if the orientation of the edge $e_i$ in $g$ agrees with its orientation
in $g_{0}$ and $b_i=1$ if it is reversed. We write the graph associated to a vector $\mathbf b$ as $g_{\mathbf b}$. 

We can reverse the direction of an edge $e_i$ by adding $1$ to the corresponding entry in $\mathbf b$ (modulo 2) which we denote as $\mathbf b\oplus\boldsymbol{\varepsilon}_i$. 
Thus the contraction identity can be written as
\begin{equation}
\omega(g_{\mathbf{b}})
+\omega(g_{\mathbf b\oplus\boldsymbol{\varepsilon}_i})
=
c_i(\mathbf b),
\qquad
c_i(\mathbf b)
\equiv
\omega(g_{\mathbf b}/e_i).
\label{eq:tree-contraction-system}
\end{equation}
Here $g_{\mathbf b}/e_i$ is obtained by contracting $e_i$: its two
endpoints are identified and the contracted edge is removed. Since
$g_{\mathbf b}/e_i$ is a tree with fewer  vertices we know the coefficient $c_i(\mathbf{b})$ by induction. 

The contraction equations do not, in general, determine all the
coefficients uniquely. Suppose $\omega(g_{\mathbf b})$ and
$\omega'(g_{\mathbf b})$ are two solutions of
\eqref{eq:tree-contraction-system}. Their difference
\begin{equation}
h_{\mathbf b}=\omega(g_{\mathbf b})-\omega'(g_{\mathbf b})
\end{equation}
satisfies the homogeneous equations
\begin{equation}
h_{\mathbf b}
+h_{\mathbf b\oplus\boldsymbol{\varepsilon}_i}=0.
\end{equation}
and thus reversing any edge changes the sign of $h_{\mathbf b}$:
\begin{equation}
    h_{\mathbf b\oplus\boldsymbol{\varepsilon}_i}
    =-h_{\mathbf b}.
\end{equation}
Crucially this implies we can relate every $h_{\mathbf b}$ to the reference orientation $g_0$ by a simple sign
\begin{equation}
h_{\mathbf b}
= 
(-1)^{\lvert\mathbf b\rvert} h_{0},
\label{eq:tree-homogeneous-mode}
\end{equation}
where
\[
\lvert\mathbf b\rvert=\sum_{i=1}^E b_i
\]
is the number of edges whose orientations differ from the reference
orientation. Thus the kernel of the contraction equations is
one-dimensional. 
The remaining unfixed $h_0$ can be fixed using the identity \eqref{eq:magic-identity-from-linear-extensions}, which we can write in terms of the $h_{\mathbf{b}}$ as
\begin{equation}
\sum_{\mathbf b\in\{0,1\}^E}
(-1)^{\lvert\mathbf b\rvert}
\phi(g_{\mathbf b})\,h_{\mathbf{b}}
=h_{0}\sum_{\mathbf b\in\{0,1\}^E}
\phi(g_{\mathbf b})=0\,,
\label{eq:tree-magic-identity}
\end{equation}
Since the number of linear extensions of tree graphs are all positive,  this implies $h_0=0$, 
and so finally
\begin{equation}
    \omega(g_{{\mathbf b}})
=
\omega'({g_{\mathbf b}})\,.
\end{equation}
Thus the Murua coefficients are fixed uniquely.

\newpage

\bibliographystyle{JHEP}
\bibliography{ScatEq.bib}

@article{Ochirov:2026ixx,
    author = "Ochirov, Alexander and Shi, Canxin",
    title = "{Classical dynamics from QFT via Stratonovich-Weyl correspondence}",
    eprint = "2608.21328",
    archivePrefix = "arXiv",
    primaryClass = "hep-th",
    month = "8",
    year = "2026"
}

@article{Bautista:2026qse,
    author = "Bautista, Yilber Fabian and Driesse, Mathias and Haddad, Kays and Jakobsen, Gustav Uhre",
    title = "{Gravitational wave scattering in spinless WQFT}",
    eprint = "2602.06125",
    archivePrefix = "arXiv",
    primaryClass = "hep-th",
    reportNumber = "HU-EP-26/05",
    doi = "10.1007/JHEP05(2026)252",
    journal = "JHEP",
    volume = "05",
    pages = "252",
    year = "2026"
}

@article{Brunello:2026rdk,
    author = "Brunello, Giacomo and Meo, Mario and Smith, Sid",
    title = "{Gravitational Compton scattering at the fourth post-Minkowskian order}",
    eprint = "2606.28239",
    archivePrefix = "arXiv",
    primaryClass = "hep-th",
    month = "6",
    year = "2026"
}

@article{Brunello:2026lzf,
    author = "Brunello, Giacomo and Meo, Mario and Smith, Sid",
    title = "{Gravitational Compton scattering at the fifth post-Minkowskian order}",
    eprint = "2608.17946",
    archivePrefix = "arXiv",
    primaryClass = "hep-th",
    month = "8",
    year = "2026"
}

@article{Kim:2025ebl,
    author = "Kim, Joon-Hwi",
    title = "{Phase Space Formulation of S-matrix}",
    eprint = "2512.23100",
    archivePrefix = "arXiv",
    primaryClass = "hep-th",
    reportNumber = "CALT-TH 2025-033",
    month = "12",
    year = "2025"
}

@article{Guo:2026hyi,
    author = "Guo, Li and Kim, Joon-Hwi and Kim, Jung-Wook and Kim, Sungsoo and Lee, Sangmin and Li, Jian-Rong",
    title = "{A Hopf Algebraic Theory of the Quantum Magnusian}",
    eprint = "2609.33587",
    archivePrefix = "arXiv",
    primaryClass = "math.CO",
    reportNumber = "CERN-TH-2026-234, KIAS-P26047",
    month = "9",
    year = "2026"
}

@book{Bapat2010,
  author    = {R. B. Bapat},
  title     = {Graphs and Matrices},
  publisher = {Springer London},
  year      = {2010},
  isbn      = {9781848829817},
  url       = {https://books.google.co.uk/books?id=w5oXUgN5xw0C}
}

@article{Guo:2026xaw,
    author = "Guo, Li and Kim, Joon-Hwi and Kim, Jung-Wook and Kim, Sungsoo and Lee, Sangmin and Li, Jian-Rong",
    title = "{The Diagrammar of Quantum Magnusian}",
    eprint = "2605.25473",
    archivePrefix = "arXiv",
    primaryClass = "hep-th",
    reportNumber = "CALT-TH-2026-020, CERN-TH-2026-115, KIAS-P26031",
    month = "5",
    year = "2026"
}

@article{Bautista:2026fcp,
    author = "Bautista, Yilber Fabian and Driesse, Mathias and Haddad, Kays and Jakobsen, Gustav Uhre",
    title = "{Gravitational wave scattering at $\mathcal{O}(G^4)$: Murua construction and elliptics}",
    eprint = "2606.27544",
    archivePrefix = "arXiv",
    primaryClass = "hep-th",
    reportNumber = "HU-EP-26/19",
    month = "6",
    year = "2026"
}

@article{SOLOMON1968363,
title = {On the {P}oincar\'{e}-{B}irkhoff-{W}itt theorem},
journal = {Journal of Combinatorial Theory},
volume = {4},
number = {4},
pages = {363-375},
year = {1968},
issn = {0021-9800},
doi = {https://doi.org/10.1016/S0021-9800(68)80062-6},
url = {https://www.sciencedirect.com/science/article/pii/S0021980068800626},
author = {Louis Solomon}
}

@article{Arnal_2018,
    author        = {Arnal, Ana and Casas, Fernando and Chiralt, Cristina},
    title         = {A general formula for the {M}agnus expansion in terms of iterated integrals of right-nested commutators},
    journal       = {Journal of Physics Communications},
    volume        = {2},
    number        = {3},
    pages         = {035024},
    year          = {2018},
    month         = mar,
    publisher     = {IOP Publishing},
    doi           = {10.1088/2399-6528/aab291},
    eprint        = {1710.10851},
    archivePrefix = {arXiv},
    primaryClass  = {math-ph}
}

@phdthesis{Malvenuto:1993,
    author  = "Malvenuto, Claudia",
    title   = "{Produits et coproduits des fonctions quasi-sym\'etriques et de l'alg\`ebre des descentes}",
    school  = "Universit\'e du Qu\'ebec \`a Montr\'eal, LACIM Publication Vol.~16, \href{https://lacim.uqam.ca/les-parutions/LACIM-Publications-Volume-16.pdf}{available online}",
    type    = "{Ph.D. thesis}",
    address = "Montr\'eal, Canada",
    year    = "1993",
    month   = "11"
}

@article{MalvenutoReutenauer:1995,
    author = "Malvenuto, Claudia and Reutenauer, Christophe",
    title = "{Duality between quasi-symmetric functions and the Solomon descent algebra}",
    journal = "J. Algebra",
    volume = "177",
    number = "3",
    pages = "967--982",
    year = "1995",
    doi = "10.1006/jabr.1995.1336"
}

@article{Dukes:2013wa,
    author = "Dukes, Mark and Gardi, Einan and Steingrimsson, Einar and White, Chris D.",
    title = "{Web worlds, web-colouring matrices, and web-mixing matrices}",
    eprint = "1301.6576",
    archivePrefix = "arXiv",
    primaryClass = "math.CO",
    doi = "10.1016/j.jcta.2013.02.001",
    journal = "J. Comb. Theor. A",
    volume = "120",
    pages = "1012--1037",
    year = "2013"
}

@article{Dukes:2013gea,
    author = "Dukes, Mark and Gardi, Einan and McAslan, Heather and Scott, Darren J. and White, Chris D.",
    title = "{Webs and Posets}",
    eprint = "1310.3127",
    archivePrefix = "arXiv",
    primaryClass = "hep-th",
    reportNumber = "EDINBURGH-2013-27, IPPP-13-84, DCPT-13-168",
    doi = "10.1007/JHEP01(2014)024",
    journal = "JHEP",
    volume = "01",
    pages = "024",
    year = "2014"
}

@article{Gardi:2013ita,
    author = "Gardi, Einan and Smillie, Jennifer M. and White, Chris D.",
    title = "{The Non-Abelian Exponentiation theorem for multiple Wilson lines}",
    eprint = "1304.7040",
    archivePrefix = "arXiv",
    primaryClass = "hep-ph",
    reportNumber = "EDINBURGH-2013-07",
    doi = "10.1007/JHEP06(2013)088",
    journal = "JHEP",
    volume = "06",
    pages = "088",
    year = "2013"
}

@book{reutenauer2003free,
  title     = {Free Lie algebras},
  author    = {Christophe Reutenauer},
  publisher = {Clarendon Press; Oxford University Press},
  isbn      = {0198536798,9780198536796},
  year      = {1993}
}

@article{Gonzo:2026yha,
    author = "Gonzo, Riccardo and Mogull, Gustav",
    title = "{Canonical Quantisation of Bound and Unbound WQFT}",
    eprint = "2603.05237",
    archivePrefix = "arXiv",
    primaryClass = "hep-th",
    reportNumber = "QMUL-PH-26-06, HU-EP-26/13",
    month = "3",
    year = "2026"
}

@article{Brandhuber:2025igz,
    author = "Brandhuber, Andreas and Brown, Graham R. and Pichini, Paolo and Travaglini, Gabriele and Vives Matasan, Pablo",
    title = "{The Magnus expansion in relativistic quantum field theory}",
    eprint = "2512.05017",
    archivePrefix = "arXiv",
    primaryClass = "hep-th",
    reportNumber = "QMUL-PH-25-27",
    doi = "10.1007/JHEP07(2026)151",
    journal = "JHEP",
    volume = "07",
    pages = "151",
    year = "2026"
}

@article{Argeri:2014qva,
    author = "Argeri, Mario and Di Vita, Stefano and Mastrolia, Pierpaolo and Mirabella, Edoardo and Schlenk, Johannes and Schubert, Ulrich and Tancredi, Lorenzo",
    title = "{Magnus and Dyson Series for Master Integrals}",
    eprint = "1401.2979",
    archivePrefix = "arXiv",
    primaryClass = "hep-ph",
    doi = "10.1007/JHEP03(2014)082",
    journal = "JHEP",
    volume = "03",
    pages = "082",
    year = "2014"
}

@article{Blanes_2010,
doi = {10.1088/0143-0807/31/4/020},
url = {https://doi.org/10.1088/0143-0807/31/4/020},
year = {2010},
month = {jun},
publisher = {},
volume = {31},
number = {4},
pages = {907},
author = {Blanes, S and Casas, F and Oteo, J A and Ros, J},
title = {A pedagogical approach to the Magnus expansion},
journal = {European Journal of Physics}
}

@article{Kim:2025olv,
    author = "Kim, Sungsoo and Lee, Hojin and Lee, Sangmin",
    title = "{Classical eikonal in relativistic scattering}",
    eprint = "2509.01922",
    archivePrefix = "arXiv",
    primaryClass = "hep-th",
    reportNumber = "KIAS-P25044",
    doi = "10.1007/JHEP11(2025)032",
    journal = "JHEP",
    volume = "11",
    pages = "032",
    year = "2025"
}

@article{Capatti:2024bid,
    author = "Capatti, Zeno and Zeng, Mao",
    title = "{Classical worldlines from scattering amplitudes}",
    eprint = "2412.10864",
    archivePrefix = "arXiv",
    primaryClass = "hep-th",
    doi = "10.1103/PhysRevD.111.125002",
    journal = "Phys. Rev. D",
    volume = "111",
    number = "12",
    pages = "125002",
    year = "2025"
}

@article{Haddad:2025cmw,
    author = "Haddad, Kays and Jakobsen, Gustav Uhre and Mogull, Gustav and Plefka, Jan",
    title = "{Unitarity and the On-Shell Action of Worldline Quantum Field Theory}",
    eprint = "2510.00988",
    archivePrefix = "arXiv",
    primaryClass = "hep-th",
    reportNumber = "HU-EP-25/33, QMUL-PH-25-29",
    month = "10",
    year = "2025"
}

@article{Arnal:2020xpt,
    author = "Arnal, Ana and Casas, Fernando and Chiralt, Cristina",
    title = "{A Note on the Baker{\textendash}Campbell{\textendash}Hausdorff Series in Terms of Right-Nested Commutators}",
    eprint = "2006.15869",
    archivePrefix = "arXiv",
    primaryClass = "math-ph",
    doi = "10.1007/s00009-020-01681-6",
    journal = "Mediterranean J. Math.",
    volume = "18",
    number = "2",
    pages = "53",
    year = "2021"
}

@article{Murua_2006,
	author = {Murua, A. },
	date = {2006/11/01},
	doi = {10.1007/s10208-003-0111-0},
	id = {Murua2006},
	isbn = {1615-3383},
	journal = {Foundations of Computational Mathematics},
	number = {4},
	pages = {387--426},
	title = {The {Hopf} Algebra of Rooted Trees, Free {Lie} Algebras, and {Lie} Series},
	url = {https://doi.org/10.1007/s10208-003-0111-0},
	volume = {6},
	year = {2006}}

@article{Pechukas1966,
  author = {Pechukas, Philip and Light, J. C.},
  title = {On the exponential form of time-displacement operators in quantum mechanics},
  journal = {The Journal of Chemical Physics},
  volume = {44},
  number = {10},
  pages = {3897--3912},
  year = {1966},
  doi = {10.1063/1.1726550}
}

@article{Blanes2009,
  author = {Blanes, Sergio and Casas, Fernando and Oteo, J. A. and Ros, Jos\'{e}},
  title = {The {Magnus} expansion and some of its applications},
  journal = {Physics Reports},
  volume = {470},
  number = {5-6},
  pages = {151--238},
  year = {2009},
  doi = {10.1016/j.physrep.2008.11.001}
}

@article{Dyson:1949bp,
    author = "Dyson, F. J.",
    title = "{The Radiation theories of Tomonaga, Schwinger, and Feynman}",
    doi = "10.1103/PhysRev.75.486",
    journal = "Phys. Rev.",
    volume = "75",
    pages = "486--502",
    year = "1949"
}

@article{Magnus:1954zz,
    author = "Magnus, Wilhelm",
    title = "{On the exponential solution of differential equations for a linear operator}",
    doi = "10.1002/cpa.3160070404",
    journal = "Commun. Pure Appl. Math.",
    volume = "7",
    pages = "649--673",
    year = "1954"
}

@article{DiVecchia:2023frv,
    author = "Di Vecchia, Paolo and Heissenberg, Carlo and Russo, Rodolfo and Veneziano, Gabriele",
    title = "{The gravitational eikonal: From particle, string and brane collisions to black-hole encounters}",
    eprint = "2306.16488",
    archivePrefix = "arXiv",
    primaryClass = "hep-th",
    reportNumber = "CERN-TH-2023-108, NORDITA 2023-026, QMUL-PH-23-09, UUITP-14/23",
    doi = "10.1016/j.physrep.2024.06.002",
    journal = "Phys. Rept.",
    volume = "1083",
    pages = "1--169",
    year = "2024"
}

@article{Damgaard:2023ttc,
    author = "Damgaard, Poul H. and Hansen, Elias Roos and Plant\'e, Ludovic and Vanhove, Pierre",
    title = "{Classical observables from the exponential representation of the gravitational S-matrix}",
    eprint = "2307.04746",
    archivePrefix = "arXiv",
    primaryClass = "hep-th",
    reportNumber = "CERN-TH-2023-135, IPhT-T23/041, LAPTh-029/23",
    doi = "10.1007/JHEP09(2023)183",
    journal = "JHEP",
    volume = "09",
    pages = "183",
    year = "2023"
}

@article{Kalin:2019rwq,
    author = {K\"alin, Gregor and Porto, Rafael A.},
    title = "{From Boundary Data to Bound States}",
    eprint = "1910.03008",
    archivePrefix = "arXiv",
    primaryClass = "hep-th",
    reportNumber = "DESY 19-167, UUITP-40/19, DESY-19-167",
    doi = "10.1007/JHEP01(2020)072",
    journal = "JHEP",
    volume = "01",
    pages = "072",
    year = "2020"
}

@article{Mogull:2020sak,
    author = "Mogull, Gustav and Plefka, Jan and Steinhoff, Jan",
    title = "{Classical black hole scattering from a worldline quantum field theory}",
    eprint = "2010.02865",
    archivePrefix = "arXiv",
    primaryClass = "hep-th",
    reportNumber = "UUITP-37/20, HU-EP-20/22-RTG",
    doi = "10.1007/JHEP02(2021)048",
    journal = "JHEP",
    volume = "02",
    pages = "048",
    year = "2021"
}

@article{Jakobsen:2022psy,
    author = "Jakobsen, Gustav Uhre and Mogull, Gustav and Plefka, Jan and Sauer, Benjamin",
    title = "{All things retarded: radiation-reaction in worldline quantum field theory}",
    eprint = "2207.00569",
    archivePrefix = "arXiv",
    primaryClass = "hep-th",
    reportNumber = "HU-EP-22/24-RTG",
    doi = "10.1007/JHEP10(2022)128",
    journal = "JHEP",
    volume = "10",
    pages = "128",
    year = "2022"
}

@article{Landshoff:1969yyn,
    author = "Landshoff, P. V. and Polkinghorne, J. C.",
    title = "{Iterations of Regge cuts}",
    doi = "10.1103/PhysRev.181.1989",
    journal = "Phys. Rev.",
    volume = "181",
    pages = "1989--1995",
    year = "1969"
}

@article{Damgaard:2021ipf,
    author = "Damgaard, Poul H. and Plante, Ludovic and Vanhove, Pierre",
    title = "{On an exponential representation of the gravitational S-matrix}",
    eprint = "2107.12891",
    archivePrefix = "arXiv",
    primaryClass = "hep-th",
    reportNumber = "IPhT-t21/037, CERN-TH-2021-111",
    doi = "10.1007/JHEP11(2021)213",
    journal = "JHEP",
    volume = "11",
    pages = "213",
    year = "2021"
}

@article{Jakobsen:2021smu,
    author = "Jakobsen, Gustav Uhre and Mogull, Gustav and Plefka, Jan and Steinhoff, Jan",
    title = "{Classical Gravitational Bremsstrahlung from a Worldline Quantum Field Theory}",
    eprint = "2101.12688",
    archivePrefix = "arXiv",
    primaryClass = "gr-qc",
    reportNumber = "HU-EP-21/03-RTG",
    doi = "10.1103/PhysRevLett.126.201103",
    journal = "Phys. Rev. Lett.",
    volume = "126",
    number = "20",
    pages = "201103",
    year = "2021"
}

@article{Bern:2021dqo,
  author        = {Bern, Zvi and Parra-Martinez, Julio and Roiban, Radu and Ruf, Michael S. and Shen, Chia-Hsien and Solon, Mikhail P. and Zeng, Mao},
  title         = {{Scattering Amplitudes and Conservative Binary Dynamics at ${\cal O}(G^4)$}},
  eprint        = {2101.07254},
  archiveprefix = {arXiv},
  primaryclass  = {hep-th},
  reportnumber  = {CALT-TH-2021-004, FR-PHENO-2021-03, OUTP-21-03P},
  doi           = {10.1103/PhysRevLett.126.171601},
  journal       = {Phys. Rev. Lett.},
  volume        = {126},
  number        = {17},
  pages         = {171601},
  year          = {2021}
}

@article{Kosower:2018adc,
  author        = {Kosower, David A. and Maybee, Ben and O'Connell, Donal},
  title         = {{Amplitudes, Observables, and Classical Scattering}},
  journal       = {JHEP},
  volume        = {02},
  year          = {2019},
  pages         = {137},
  doi           = {10.1007/JHEP02(2019)137},
  eprint        = {1811.10950},
  archiveprefix = {arXiv},
  primaryclass  = {hep-th},
  slaccitation  = {%%CITATION = ARXIV:1811.10950;%%}
}

@article{Parra-Martinez:2020dzs,
  author        = {Parra-Martinez, Julio and Ruf, Michael S. and Zeng, Mao},
  title         = {{Extremal black hole scattering at $\mathcal{O}(G^3)$: graviton dominance, eikonal exponentiation, and differential equations}},
  eprint        = {2005.04236},
  archiveprefix = {arXiv},
  primaryclass  = {hep-th},
  reportnumber  = {FR-PHENO-2020-007, UCLA/TEP/2020/103},
  doi           = {10.1007/JHEP11(2020)023},
  journal       = {JHEP},
  volume        = {11},
  pages         = {023},
  year          = {2020}
}

@article{Saotome:2012vy,
  author        = {Saotome, Ryo and Akhoury, Ratindranath},
  title         = {{Relationship Between Gravity and Gauge Scattering in the High Energy Limit}},
  eprint        = {1210.8111},
  archiveprefix = {arXiv},
  primaryclass  = {hep-th},
  doi           = {10.1007/JHEP01(2013)123},
  journal       = {JHEP},
  volume        = {01},
  pages         = {123},
  year          = {2013}
}

@article{Kim:2024svw,
    author = "Kim, Joon-Hwi and Kim, Jung-Wook and Kim, Sungsoo and Lee, Sangmin",
    title = "{Classical eikonal from Magnus expansion}",
    eprint = "2410.22988",
    archivePrefix = "arXiv",
    primaryClass = "hep-th",
    doi = "10.1007/JHEP01(2025)111",
    journal = "JHEP",
    volume = "01",
    pages = "111",
    year = "2025"
}

\end{document}